\documentclass[useAMS, usenatbib]{mn2e}

\usepackage{journalnames}
\usepackage{amssymb}
\usepackage{epsfig}

\def\Msun{{\rm\thinspace M_{\odot}}}
\def\Mdot{\dot{M}}
\def\ergs{{\rm\thinspace ergs}}

\title[The evolution of embedded star clusters]{The Evolution of Embedded Star Clusters}
\author[F.I. Pelupessy and S. Portegies Zwart]{
 F. I. Pelupessy$^1$\thanks{E-mail: pelupes@strw.leidenuniv.nl} and 
 S. Portegies Zwart$^1$ \\
 $^1$Leiden Observatory, Leiden University, PO Box 9513,
2300 RA, Leiden, The Netherlands \\
}

\begin{document}

\date{}

\pagerange{ -- } \pubyear{2011} 

\maketitle

\begin{abstract}

We study the evolution of embedded clusters. The
equations of motion of the stars in the cluster are solved by direct
N-body integration while taking the effects of stellar evolution and
the hydrodynamics of the natal gas content into account. The gravity
of the stars and the surrounding gas are coupled self
consistently to allow the realistic dynamical evolution of the
cluster. While the equations of motion are solved, a stellar
evolution code keeps track of the changes in stellar mass,
luminosity and radius. The gas liberated by the stellar winds and
supernovae deposits mass and energy into the gas reservoir in which
the cluster is embedded. We examine cluster models with 1000 stars, but
we varied the star formation efficiency (between 0.05-0.5), cluster
radius (0.1-1.0 parsec), the degree of virial support of the initial
population of stars (0-100\%) and the strength of the feedback. We
find that an initial star fraction $M_\star/M_{\rm
tot} > 0.05$ is necessary for cluster survival. Survival is 
more likely if gas is not blown out violently by a supernova and if
the cluster has time to approach virial equilibrium during out-gassing. 
While the cluster is embedded, dynamical friction drives early and
efficient mass segregation in the cluster. Stars of $m \gtrsim
2\,M_\odot$ are preferentially retained, at the cost of the loss of less
massive stars. We conclude that the degree of mass segregation in
open clusters such as the Pleiades is not the result of secular
evolution but a remnant of its embedded stage.

\end{abstract}


\section{Introduction}

The loss of gas that signifies the end of star formation in an
embedded proto-cluster may also cause the young cluster to dissolve
\citep{Hills1980, Lada1984}.
The energy output from all the massive stars typically exceeds the
total binding energy of the embedded star cluster. The survivability
of the cluster therefore depends on the efficiency with which the
radiative, thermal and mechanical energy of the stellar outflows
couple to the inter-cluster gas \citep[and as long as this energy is 
not carried away by outflows, ][]{Dale2005}. The time scale on 
which gas is removed from the embedded cluster (and any spatial dependence 
of this process) affects the mass and number of stars in the surviving 
cluster, but also its
degree of mass segregation and the density profile. The parameters
that strongly affect the outcome of the out-gassing phase are the star
formation efficiency  \citep[SFE, which by itself could depend on  
feedback processes, ][]{Dale2008,Vazquez-Semadeni2010}, the efficiency of 
the radiative coupling and the population of the most massive stars in the proto cluster.
Traditionally the critical value for the star formation
efficiency is taken to be ${\rm SFE} \approx 0.5$ on the 
basis of virial arguments \citep[e.g.][]{Hills1980,
Mathieu1983}, but later work showed that SFEs as low as 0.1 can still lead
to bound clusters, albeit with a reduced mass \citep{Lada1984, Adams2000,
Geyer2001, Baumgardt2007, Goodwin2009}. 
Observed star formation efficiencies fall into the range 0.05-0.3
\citep{Williams1997}. It is therefore expected that clusters must 
experience considerable structural disruption before settling.

From an observational point of view, the number of embedded clusters is 
too high with respect to the number of observed gas-free clusters. Within 
the solar neighborhood  the observed distribution of cluster ages suggests
that more than 95\% of embedded clusters dissolve into the field within  100
Myr \citep{Lada2003}.  The disruption of young clusters by the loss of the
remnant gas provides a natural explanation, although alternatively the 
young clusters may be disrupted by tidal forces \citep{Elmegreen2010,
Kruijssen2011}. In this scenario it is not the internal evolution of the
cluster that disrupts the structure, but tidal shocks from passing gas
clouds   \citep[see also][]{Gieles2006}. In order to find out  what the
relative contribution is from these two processes we need to understand the
effects of gas loss quantitatively. Ultimately the modeling of the formation
of cluster populations is needed to interpret the observations as they play
a pivotal role in galaxy evolution studies  \citep[e.g.][]{Schweizer1998,
Larsen2001, Bastian2005}. Without understanding early cluster evolution we 
cannot relate the young cluster  populations (which are easily observed to
large distances)  to the older stars and the stellar population in the
field. Contrary to stars, clusters are sensitive to the
galactic environment and if we would understand the starting conditions of
the cluster population we could search for the signatures of later galactic
evolutionary events such as galaxy mergers and major gas inflows.

The expulsion of gas from young clusters has been studied by applying a
time-dependent background potential within which the N-body system is
integrated \citep{Lada1984, Geyer2001, Boily2003b, Baumgardt2007, Chen2009,
Goodwin2009, Smith2011}. The time scale for out-gassing can then be controlled
rather straightforwardly by the decay time scale of the background potential. 
This parametrization allows the study of different
mass loss scenarios without the complications of modeling gas dynamics and
feedback. \cite{Geyer2001} examined the evolution of a cluster both using
the external potential approximation and for a limited set of combined
N-body/ SPH simulations (using simplified feedback prescriptions). They found general
agreement between the simulations with a dynamic gas component and a static gas
potential, and found that relatively long timescales of  gas loss are necessary for
clusters to survive. To reconcile observed  low SFEs with their simulations,
they point out that  although the global SFE may be low in a star forming
region, locally the  SFE may be much higher and allow  for bound clusters
\citep[also][]{Adams2000}. Furthermore they proposed  sub-virial stellar
velocity dispersions as an alternative possibility  \citep[see
also][]{Goodwin2009}. 

\cite{Baumgardt2007} conducted a grid of simulations in which they varied
the SFE, gas expulsion timescale and external tidal field. They confirmed 
that clusters with a SFE as low as 10\% could survive, although they found a 
strong dependence on the timescale (parametrized by an e-folding time) on 
which the gas potential was removed. Surviving clusters had expanded 
drastically; for SFEs of around 25\% this  was a factor 3-4, nearly independent of the gas
loss timescale. The velocity distributions became strongly radially 
anisotropic in the cases where the gas was removed on timescales much 
shorter than a crossing time. 

It has become clear from observations and simulations that young stars 
start out in a structured, clumpy, state \citep{Lada2003,
PortegiesZwart2010}.  Recent work examining hierarchical cluster formation
\citep{Smith2011, Kruijssen2011} suggests that sub-clustering increases the 
survivability. If clusters tend to be born sub-virial, they can survive even 
the loss of large fractions of intercluster gas, in which case the SFE  may
not be a good indicator of survivability.  \citet{Smith2011} \citep[see
also][]{Kruijssen2011} found that the local gas fraction immediately prior
to gas loss is a better diagnostic for the cluster survival than the global 
SFE. By studying sub-clusters in hydrodynamical simulations of collapsing 
molecular clouds where sink particles represent stars, \citet{Kruijssen2011}
found that the environment around $\sim 40\,\Msun$ clumps has a gas content of
only 10\%-30\%. The
relative lack of gas resulted from the accretion of gas onto stars (which
depletes the gas reservoir) and the shrinkage of the virialized sub clumps 
(because the dynamics of these clumps decouples from the gas dynamics). 
For such low gas fractions no major disruption is expected when feedback 
clears the gas. On the larger scales gas fractions are still considerably 
higher. \cite{Kruijssen2011} argues that a similar effect will also cause
the larger scales to become gas poor, if sufficient time elapses before
the onset of feedback. However, during this delay the different clumps 
would merge and relax, evolving from the highly structured state towards 
a smooth density distribution (this was also the case for the simulations for 
\cite{Smith2011}, where this happened within 3 Myr). The 
conditions at this stage would determine the outcome with regard to the 
final survival, and by this stage the local depletion of gas by accretion
may no longer be present. 

The above studies have expanded considerably on the simple picture where 
clusters dissolution is determined solely or mainly by the single SFE
parameter. Clusters seem to have a number of ways in which they can resist
the disruptive  effect of the loss of their gas content: disruption is less
or absent if the gas loss occurs on timescales longer than the crossing time
or if the gas loss is delayed enough that shrinking due to relaxation
processes can occur. On the other hand, if the stellar population of 
the cluster forms sub-virialized, some of the stars in the cluster 
can evolve into low energy orbits and these will not be ejected as easily, 
even if the SFE is low. Hence, to constrain the disruption process of clusters, 
we need to employ numerical simulations to unravel various complex physical 
processes at work. The determination of the gas loss timescale and the 
spatial distribution requires careful modeling of the response of the 
parent gas cloud to the mechanical luminosity of the young stars using 
realistic stellar evolution models (the feedback from radiation, 
stellar winds and supernovae will be neither constant nor instantaneous) coupled 
to a hydrodynamic code. Following the evolution of the stellar component 
requires that the stellar dynamics be resolved in high detail, including 
two-body relaxation processes. For this an N-body code for collisional 
gravitational dynamics is necessary. The relaxation processes also require 
that a realistic initial mass function (IMF) is adopted, which is also 
required for the stellar evolution modeling. The stellar dynamics 
and hydrodynamics code must be mutually coupled as they feel each other's 
gravity. Here we will present simulations that were
conducted using the  newly developed Astrophysical Multi-purpose Software
Environment \citep[AMUSE;][]{PortegiesZwart2009, Pelupessy}
that couple stellar dynamics of young clusters with the gas dynamics using
realistic feedback physics. Using these simulations we will explore
idealized but realistic scenarios of gas loss and analyze the trends with 
SFE, feedback strength and the dependence on the initial distribution. We
will start by presenting the AMUSE package, the solvers and
initial conditions of the runs we conducted in Section \ref{sec_method}. 
The results are presented in Section \ref{sec_results} and discussed  in
\ref{sec_disc}. We summarize our main conclusions in
Section~\ref{sec_concl}.

\section{Method}
\label{sec_method}

We study the evolution of embedded star clusters using a hybrid
simulation environment. Our models combine the effects of the
self-gravity of the stars, their nuclear evolution and the
hydrodynamics of the gas. The latter has contributions of the
primordial gas content and of the gas liberated by the stars in stellar
winds and supernovae. The various simulation ingredients are coupled
self-consistently using AMUSE \citep[][]{PortegiesZwart2009, Pelupessy}. 
In the code, a script in the AMUSE framework, we couple a collisional 
N-body code, a Smoothed Particles Hydrodynamics (SPH) code and a stellar 
evolution code together to form a complete description
of the physics that govern the evolution of an embedded star cluster.
The integration of the equations of motion of the stars was conducted
using the fourth-order predictor-corrector Hermite N-body solver
phiGRAPE \citep{Harfst2007}.  The dynamics of the gas was resolved
using the SPH code Gadget-2 \citep{Springel2005}. For the gravitational 
coupling between the stars and the self-gravity of the gas we used the 
gravitational treecode Octgrav \citep{Gaburov2010}. The stars in our 
simulations were evolved using parametrized stellar evolution tracks
\citep[using the SSE code,][]{Hurley2000}.  Each of the numerical 
simulation codes is embedded in the AMUSE framework, and the code
that realizes the self-consistent coupling between each of the
physical sub-domains turns out to be a relatively straightforward
script. Within the AMUSE environment the individual sub-solvers are
referred to as \emph{community codes}. The algorithmes used
by these community codes remain the same as described by their respective 
authors. The codes are written in different 
languages  and have a wide range of programming styles. Because the adopted 
community codes are untouched, we will not further explain their operations, 
but simply refer the reader to the appropriate literature. The AMUSE framework
that stitches each of the community codes together is one of the major 
ingredients that enables us to perform a consistent coupling between the 
various scales and physical domains. Therefore we give a short overview
of the framework in the following section.

\subsection{AMUSE}
\label{sec_AMUSE}

The AMUSE environment that we use to conduct the simulations is a package 
that allows astrophysical codes from different domains to be combined to
conduct numerical experiments (see Pelupessy et al. 2011  for a more
complete description). AMUSE is a development of the  Multi-physics and
Multi-scale Software Environment  \citep[MUSE, ][]{PortegiesZwart2009} and is
freely available for download\footnote{\texttt{www.amusecode.org}}. The
fundamental idea of AMUSE is the abstraction of the functionality of simulation
codes into physically motivated interfaces that hide the complexity and
numerical implementation of the codes. AMUSE presents the user with building
blocks that can be combined into applications and numerical
experiments. The binding language that stitches the codes together is
Python\footnote{\texttt{www.python.org}}, as the focus in these high level
interactions is not so much performance (the computational cost being 
concentrated in the component codes) but on algorithmic flexibility and
ease of programming to allow rapid prototyping. 

\begin{figure*}
\centering
\includegraphics[width=0.99\textwidth]{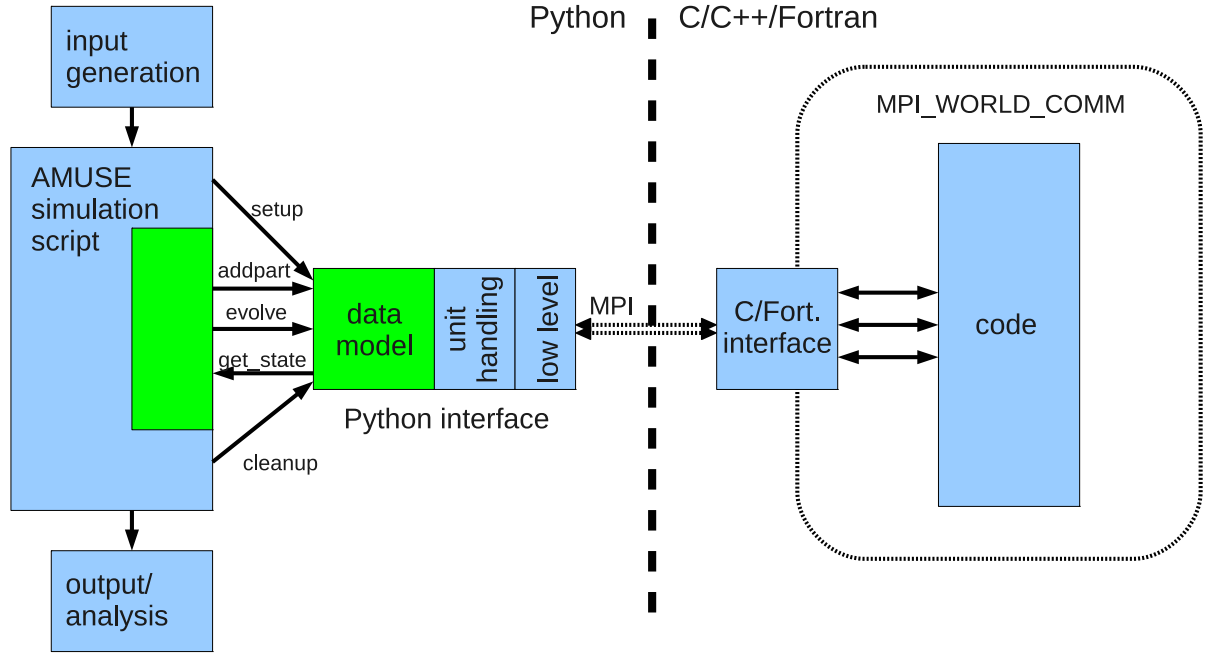}
\caption{The AMUSE interface design. This diagram represents the way in
which a community code (``code'') is accessed from the AMUSE framework. The
code has a thin layer of interfaces functions in its native language which
communicate through an MPI message channel with the python host process. On
the python side the user script (``AMUSE simulation script'')
only accesses generic calls (``setup,'' ``evolve'' etc.) to a high level 
interface. This high level interface calls the low level interface functions,
hiding details about units and the code implementation (together these form
the \emph{community module} of the code). The communication
through the MPI channel does not interfere with the code's own
parallelization.}
\label{fig_interface}
\end{figure*}

An AMUSE application consists, roughly speaking, of a \emph{user
script}, one or more \emph{community modules} and the \emph{community code} 
(fig \ref{fig_interface}). The user script specifies the initial conditions
and selects the simulation codes. The relationships between the community codes
define the solver and for our case we construct a combined  N-body/hydrodynamic
solver using a direct N-body code and an SPH solver. 
The setup and communication with a community code is handled by
the community module. This consists of an 
MPI\footnote{\texttt{http://www.mpi-forum.org/}} based communication
interface with the code as well as unit handling facilities and an
object oriented data model. The heart of an AMUSE application consists of
the solvers employed: here these are the Gadget-2 SPH code
\citep{Springel2005}, the PhiGRAPE Hermite N-body code \citep{Harfst2007},
the tree gravity code Octgrav \citep{Gaburov2010} and the stellar evolution 
code SSE \citep{Hurley2000}.

\subsection{The combined solver}
\label{sec_solver}

\begin{figure}
\centering
\includegraphics[width=.5\textwidth]{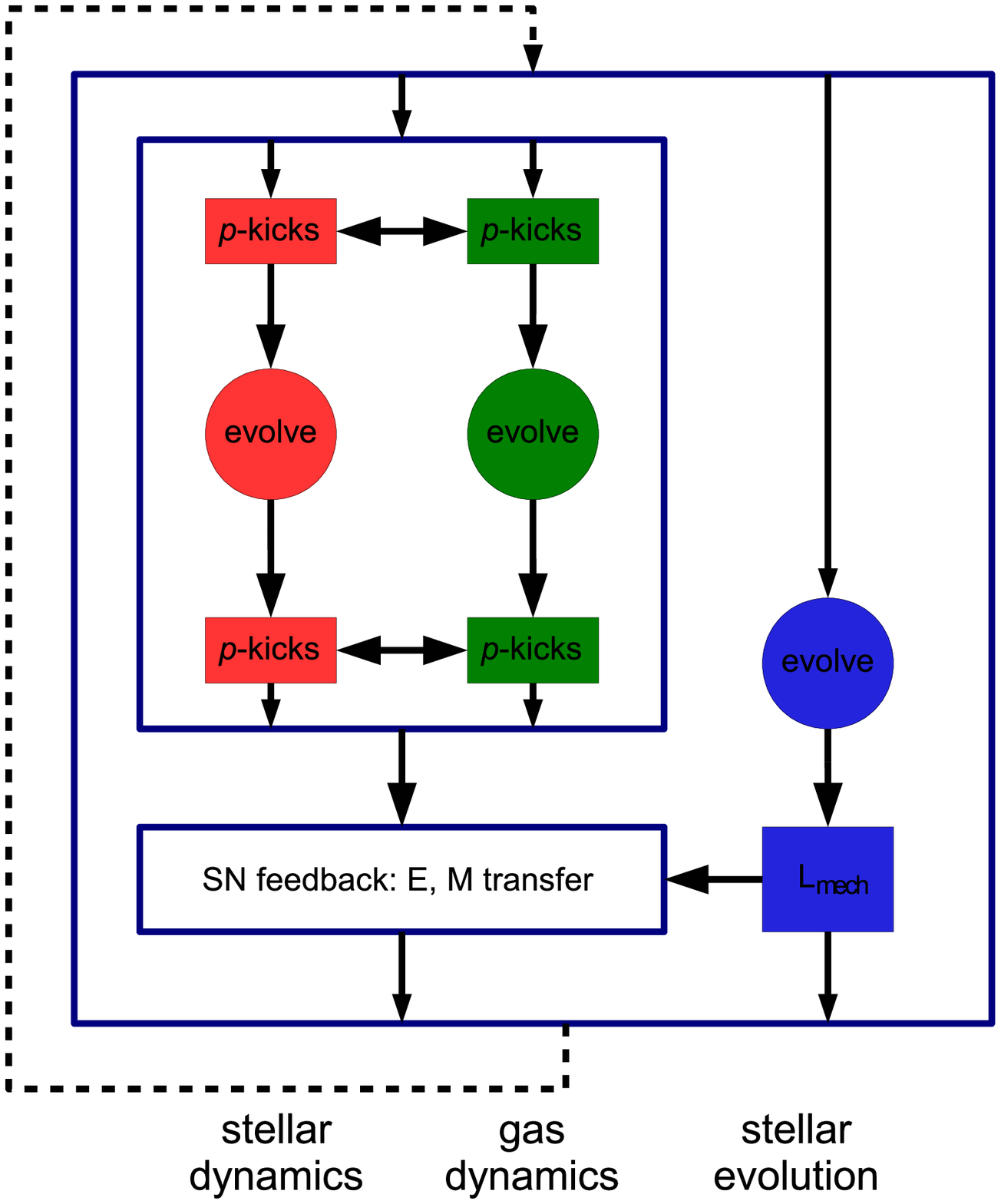}
\caption{The AMUSE gravitational/hydro/stellar evolution integrator. 
This diagram shows the calling sequence of the different AMUSE elements in
the combined gravitational/hydro/stellar solver during a time step of the 
combined solver. Circles indicate calls to the (optimized) component solvers, 
while rectangles indicate parts of the solver implemented in Python 
within AMUSE. }
\label{fig_int}
\end{figure}

An overview of the calling sequence of our combined solver is given in
figure~\ref{fig_int}. It consists of a combined integrator for coupled
gas/gravitational dynamics systems and a feedback prescription for
mechanical energy input.

\subsubsection{BRIDGE integrator}

The BRIDGE integrator \citep{Fujii2007} provides a semi-symplectic
mapping for the gravitational evolution in cases were the dynamics of
a system can be split in different regimes. A typical application
would be a dense star cluster in a galaxy where the internal dynamics
of the former evolves on a relatively short timescale compared to the
dynamics of the host galaxy.  A similar idea was applied by
\cite{Saitoh2010} by splitting the gravitational and hydrodynamic
evolution operators for simulating gas-rich galaxy mergers. They
implemented the algorithm in a single monolithic code, whereas we adopt
the concept of operator splitting within AMUSE to couple different
codes. In our case we will also employ a split between the gravitational
evolution operator of the stellar component (which is governed by
collisional dynamics) and the hydrodynamic evolution operator (governed by
SPH dynamics).

The BRIDGE formulation is derived by applying the splitting  method
developed in planetary dynamics \citep{Wisdom1991,Duncan1998}  to the
Hamiltonian 
\begin{eqnarray}
\label{eq:H}
H & = & \sum_{i \in A \cup B} \frac{p_i^2}{2 m_i} + 
        \sum_{i \ne j \in A \cup B} \frac{G m_i m_j}{\| {r_i-r_j} \| }     
\end{eqnarray}
of a system of particles which consists of subsystems A and B 
by dividing the Hamiltonian into three parts
\begin{eqnarray}
\label{eq:splitH}
H & = & \sum_{i \in A} \frac{p_i^2}{2 m_i} + 
        \sum_{i \ne j \in A} \frac{G m_i m_j}{\| {r_i-r_j} \|} + \nonumber \\
  & &   \sum_{i \in B} \frac{p_i^2}{2 m_i} +
        \sum_{i \ne j \in B} \frac{G m_i m_j}{\|{r_i-r_j} \|} + \\
  &   & \sum_{i \in A, j \in B} \frac{G m_i m_j}{\| {r_i-r_j} \| } \nonumber \\ 
  & = & H_A + H_B + H^{\rm int}_{A,B}    
\end{eqnarray}
where $H_A$ and $H_B$ are the Hamiltonians of subsystems A and B
respectively and the cross terms are collected in $H^{\rm int}$. The formal 
time evolution operator of the system can then be written as ($\mathbb{H}$
is the Hamiltonian vector field corresponding to a Hamiltonian $H$):
\begin{eqnarray}
e^{\tau \mathbb{H}} & = &
e^{\tau/2 \mathbb{H}^{\rm int}} e^{\tau \left(\mathbb{H}_A+\mathbb{H}_B\right)} e^{\tau/2 \mathbb{H}^{\rm int}} \nonumber \\
& = & e^{\tau/2 \mathbb{H}^{\rm int}} e^{\tau \mathbb{H}_A} e^{\tau \mathbb{H}_B} e^{\tau/2 \mathbb{H}^{\rm int}}.
\end{eqnarray}
The $e^{\tau/2 \mathbb{H}^{\rm int}}$ operator represents momentum
kicks because $H^{\rm int}$ depends only on the position of the particles. 
The operator  $e^{ \tau (\mathbb{H}_A + \mathbb{H}_B)}$ splits 
into secular evolution operators, because $H_A$ and $H_B$ are  independent.  The
evolution of the total system over a time step $\tau$ can subsequently be
calculated from the mutual forces exerted by systems A on B and by kicking
the momenta for time $\tau/2$, after which the two systems are evolved in
isolation for a time $\tau$, after which the time step is finished by another
mutual kick. This procedure is beneficial if, for example, the time step 
allowed by the mutual interactions is considerably longer than that 
required by the internal dynamics of the component systems, or if the 
gravitational dynamics of the two sub-systems need different solving strategies.

Within AMUSE we view the various gravitational and hydrodynamical
codes as implementations of the time evolution operators $e^{\tau \mathbb{H}}$
(for this the interface specification provides an {\tt evolve}
function). The momentum kicks are easy and reasonably fast to
implement within the framework in Python once the forces are known,
and for this the gravitational dynamics interface provides utility
functions to query the gravitational forces at arbitrary
positions. 

The gas dynamical SPH evolution can also be derived from a
Hamiltonian formalism and thus the above directly carries over to a split
between purely gravitational and SPH particles. The application of a
BRIDGE type integrator to SPH is computationally efficient if the
time step requirement for the hydrodynamics is more stringent than for
the gravitational dynamics \citep[which generally is the case,][]{Saitoh2010}. 
Our application within AMUSE has the additional benefit that we can 
easily swap in and out different gravitational solvers and SPH codes, even
if they use completely different algorithms (such as in our case a direct 
Hermite code and a tree based SPH integrator).  

For validation, the resulting combined gravitational-hydrodynamical solver
was checked in problems which could also be calculated using a monolithic SPH
code. For this a number of static and dynamic test runs were conducted and
the results where verified to be consistent with runs conducted purely with 
Gadget-2. First, in order to  check that the  alternation of hydrodynamic and
gravitational evolution operators does not introduce artifacts in cases
where gravitational and hydrodynamic forces are in equilibrium, we conducted
static tests where the ability of the code to maintain a Plummer sphere in
equilibrium was examined. Tests where conducted using different particle
numbers (10k-200k) and gas fractions  (50\% and 100\%). As a dynamic test we
conducted the \emph{Evrard}  \citep{Evrard1988} collapse tests with the
combined AMUSE solver. Both the dynamic and static tests 
showed satisfactory results compared to a conventional 
integrated code \citep[][]{Pelupessy}.

\subsubsection{Feedback prescription}

In our calculations we include feedback from stellar wind and
supernovae. Radiative processes are not taken into account
self-consistently, even though the ionizing
radiation from massive stars can drive violent outflows
\citep{RodriguezGaspar1995,DiazMiller1998, Dale2005}; the relative strength of
the radiative feedback is expected to be of the same
order of magnitude as the mechanical feedback from stellar winds
\citep{Fall2010}, which is included in our calculations.

Given these uncertainties (and the fact that we do not explicitly include
radiative feedback) we have chosen to simplify the modeling by taking
an adiabatic equation of state and not solving the thermal and chemical
balance of the gas.  We parametrize the energy budget in the absence of 
radiative losses with a coupling strength parameter $f_{fb}$, which is 
the fraction of the total supernova and wind energy output that ends 
up as thermal energy in the ISM. The relative strength of the feedback 
is of the order of a few percent, the rest is radiated away 
\citep{Wheeler1980, Silich1996, Freyer2003, Freyer2006}. 

For the actual implementation of the feedback within our model we
calculate the mechanical luminosity $L_{\rm mech}=\Mdot v_\infty^2$ of
the stars, from parametrized mass loss rates and terminal wind
velocities \citep{Leitherer1992, Prinja1990}.  The mass liberated in
the stellar wind of the stars is subsequently injected into the
cluster medium. All gas particles have the same mass $m_{\rm gas}$
throughout the simulation. The gas particles are released 
 at rest relative to the star releasing them in a sphere with a 
radius equal to the smoothing length of the gas particles around the star, 
and the stellar mass is reduced accordingly. 
A consequence of this prescription is that each star loses mass in 
discrete steps of at least $m_{\rm gas}$. For the runs
presented here the $m_{\rm gas}$ varies between $\approx 3\times 10^{-3} - 6
\times 10^{-2} \Msun$.  As a consequence the stellar wind feedback
will exhibit some discreteness noise if the mass loss rate of a star is lower 
than $\sim 7.5\times10^{-8} - 1.5\times 10^{-6} \Msun/ {\rm yr}$ (Note that the
 feedback will still be approximately isotropic if a low number of particles
 is released due to the SPH smoothing). We 
verified that the results of our simulations with respect to the stellar 
response  converges and therefore are not affected by this discretization. 
 For the most active wind phases ($\dot{M}\gtrsim 10^{-5} \Msun/{\rm yr}$)
the feedback prescription leads to 10s-100s of particles released per
timestep and a smooth wind (see section \ref{sec_results}).
The internal energy $u$ of the gas lost by the star is calculated from the
total energy output of the star $E_{\rm mech}$ since the last discrete 
mass loss event:
\begin{equation}
       u = f_{\rm fb} \frac{E_{\rm mech}}{m_{\rm grav}- m_{\rm evo}}
         = f_{\rm fb} \int_{t_{\rm last}}^{t} L_{\rm mech} 
           / (m_{\rm grav}- m_{\rm evo}) dt
\end{equation} 
 here $m_{\rm grav}$ is the dynamical mass 
and $m_{\rm evo}$ the mass calculated by the stellar evolution code (so 
reduced by the mass loss).
In case that a star experiences a supernova explosion we inject
$E_{\rm sn}=10^{51} \ergs$ in the escaping particles by added this
amount of energy to $E_{\rm mech}$. The feedback efficiency parameter
$f_{\rm fb}$ that accounts for the radiative losses is a free
parameter, which we varied between $f_{\rm fb}=0.01$ and $0.1$.
In order to track the rapid evolution caused by feedback we adopt a
maximum time step for the gas particles of $10^3$ years and after supernova
events the maximum hydrodynamic time step is reduced to 20 years for 30k 
years in order to prevent time-stepping artefacts \citep[a more elegant fix
is described in][but our approach has only a minor effect on the total 
running time]{Saitoh2009, Durier2011}. We summarize the various numerical 
parameters in Table~\ref{table_par}.

\begin{table*}
\centering
\begin{minipage}{140mm}
\caption{
Overview of the numerical parameters: $\Delta t_{\rm gas}$ is the
maximum time step for hydrodynamics, $\Delta t_{\rm fb}$ defines the interval
between feedback, $\Delta t_{\rm bridge}$ is the coupling time for mutual
gas - star gravity kicks, $t_{\rm sim}$ is the total simulated time, 
$\epsilon_{\rm soft}$ is the gravitational smoothing adopted  (for star-star
interactions. Interactions with gas particles are smoothed using the local
smoothing length), $\eta$ is the
usual N-body time step parameter for PhiGRAPE \citep{Aarseth2003} and 
$\theta$ the \citet{BarnesHut1986} opening angle for Octgrav.}
\label{table_par}
\begin{tabular}{ l | c  c  c  c  c  c  c}

\hline
\hline
parameter:  & $\Delta t_{\rm gas}$&  $\Delta t_{\rm fb}$  &  $\Delta t_{\rm bridge}$  &  
 $t_{\rm sim}$  &  $\epsilon_{\rm soft}$  &  $\eta$  &  $\theta$  \\
value:      & 0.001 Myr & 0.04 Myr             &  0.04 Myr                 & 
 30 Myr         &  0.001 $R_{\rm scale}$  &  0.01  & 0.5    \\
\hline
\hline

\end{tabular}
\end{minipage}
\end{table*}

\subsection{Initial conditions}
\label{sec_ic}

\begin{table*}
\centering
\caption{Overview of initial conditions.} 
\label{table_ic}
\begin{tabular}{ l | c  c  c  c  c  c  c  c l}

\hline
\hline
Model & $N_\star$ & SFE  & $M_\star$ & $N_{\rm gas}$  & $M_{\rm gas}$ & $R_{\rm scl}$ & $f_{\rm fb}$ & comment \\
      &           &      & ($\Msun$) &                & ($\Msun$)     & (pc)          &              &         \\
\hline
A1    & 1000      & 0.05 & 355       & 100k           & 6740          & 0.5           & 0.1          & \\             
A2    & 1000      & 0.3  & 350-366   & 100k           & 815-855       & 0.5           & 0.1          & $\times 4$\\
A3    & 1000      & 0.5  & 355       & 100k           & 355           & 0.5           & 0.1          & \\                     
A4    & 1000      & 0.05 & 355       & 100k           & 6740          & 0.5           & 0.01         & \\                     
A5    & 1000      & 0.3  & 355-366   & 100k           & 815-855       & 0.5           & 0.01         & $\times 4$\\                     
A6    & 1000      & 0.5  & 355       & 100k           & 355           & 0.5           & 0.01         & \\                     
A7    & 1000      & 0.3  & 355       & 100k           & 840           & 0.2           & 0.1          & \\                     
A8    & 1000      & 0.3  & 355       & 100k           & 840           & 0.2           & 0.01         & \\                     
A9    & 1000      & 0.3  & 355       & 100k           & 840           & 1.0           & 0.1          & \\                     
A10   & 1000      & 0.3  & 355       & 100k           & 840           & 1.0           & 0.01         & \\                     
B1    & 1000      & 0.05  & 355   & 100k           & 6740         & 0.5          & 0.1           & sub-virialized\\
B2    & 1000      & 0.05  & 355   & 100k           & 6740         & 0.5           & 0.01          & sub-virialized\\
B3    & 1000      & 0.05  & 355   & 100k           & 6740       & 1.           & 0.1           & sub-virialized\\
B4    & 1000      & 0.05  & 355   & 100k           & 6740       & 1.           & 0.01           & sub-virialized\\
\hline
\end{tabular}
\end{table*}

The simulated clusters in Table\,\ref{table_ic} are composed of a
mixture of gas and stars (with $N_\star=1000$) distributed in a Plummer 
sphere.  Stellar-stellar gravitational interactions are smoothed (Plummer
smoothing) with a smoothing length $0.001 \times R_{\rm scale}$ and the 
interaction between gas particles and star particles are smoothed 
(using the SPH spline kernel) with the gas particle smoothing length. 
 The stellar masses are assigned using a 
Salpeter IMF between 0.1 and 100 $\Msun$, with the additional constraint 
that the most massive star in each cluster has a mass close to the 
median maximum mass for a cluster of this size ($\approx 22 \Msun$).
This additional requirement is consistent with the expected maximum mass
for small clusters \citep{Weidner2006, PortegiesZwart2010}. The masses of 
the stars are assigned independently of the position in the cluster
(constant SFE), so no initial mass segregation is imposed.
 Stellar collisions are not taken into account and during the 
simulation no gas accretion unto existing stars or new star formation
is allowed. Note that any accretion of gas onto the stars would be very 
limited since we start from a smooth gas distribution and no cooling is 
included. 
Feedback induced star formation can also not occur in our approach, 
but is not very likely in the systems we simulate \citep[see also][]{Dale2008}.

In the series of calculations indicated with the letter ``A'' in
Table\,\ref{table_ic} we explore the effect of the formation
efficiency, the initial scale length and the feedback efficiency. We
run models with a SFE of 0.05, 0.3 and 0.5. The initial
Plummer radius of the initial conditions is either 0.1, 0.5 or 1.0\,pc.
Simulations with a feedback
efficiency of $f_{\rm fb}=0.01$ and $0.1$ are performed. For small 
clusters the number of high mass stars can vary quite substantially 
between different realizations of the IMF, and we perform the model 
A2 and A5 with realizations of the IMF generated from different random 
seeds to examine this effect. We have performed 
additional simulations in which we varied the number of gas 
particles to tests that the results are independent of the resolution of 
the gas dynamics.

We also run a series of models in which the stars are dynamically cold
(these models with a sub-virialized velocity dispersion 
are designated as model set ``B'' in Table~\ref{table_ic}). This choice 
is motivated by the fact that the initial stellar velocity dispersion
comes from turbulent velocities in the parent cloud \citep{Geyer2001}.
The stellar component may form with a lower velocity dispersion than that 
required to be dynamically stable if the turbulence in the parent gas cloud 
has a scale length smaller than the Jeans length or if the gas cloud
is (partially) pressure supported or partially supported by magnetic 
fields \citep{Verschueren1990}. For this set of simulations we decrease the
stellar velocities to $20\%$ of the virialized values.

\section{Results}
\label{sec_results}

\begin{figure*}
\includegraphics[ height=.31\textwidth]{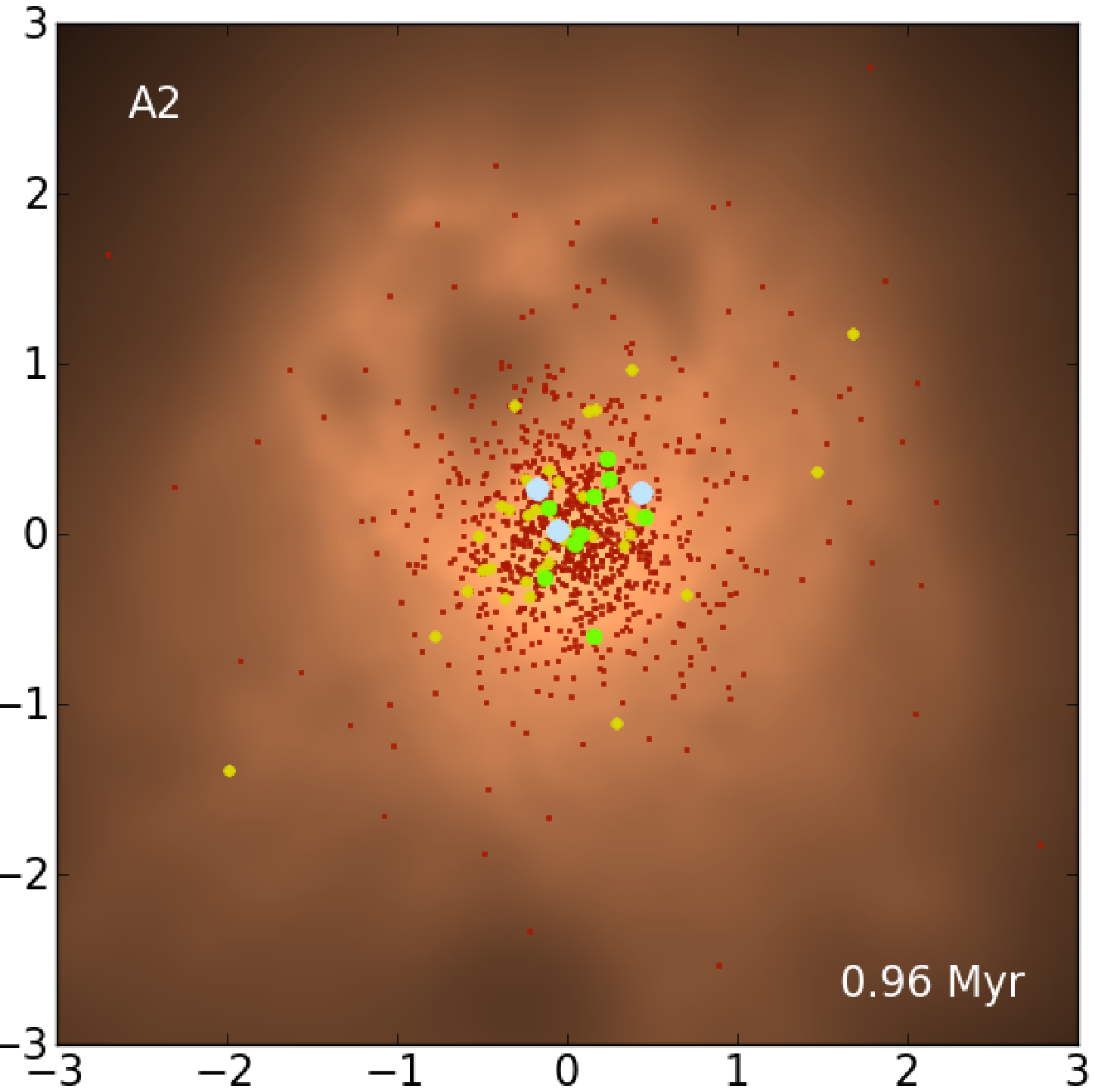}
\includegraphics[ height=.31\textwidth]{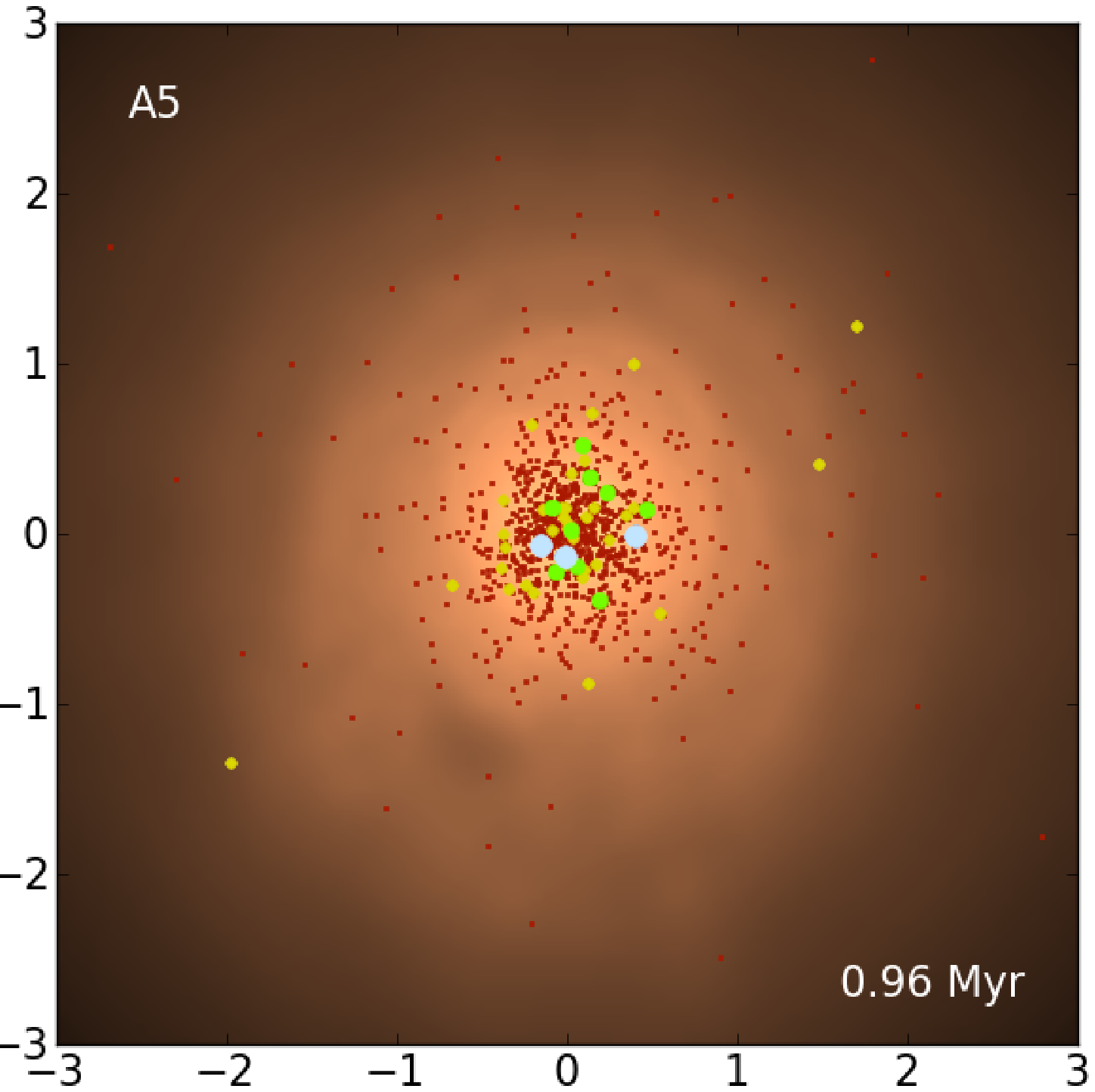}

\includegraphics[ height=.31\textwidth]{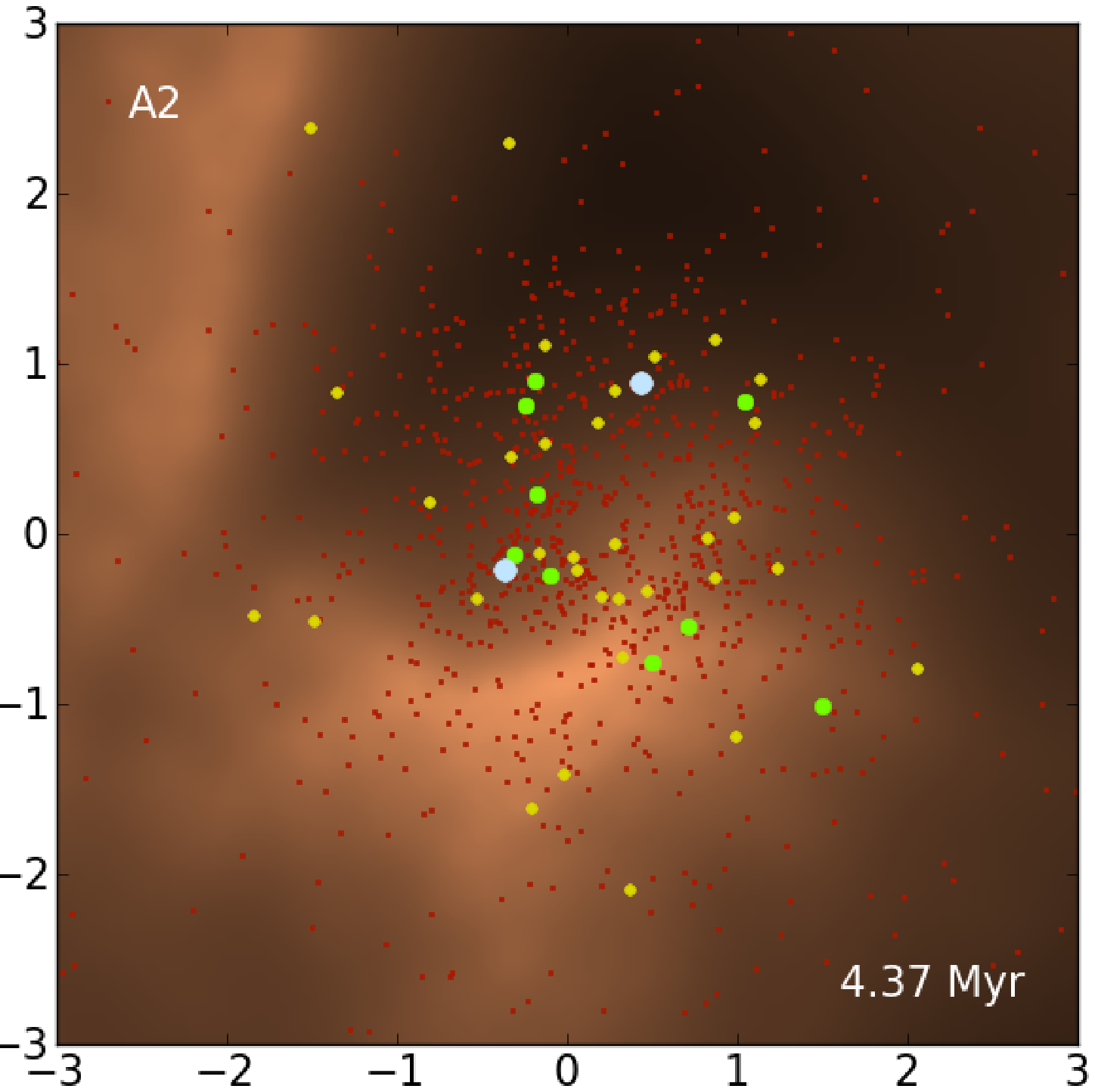}
\includegraphics[ height=.31\textwidth]{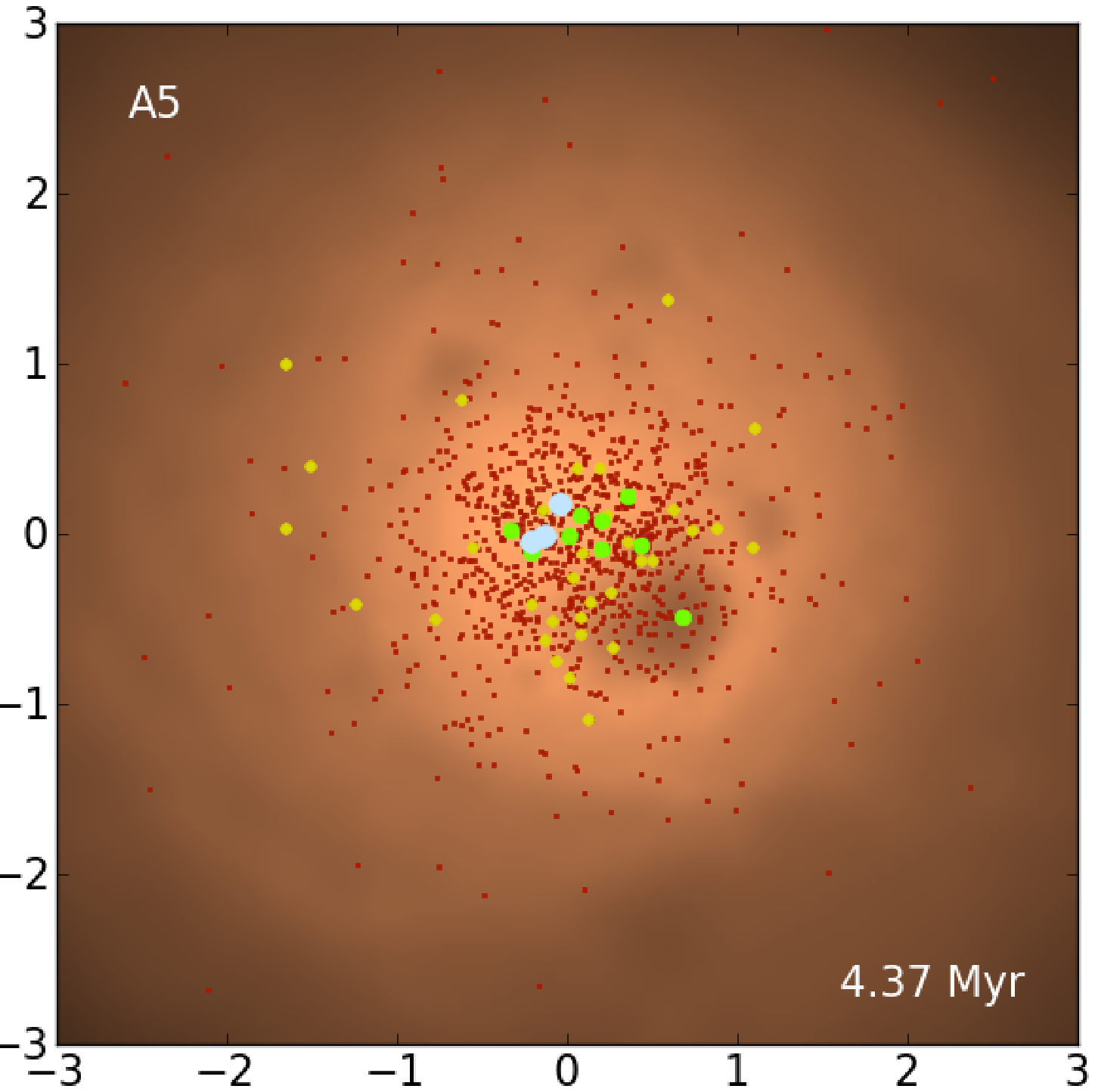}

\includegraphics[ height=.31\textwidth]{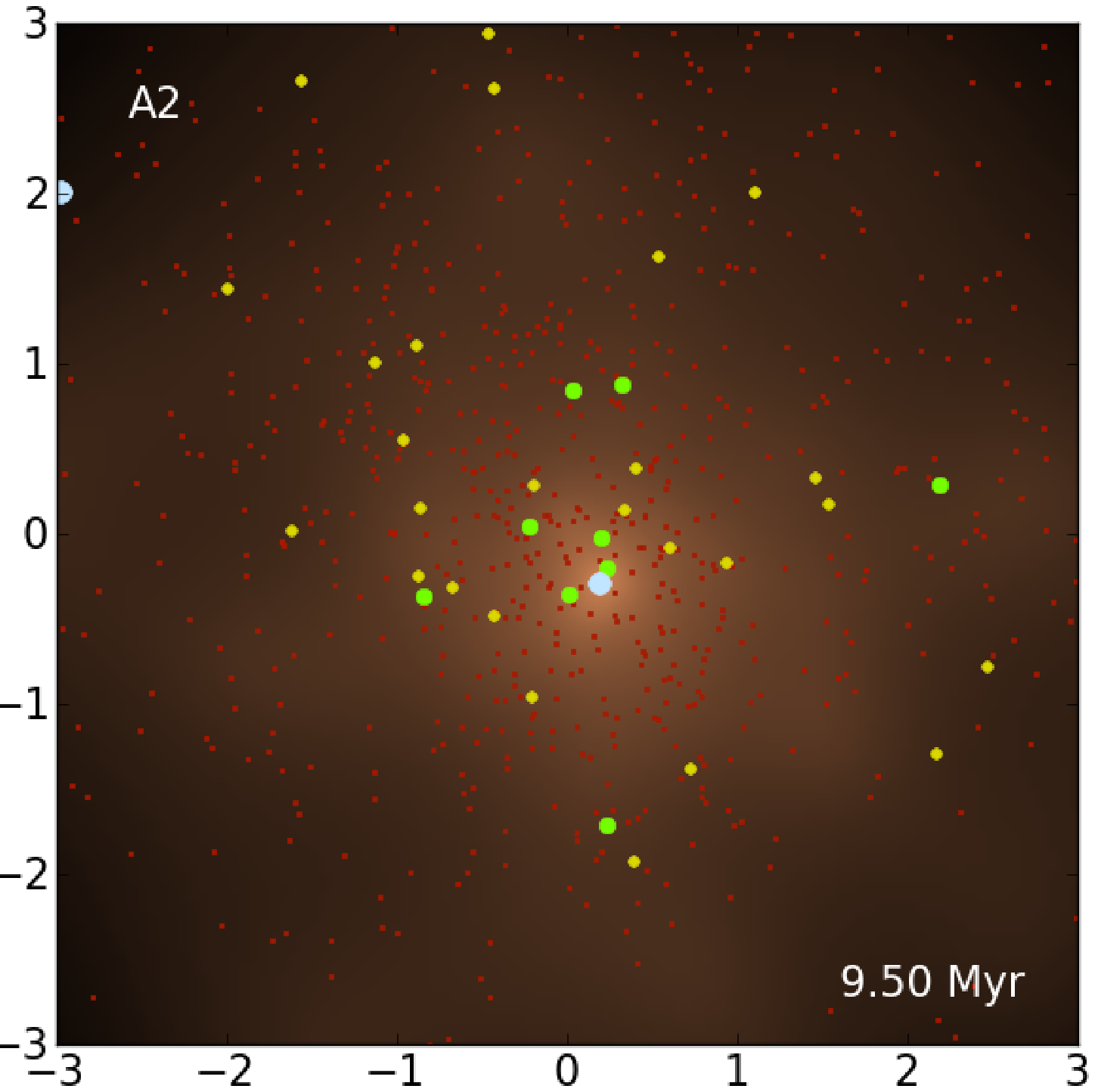}
\includegraphics[ height=.31\textwidth]{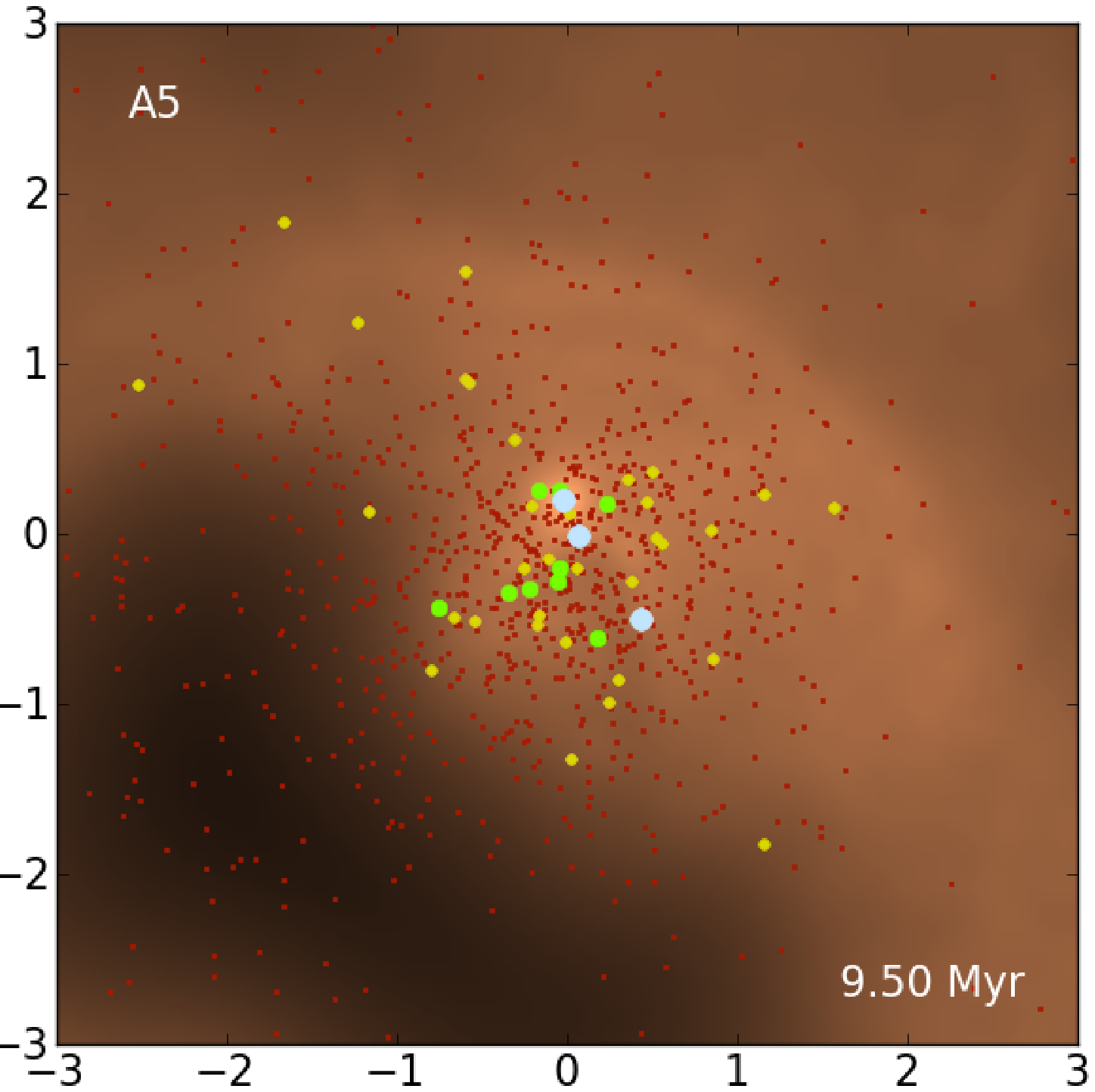}

\includegraphics[ height=.33\textwidth]{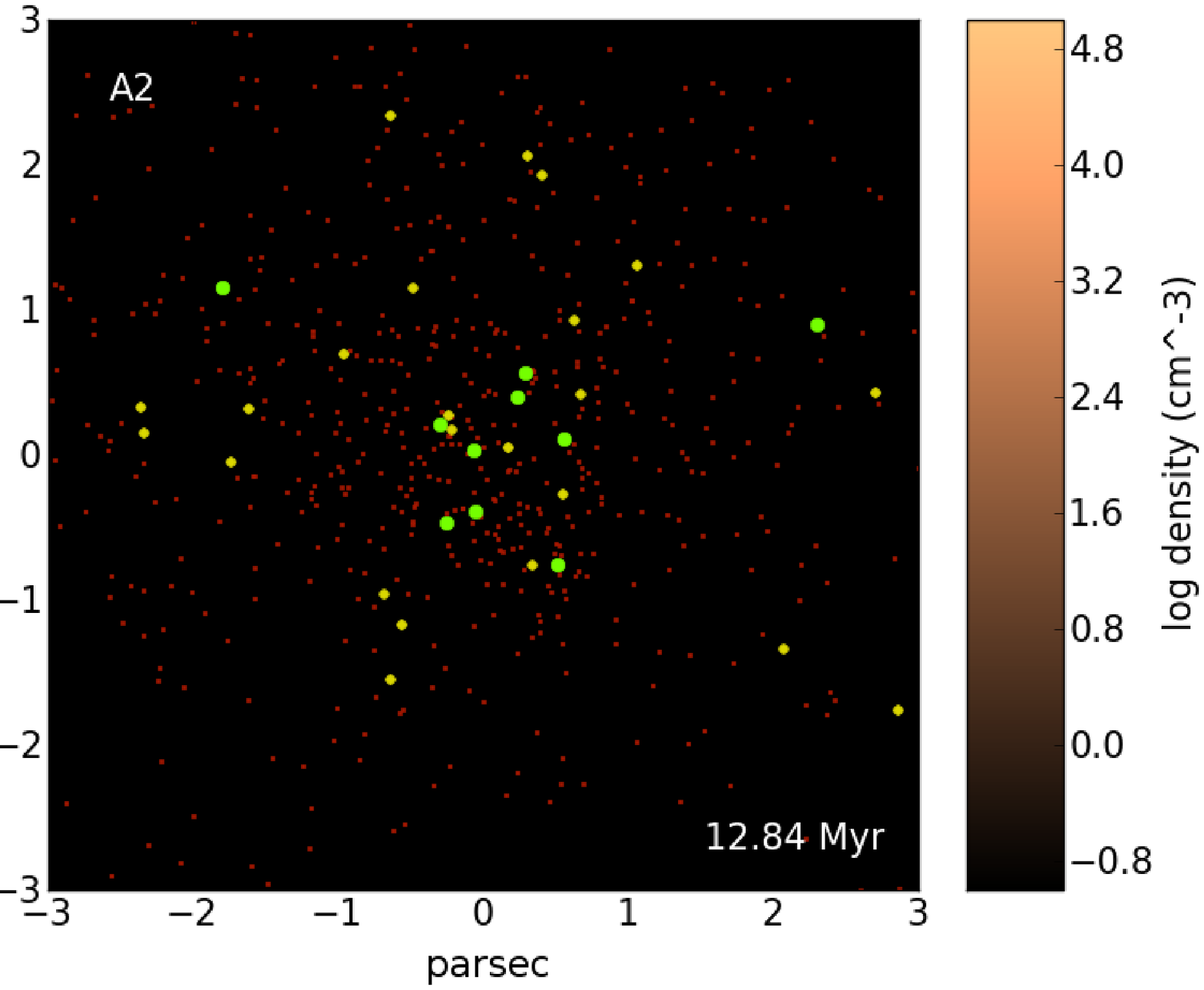}
\includegraphics[ height=.33\textwidth]{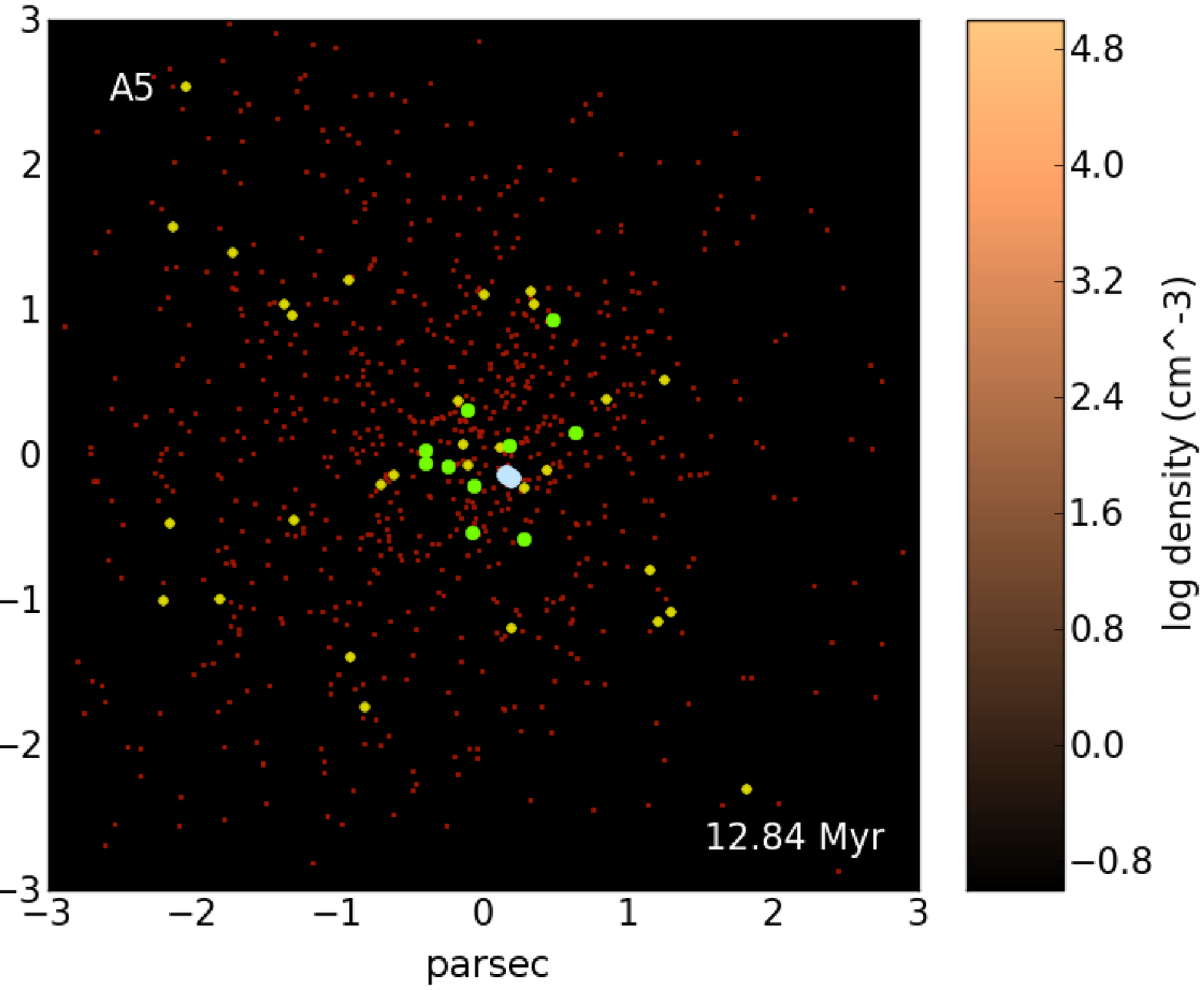}

\caption{Stellar and gas distribution of the A2 and A5 run. 
Left panels show the gas and stellar distribution of the A2 run, Right
panels of the A5 run. Snapshots are labelled with their time in the lower
right corner. Shown as a density plot is a slice through the midplane 
of the gas density. Points show the stars in 4 mass groups: 
$m_\star \le 0.9 \Msun$  (smallest red dots), 
$0.9 \Msun \le m_\star \le 2.5 \Msun $  (intermediate yellow dots),
$2.5 \Msun \le m_\star \le 10 \Msun$ (intermediate green dots)
and $m_\star \ge 10 \Msun$ (large light blue dots).
}
\label{fig_maps}
\end{figure*}

\begin{figure*}
\begin{flushleft}
\includegraphics[ width=0.32\textwidth]{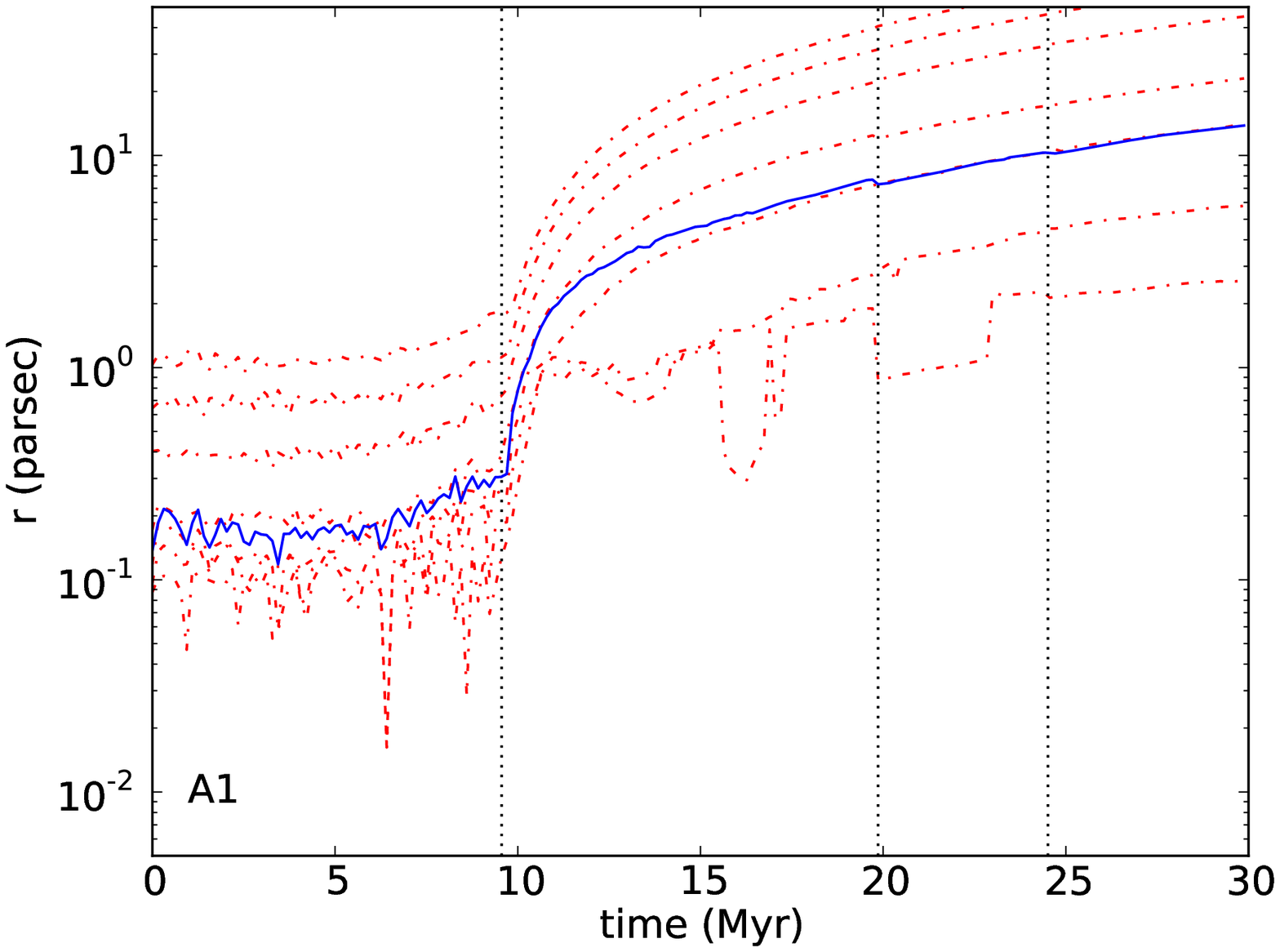}
\includegraphics[ width=0.32\textwidth]{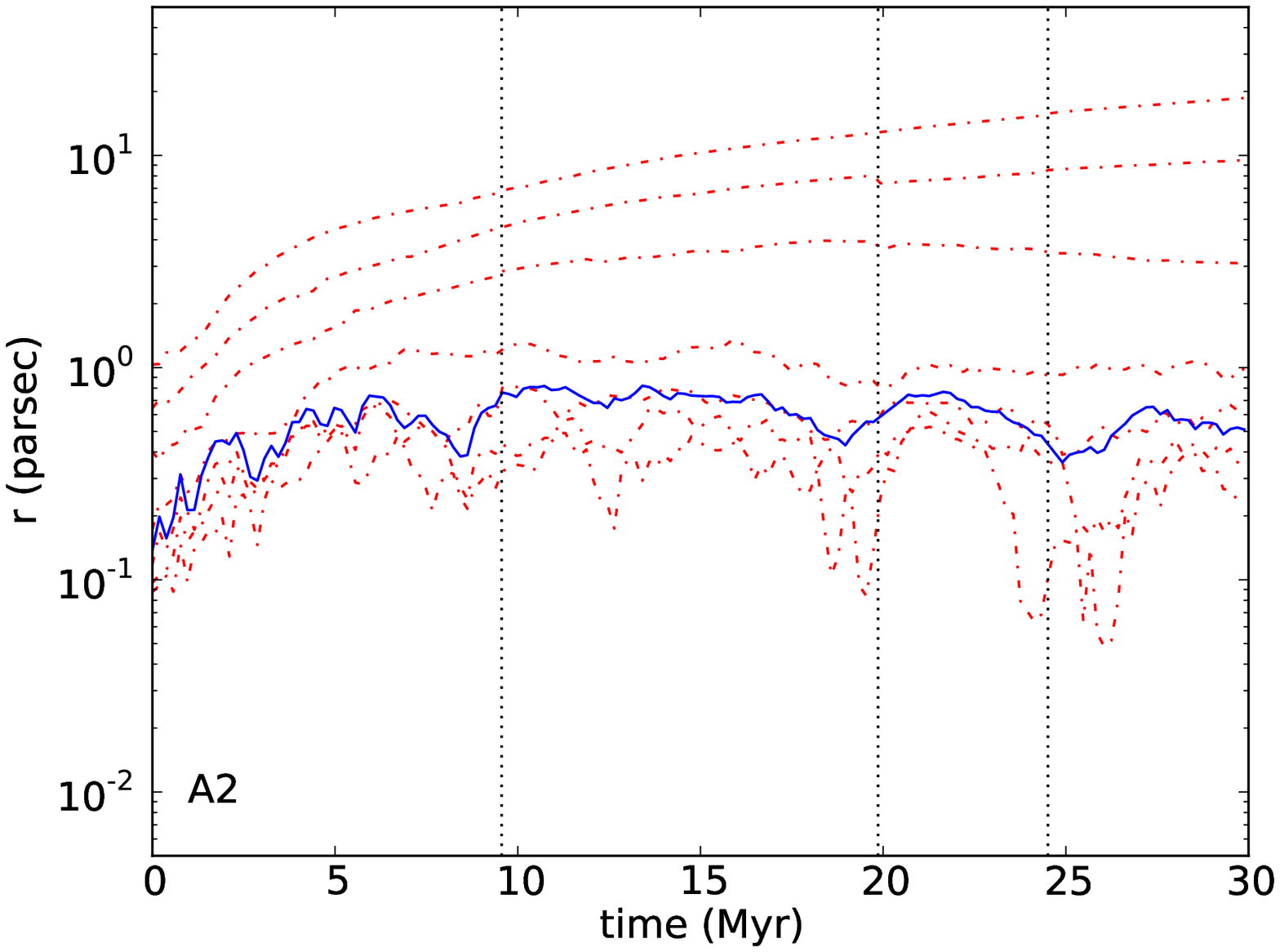}
\includegraphics[ width=0.32\textwidth]{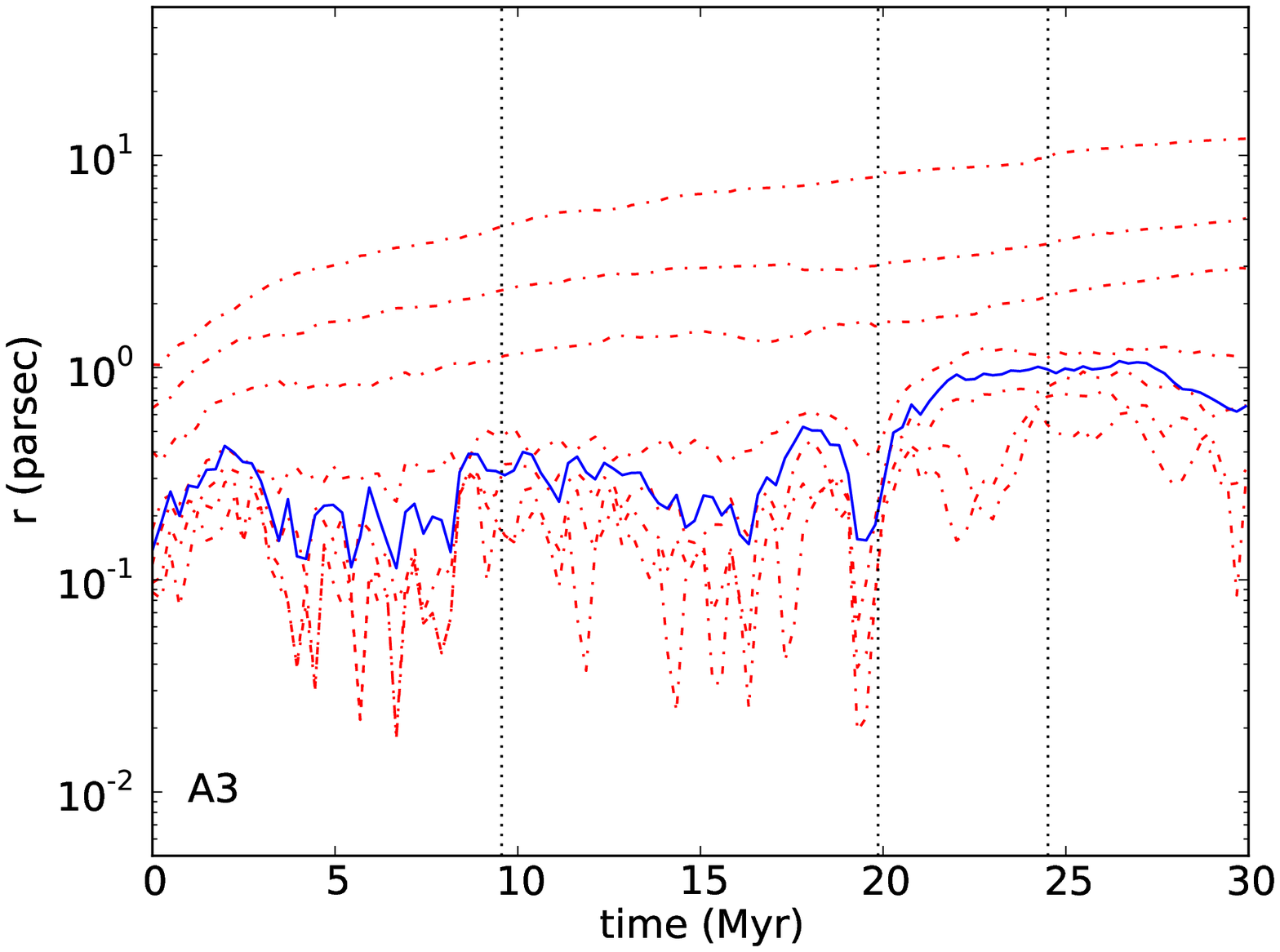}
\includegraphics[ width=0.32\textwidth]{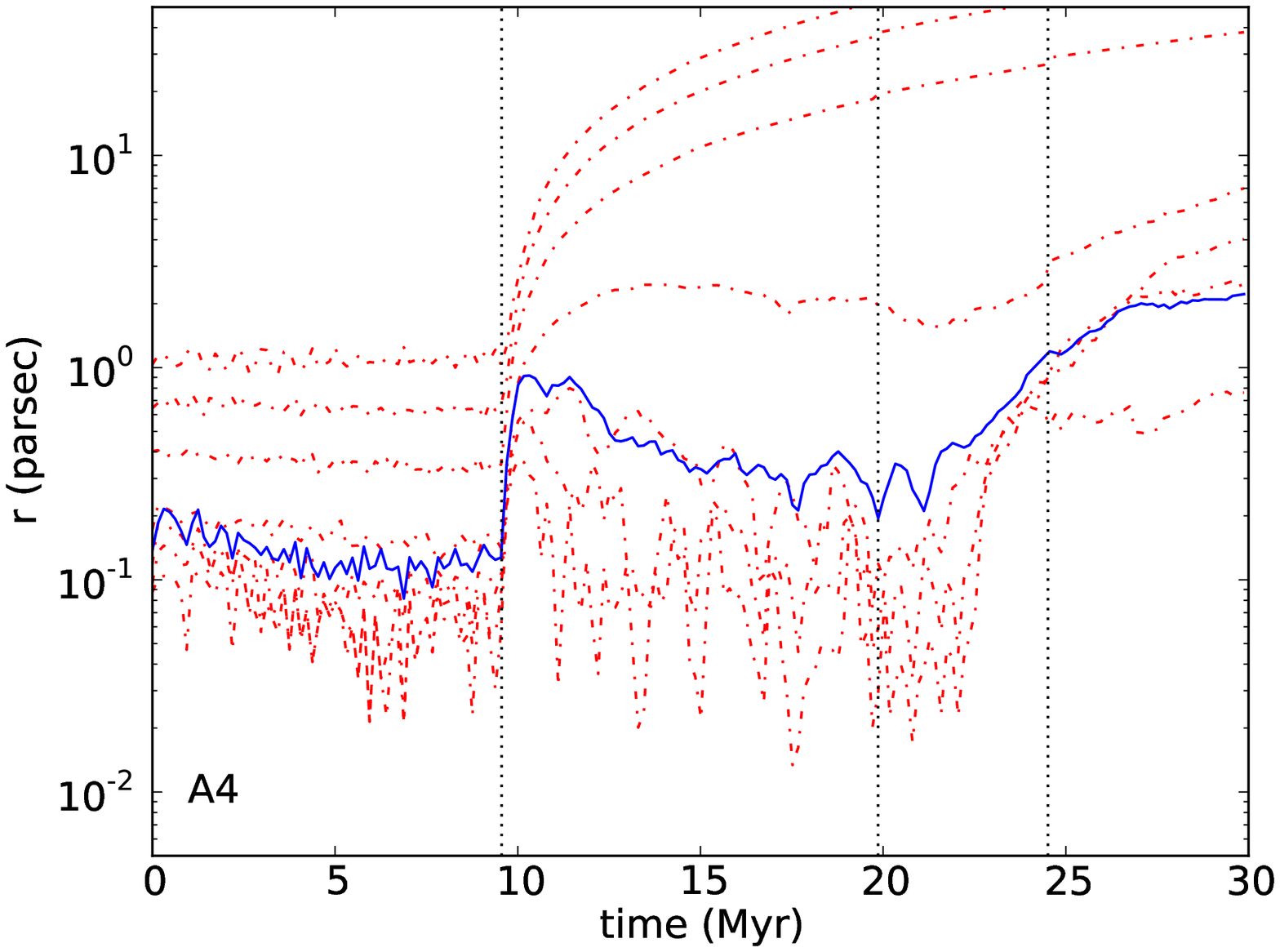}
\includegraphics[ width=0.32\textwidth]{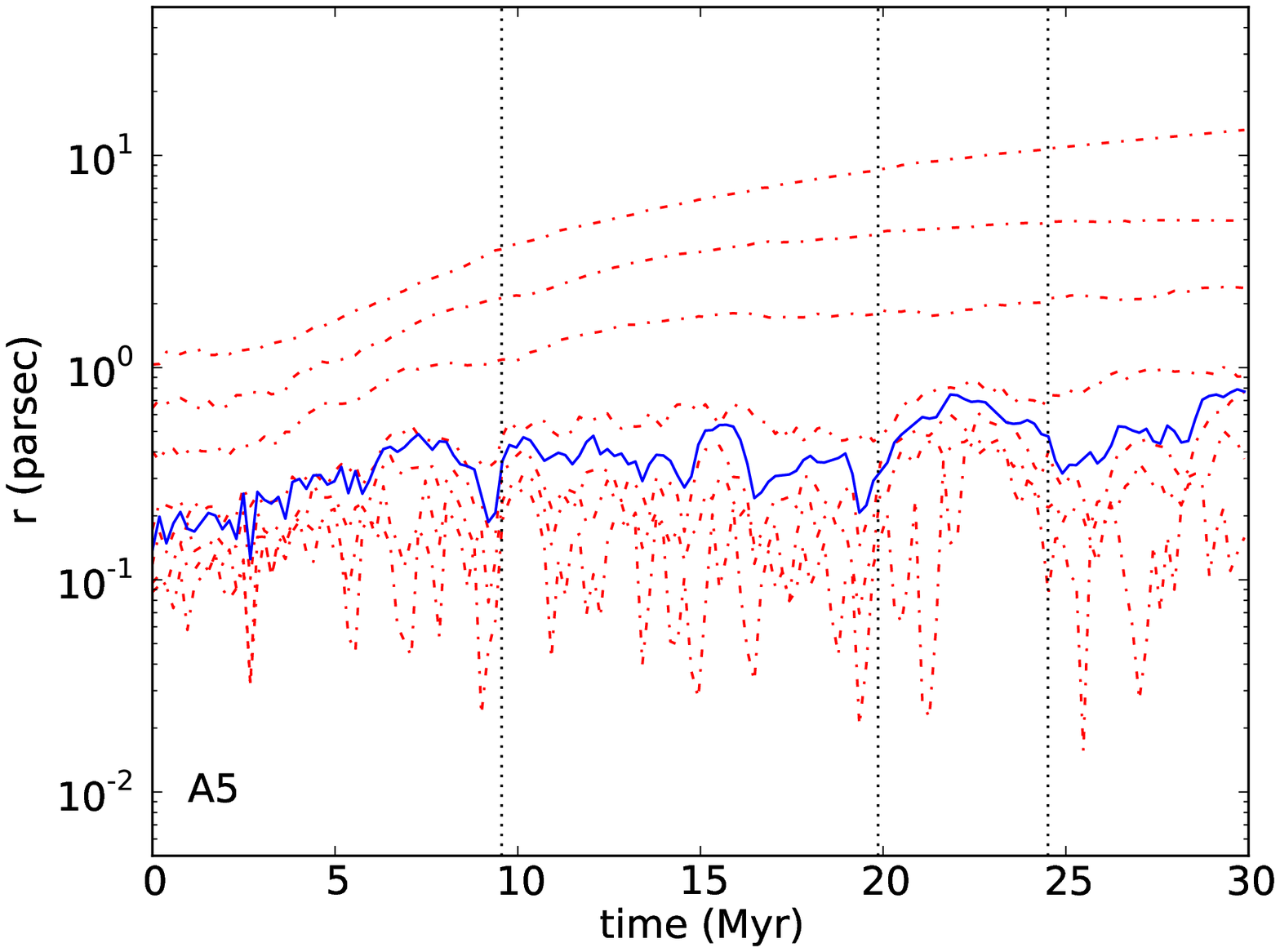}
\includegraphics[ width=0.32\textwidth]{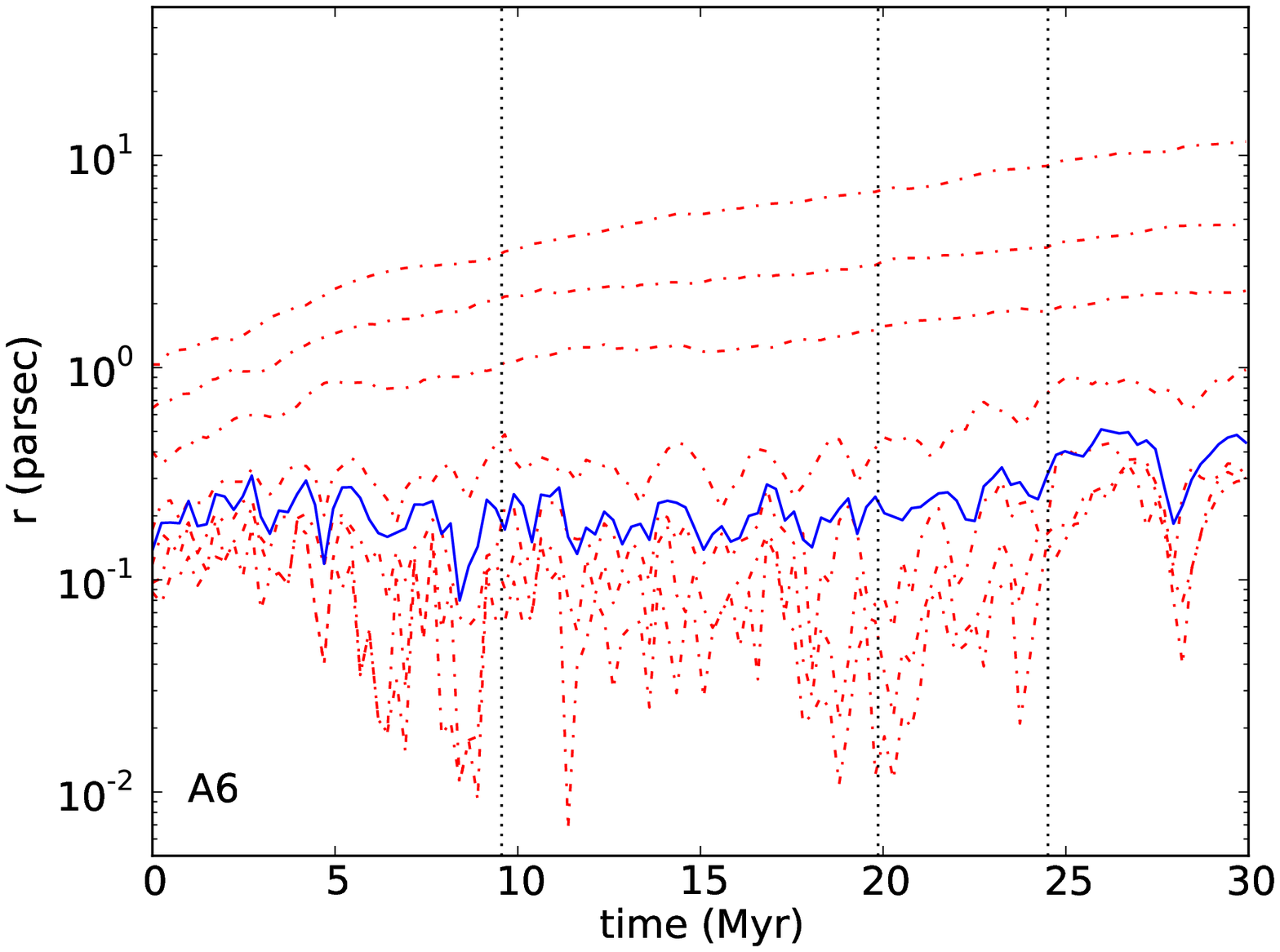}
\includegraphics[ width=0.32\textwidth]{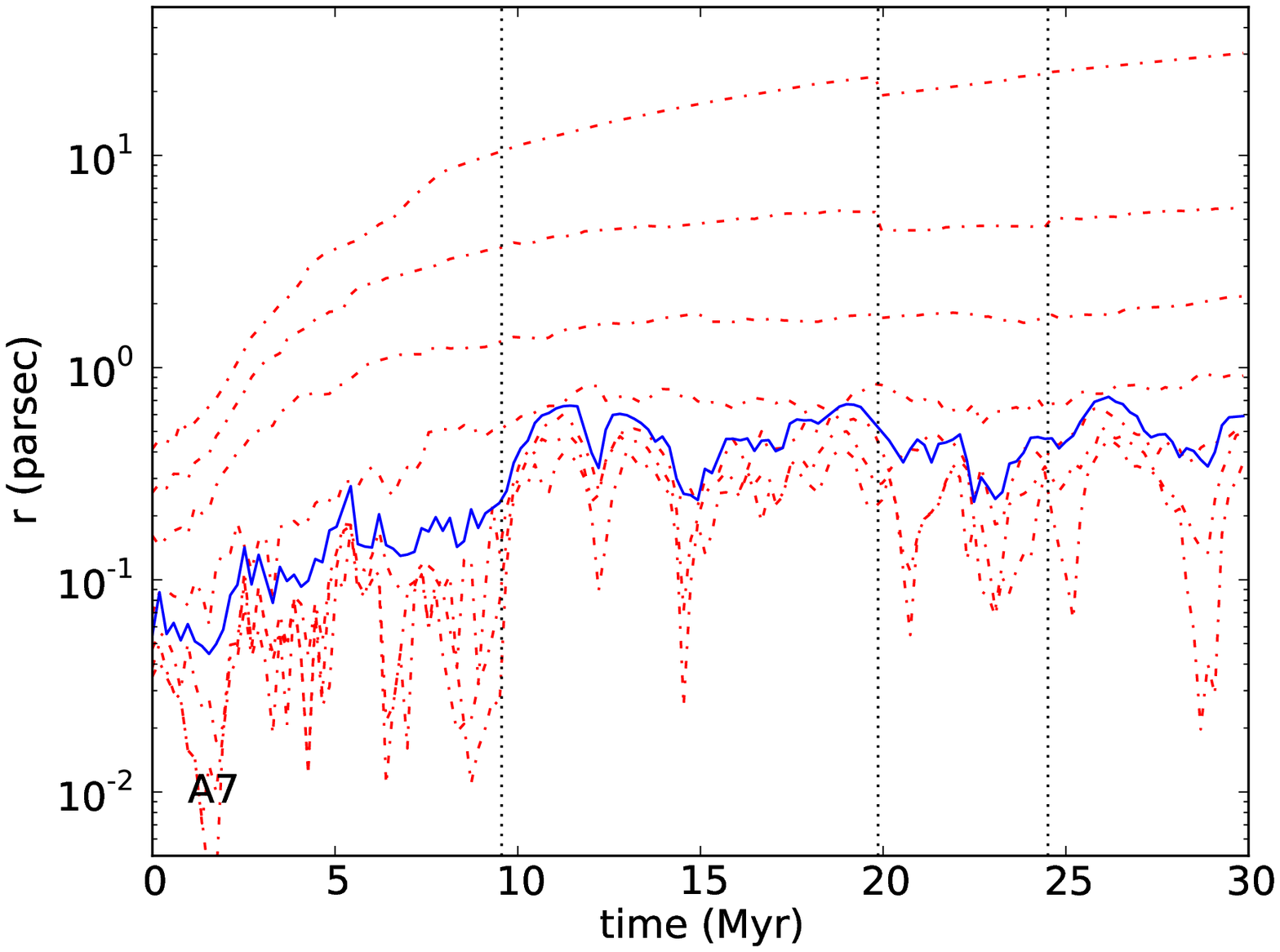}
\includegraphics[ width=0.32\textwidth]{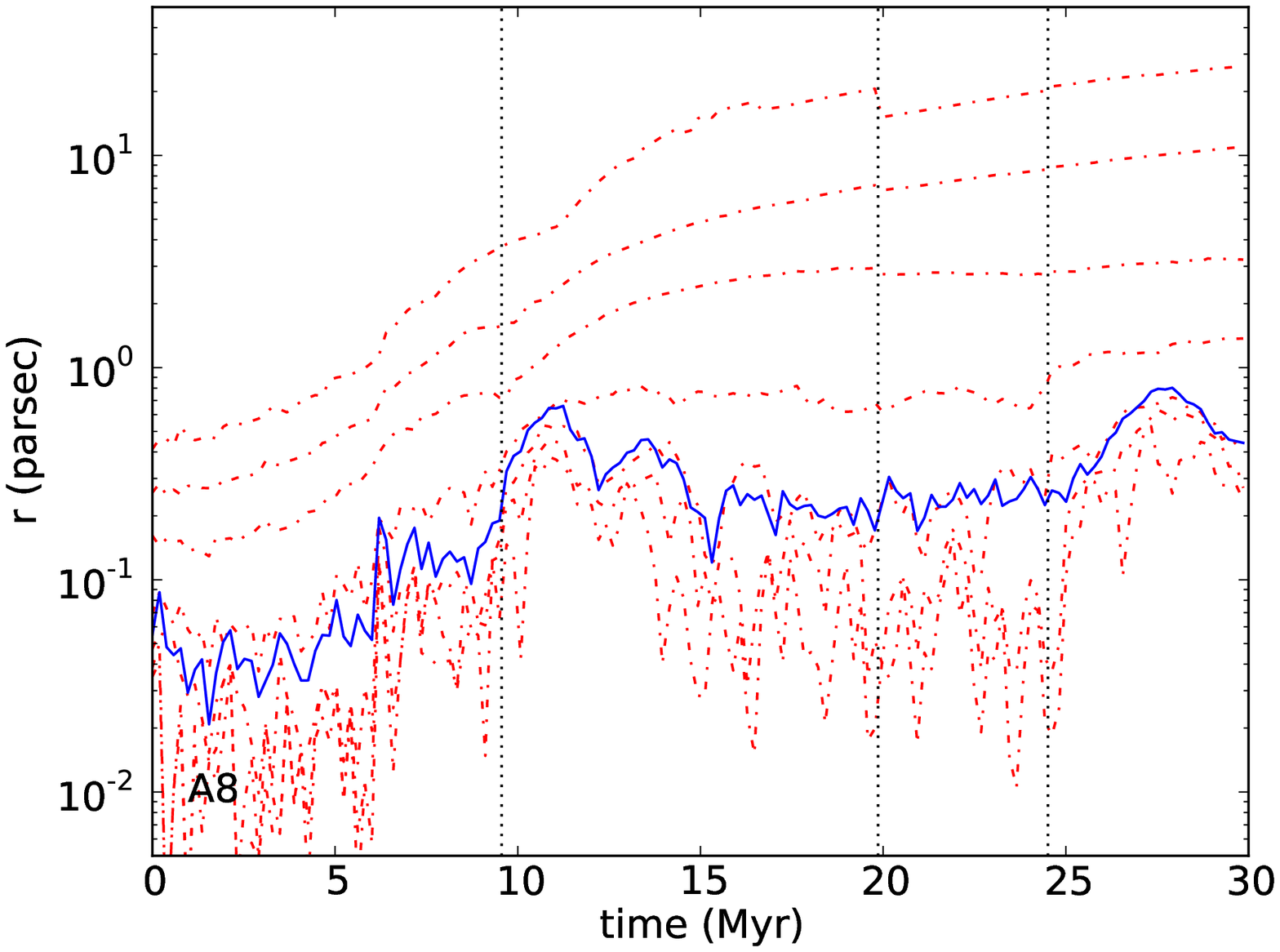}
\includegraphics[ width=0.32\textwidth]{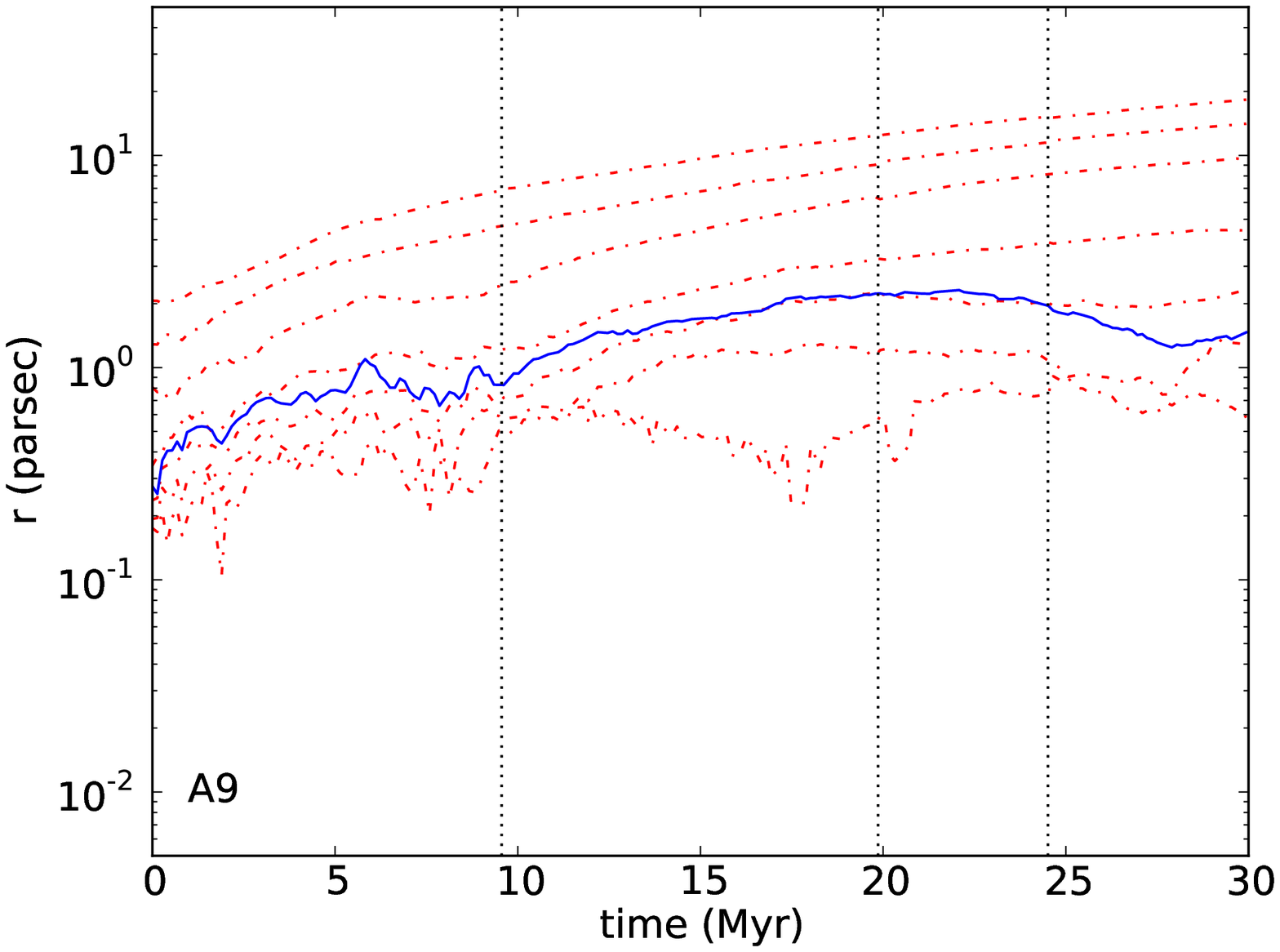}
\includegraphics[ width=0.32\textwidth]{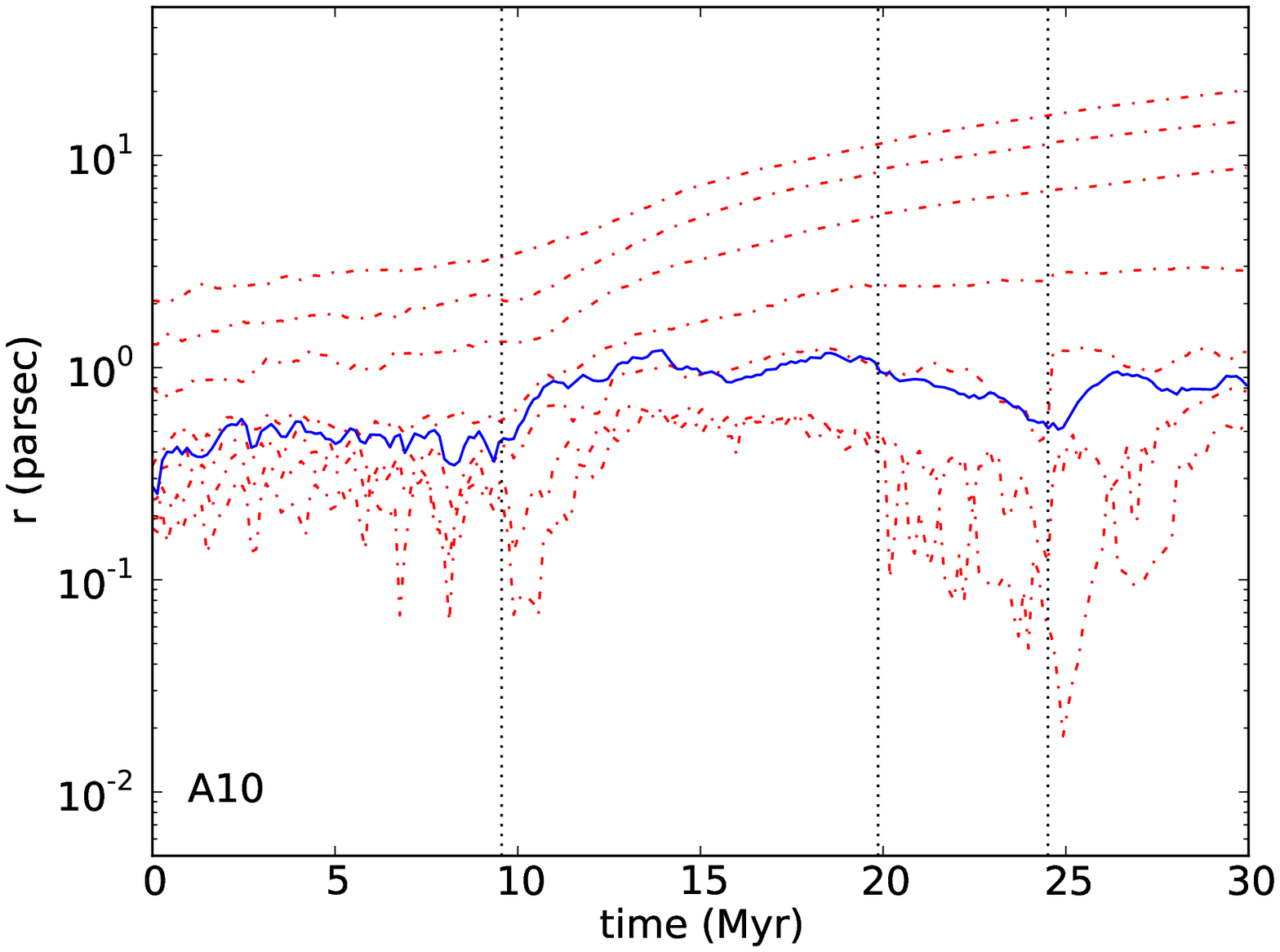}
\end{flushleft}
\caption{Lagrangian radii and core radius for runs A1-A10.
Plotted in red dashed lines are the time evolution of the 
(0.02,0.05,0.1,0.2,0.5,0.75 and 0.9 ) Lagrangian mass radii of the stellar 
mass distribution. Blue line is the core radius. Vertical dotted lines 
indicate supernova events.}
\label{fig_lr}
\end{figure*}

\begin{figure*}
\begin{flushleft}
\includegraphics[ width=0.32\textwidth]{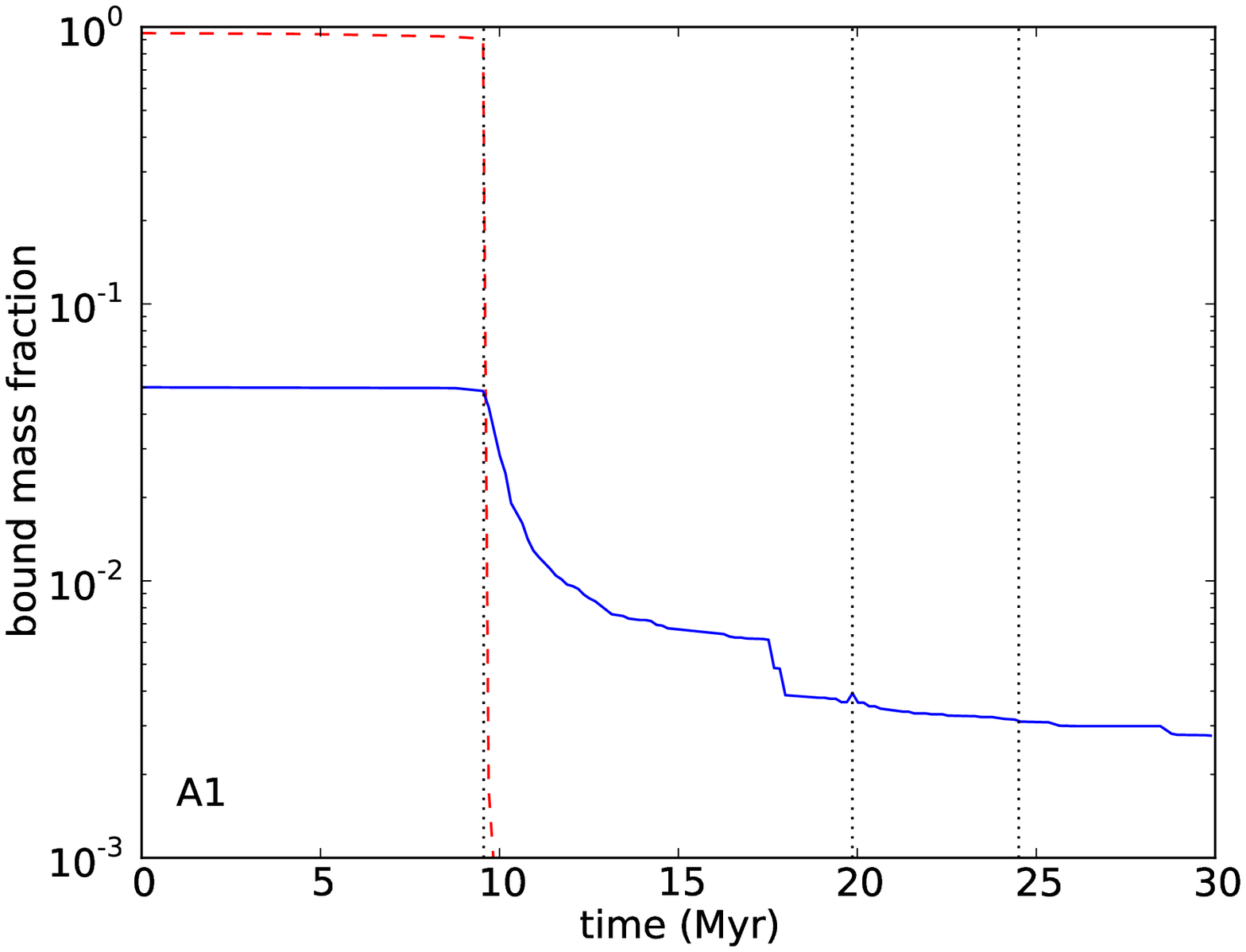}
\includegraphics[ width=0.32\textwidth]{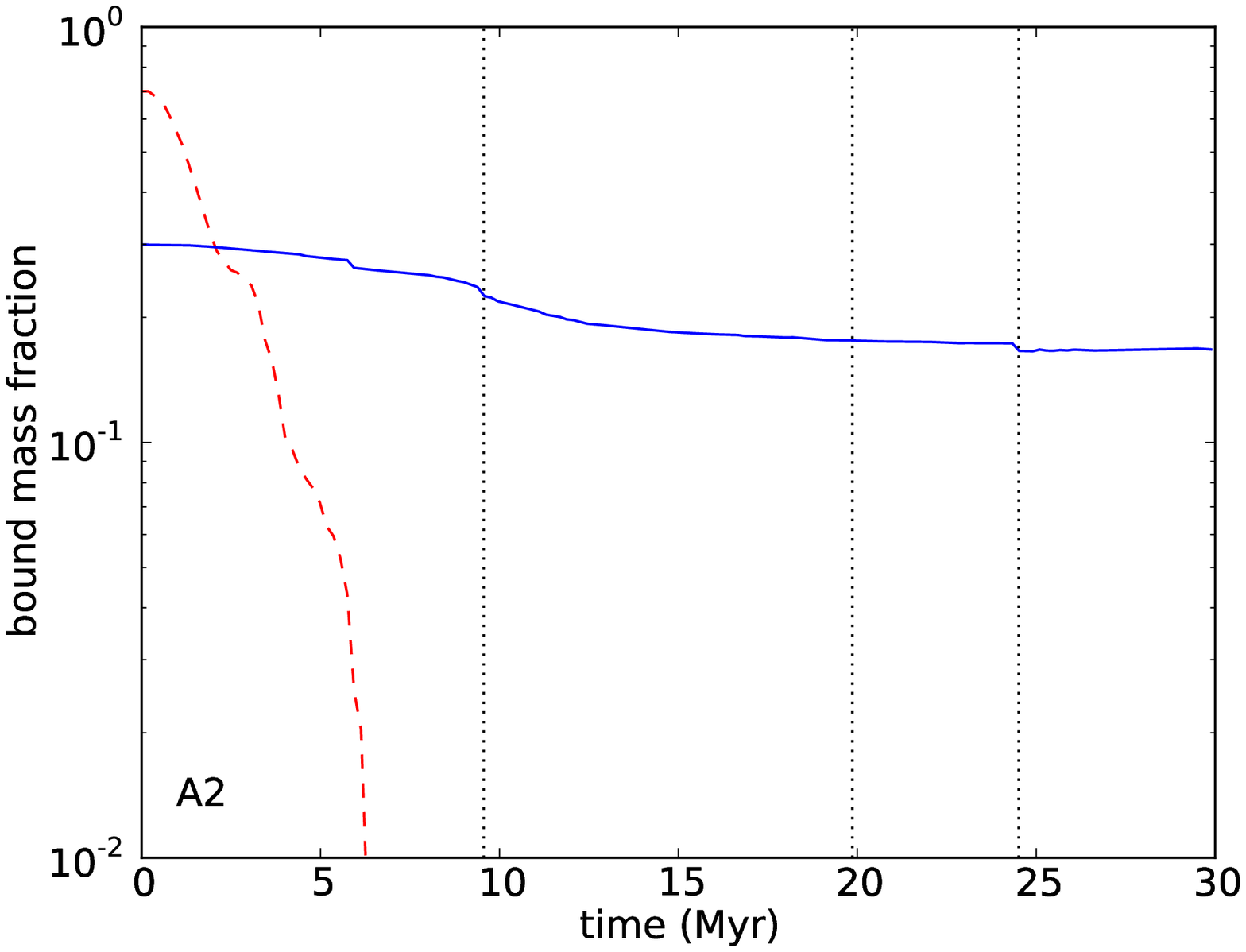}
\includegraphics[ width=0.32\textwidth]{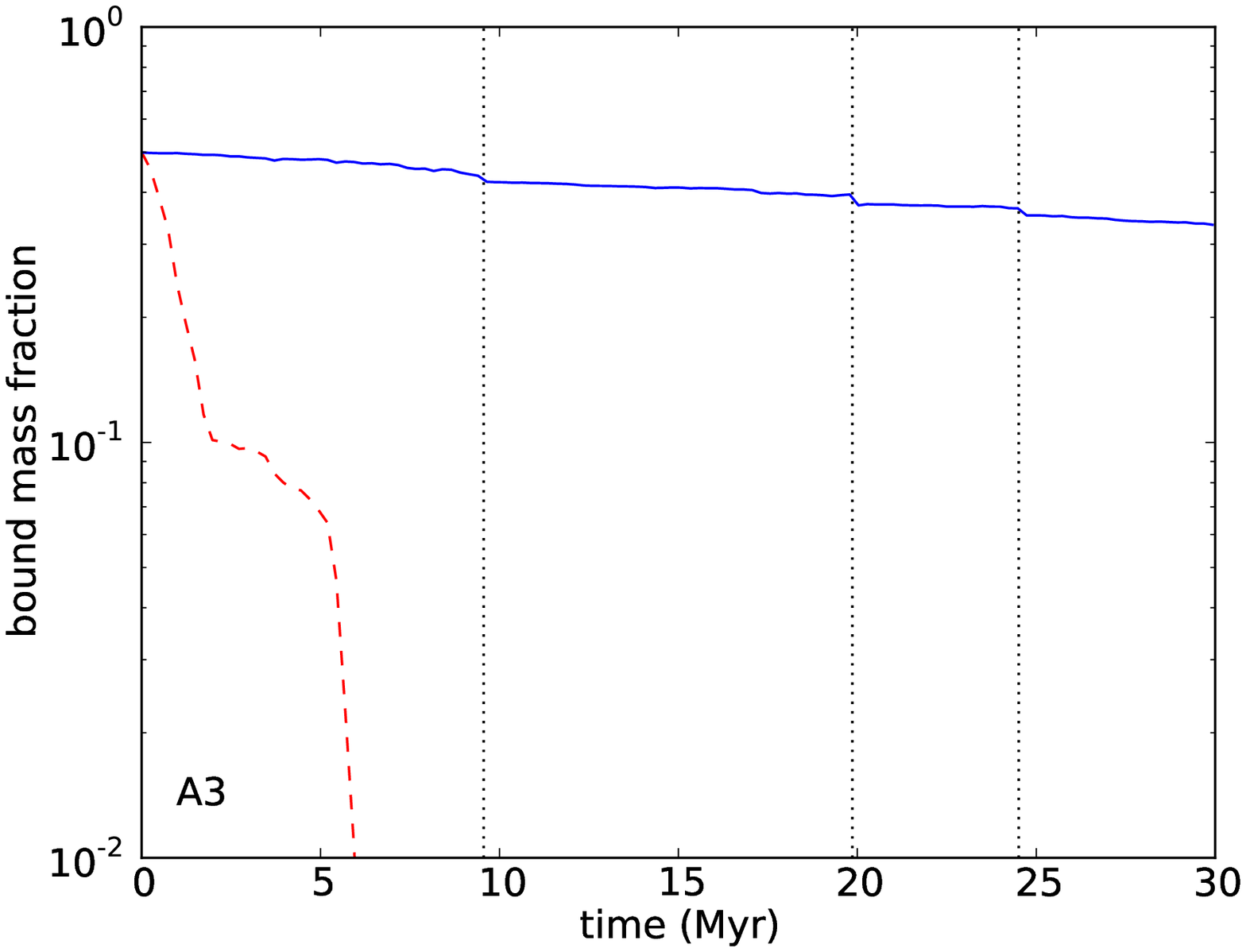}
\includegraphics[ width=0.32\textwidth]{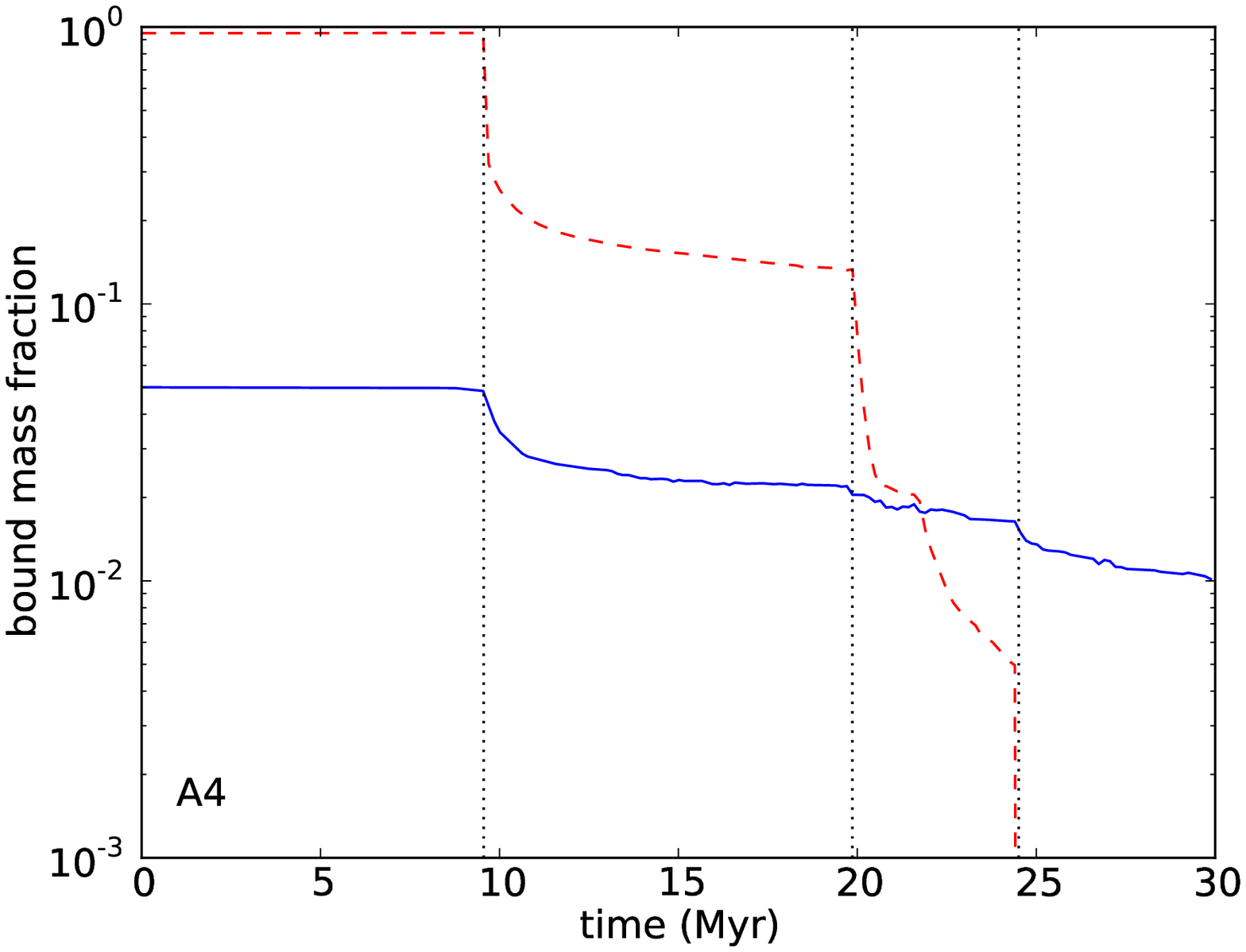}
\includegraphics[ width=0.32\textwidth]{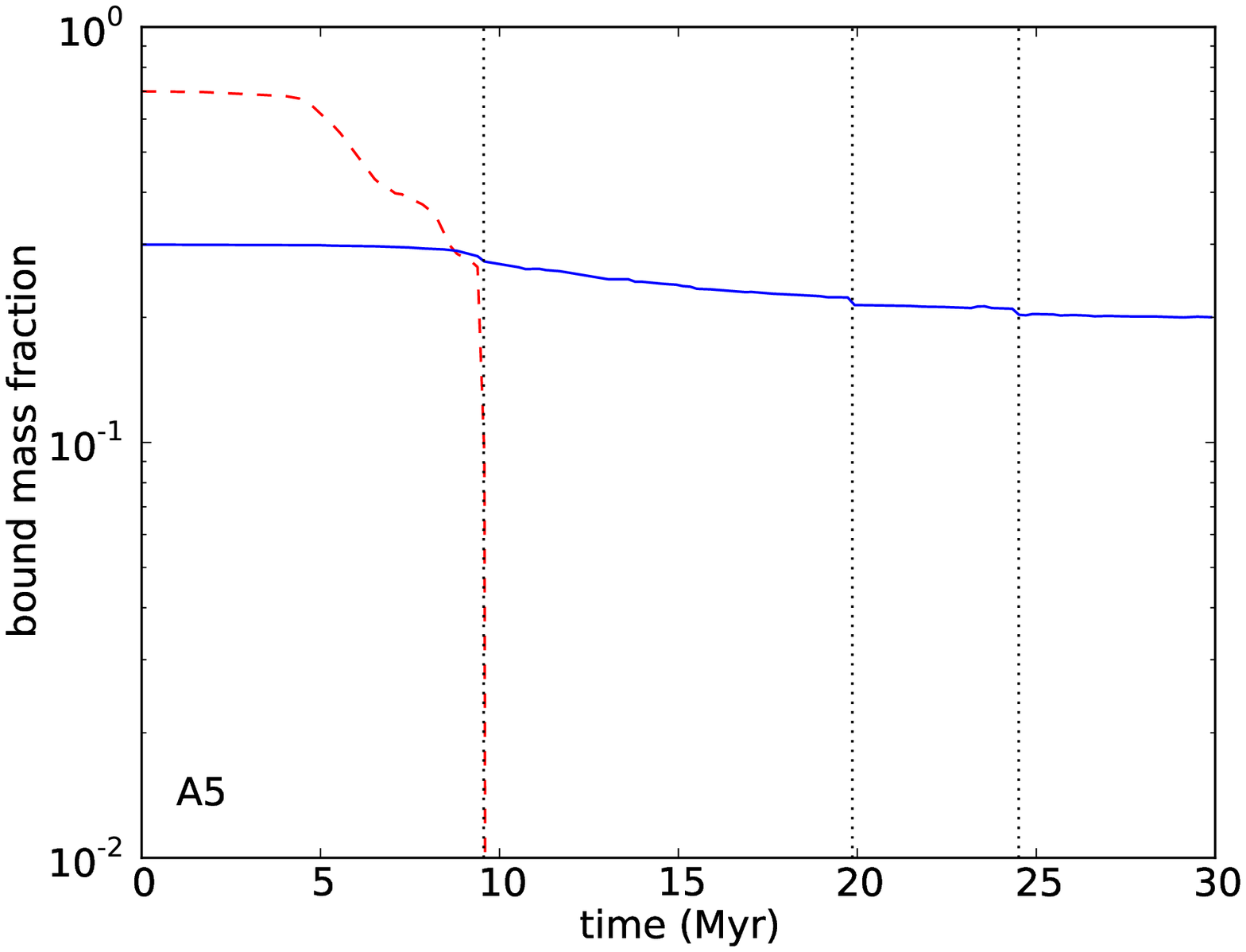}
\includegraphics[ width=0.32\textwidth]{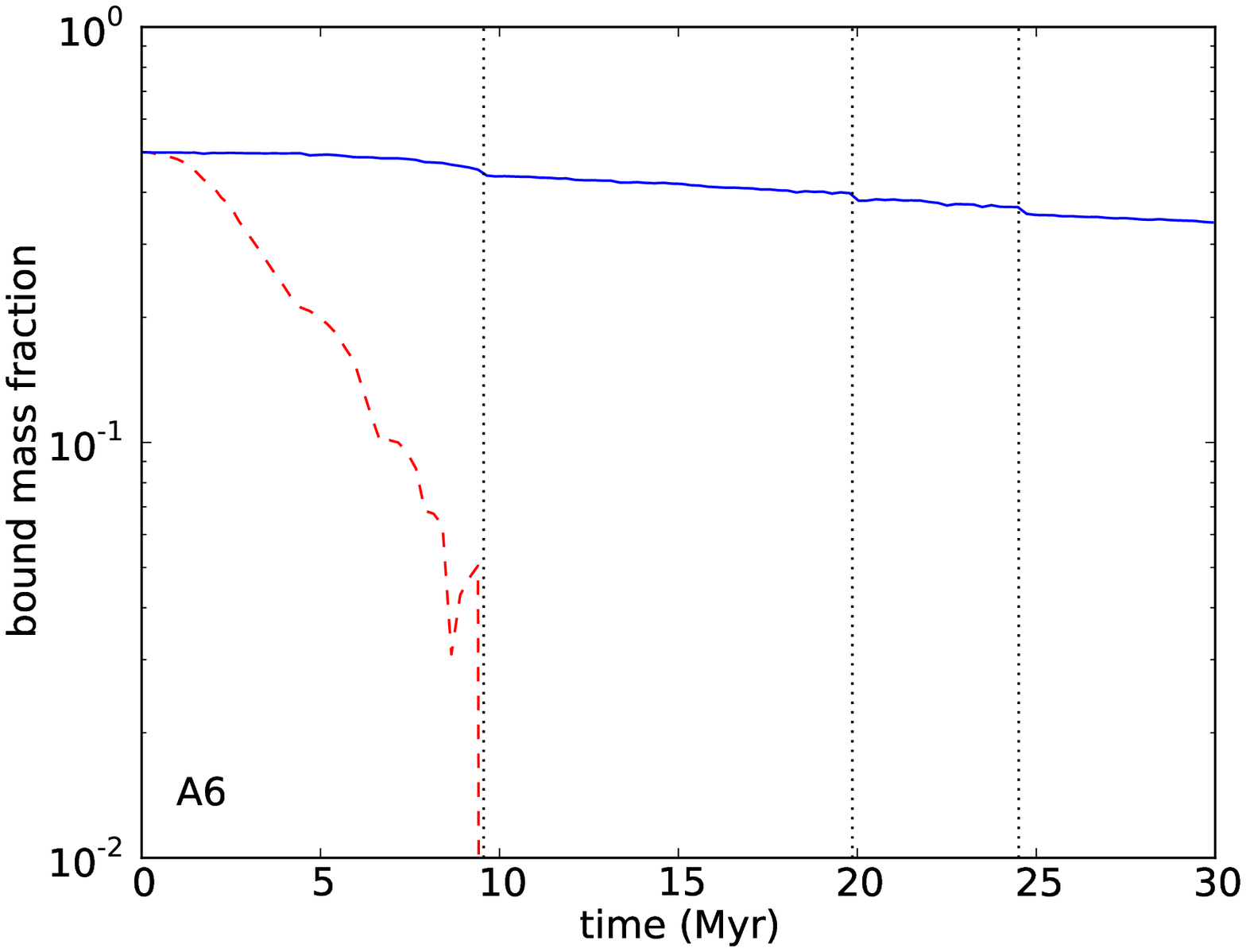}
\includegraphics[ width=0.32\textwidth]{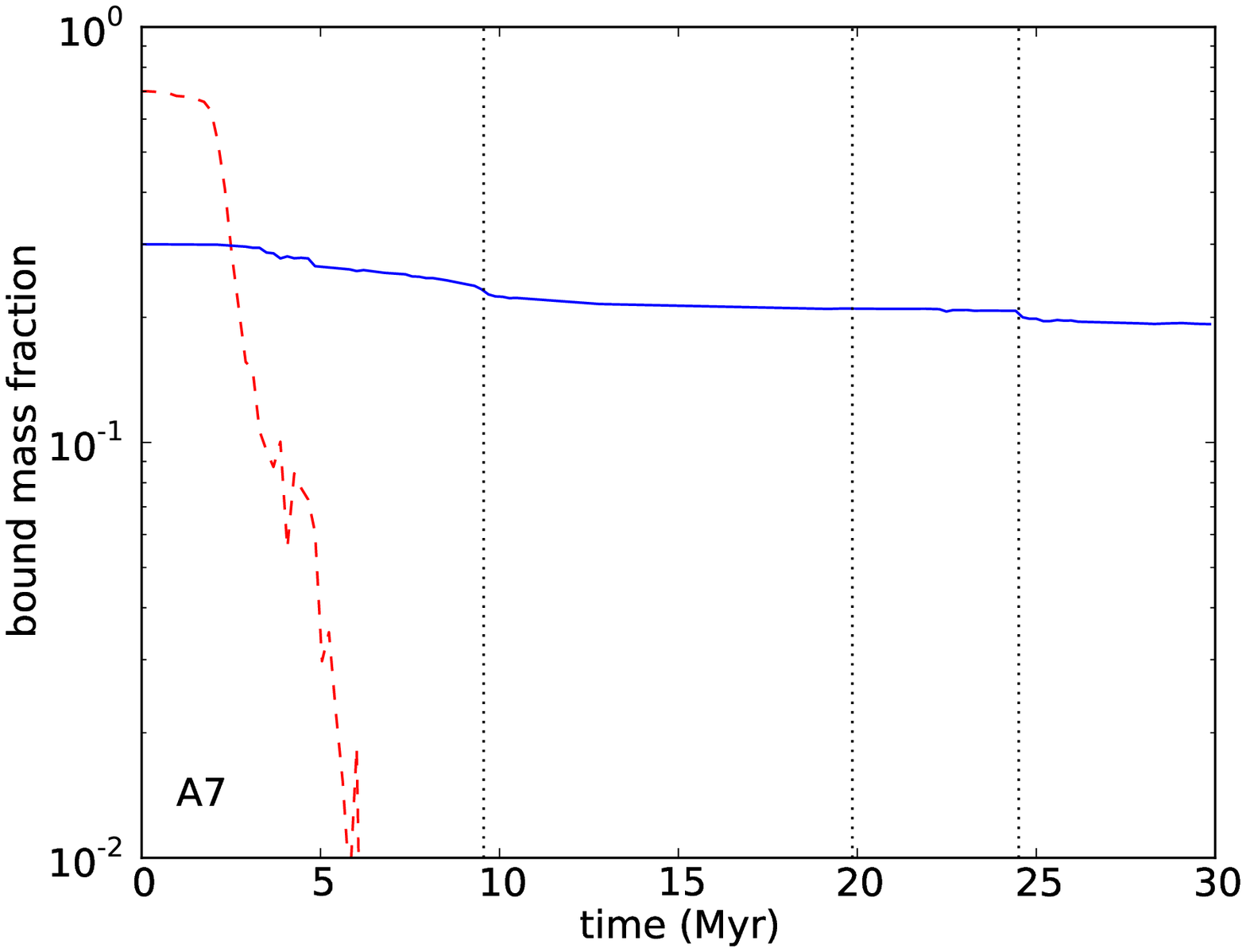}
\includegraphics[ width=0.32\textwidth]{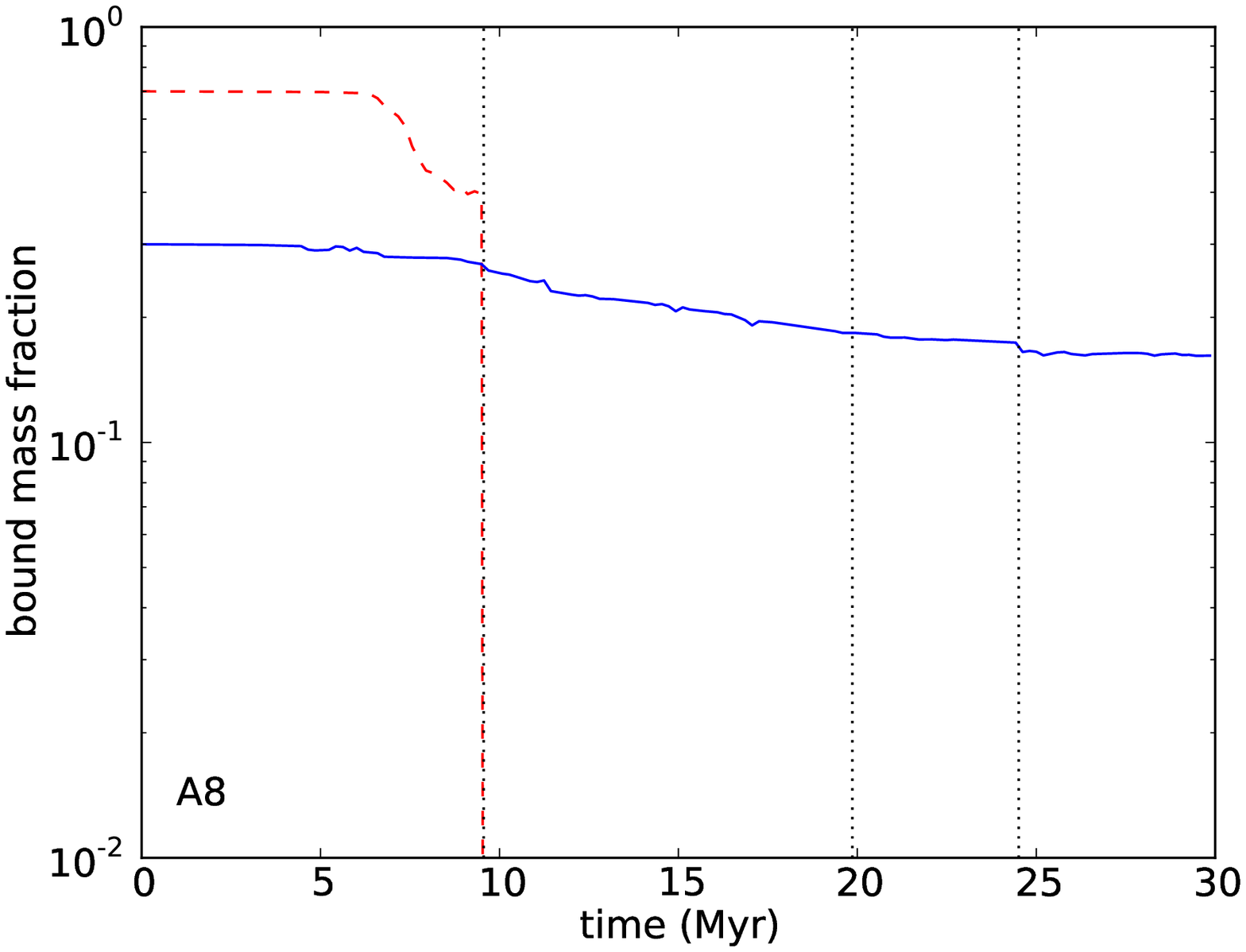}
\includegraphics[ width=0.32\textwidth]{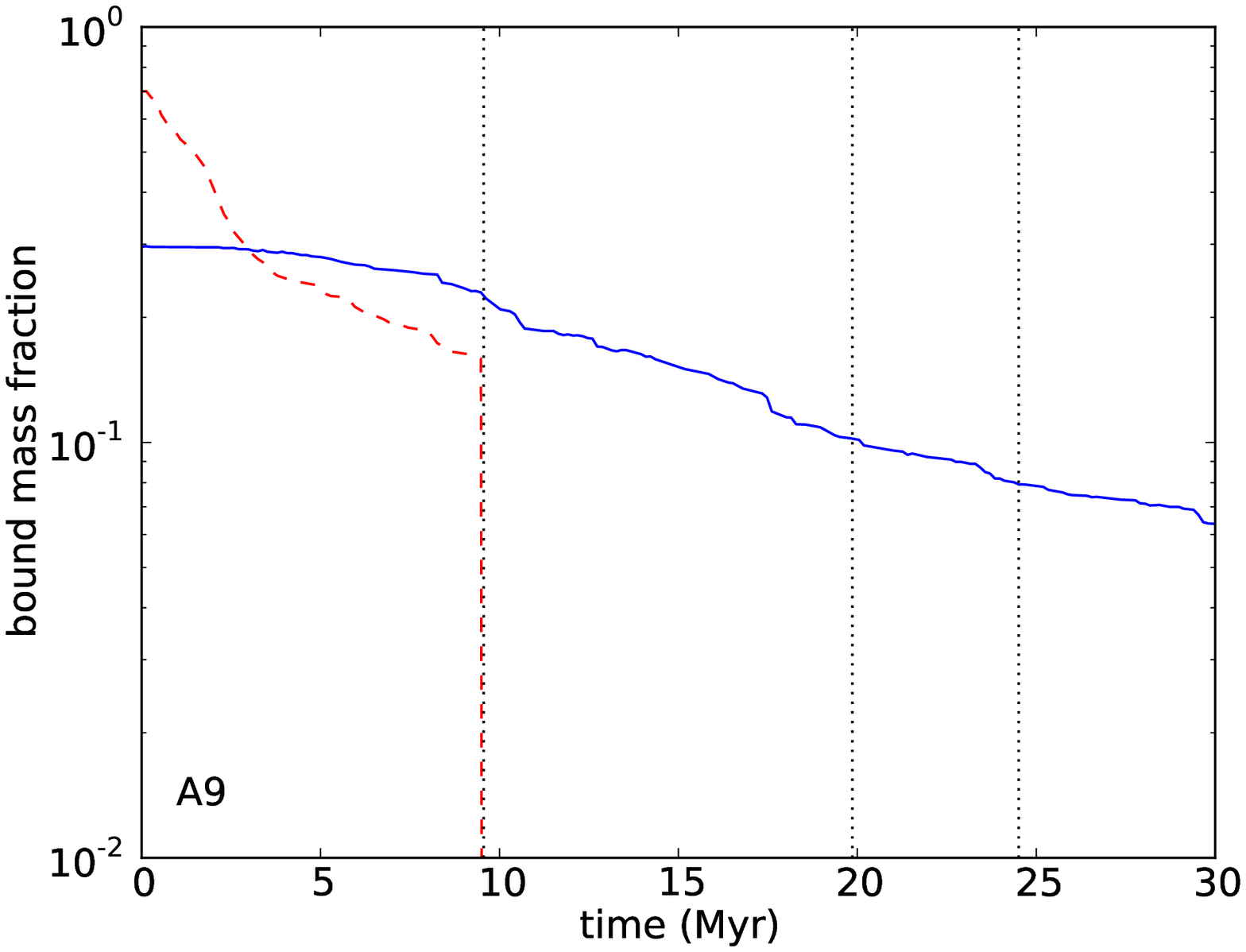}
\includegraphics[ width=0.32\textwidth]{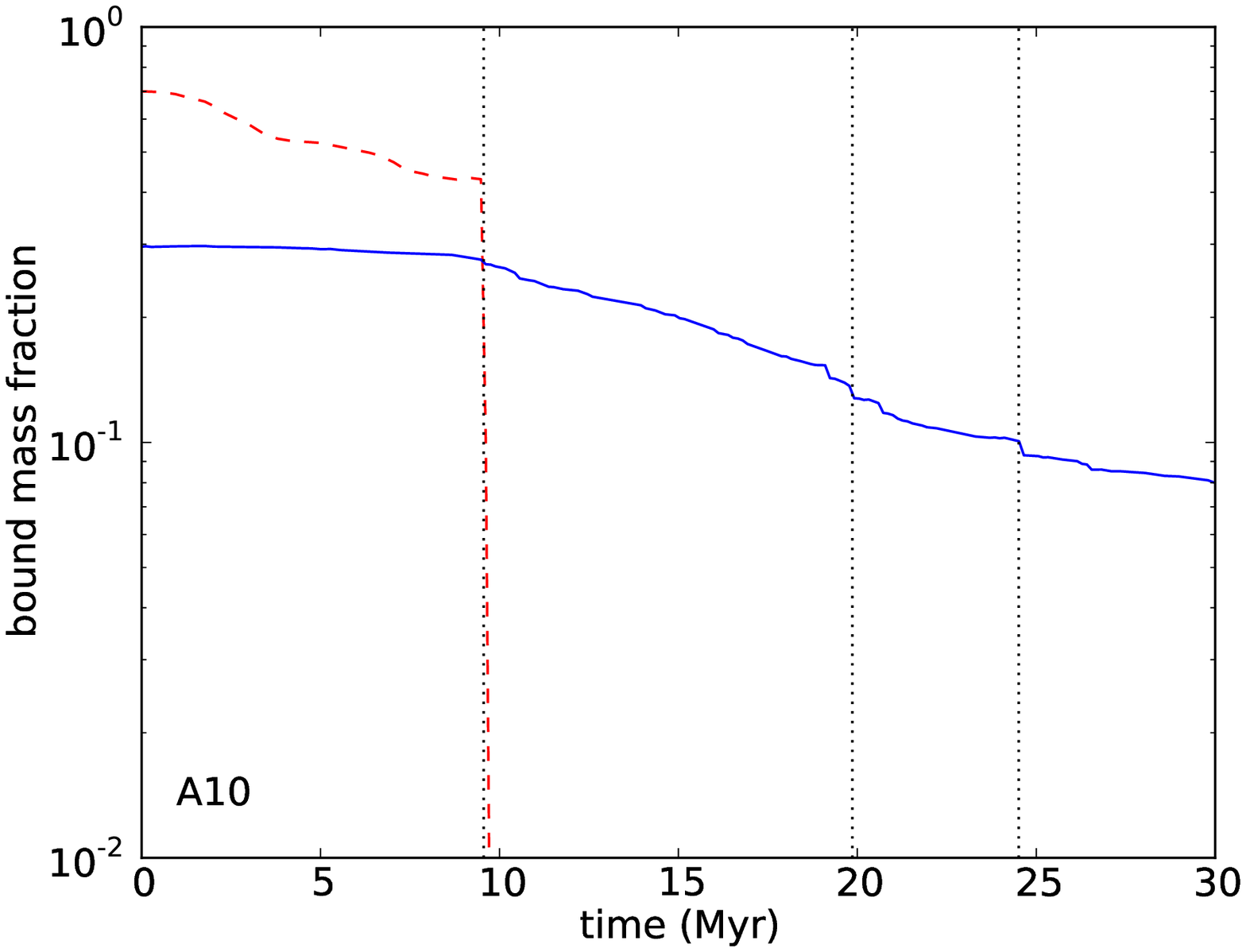}
\end{flushleft}
\caption{Stellar and gaseous bound mass fractions for runs A1-A10. Plotted is
the bound stelllar fraction (blue line) and the bound gaseous fraction (red
dashed line)  as fraction of the total initial system mass. Vertical lines 
indicate supernova events.}
\label{fig_bf}
\end{figure*}

In Figure \ref{fig_maps} we show the stellar and gas distribution of
our `baseline' run A2 and its low feedback equivalent, run A5. 
At four moments in time we plot the stars and a cut through 
the $z=0$ plane of the gas density distribution. Looking at the first A2 frame 
(at 0.96 Myr) we see the early stages where stellar winds blow bouyant 
bubbles that rise in the potential of the star cluster. As the mechanical 
luminosity increases these bubbles grow until they start blowing away sizable
fractions of the cluster medium and a free-flowing wind develops (4.37 Myr
frame). The strong feedback proceeds to unbind most of the gas of the cluster. 
At approximately  9.5 Myr the cluster ISM has been blown away - the gas that is 
visible in this frame originates from the strong AGB wind of the heaviest 
progenitor (21 $\Msun$). The supernova that explodes at 9.54 Myr leaves the
cluster devoid of gas.

For the A5 simulation ($f_{\rm fb}=0.01$) (both simulations use the same 
initial realizations for  the IMF and stellar positions) the initial wind 
stages proceed less violently, with smaller bubbles (0.96 Myr), while a 
free-flowing wind does not develop (compare the frame at 4.37 Myr) until 
just  before the supernova. The main difference between the  A2 and A5  run
is that most of the cluster gas remains present in the latter case, until the
first supernova, at which stage the remaining gas is blown away. Just before
the supernova the cluster is much more compact than in the A2 run.

The corresponding evolution of the Lagrangian radii and core
radius and central gas density (of models A1-A10) is shown in
Figure~\ref{fig_lr} and we plot in Figure~\ref{fig_bf} the (instantaneous)
bound fraction of the gas and star component. For run A2,
in the early phase (before 8 Myr) the gas is lost by stellar
winds. From the beginning of the simulation the gas starts to boil off
vigorously under the influence of the stellar winds. The stellar 
distribution reacts by expanding. Within 3 Myr the core radius increases a
factor $\approx 2.5$ (within this time approx $2/3$ of the gas is lost)
and stays more or less constant thereafter. This is also visible in the
maps of Fig.~\ref{fig_maps}, where after the 4.37 Myr frame the stellar
distribution remains qualitatively similar. Fig.~\ref{fig_bf} also shows
that when the first supernova goes off, the gas in the cluster has already 
been completely removed. Later supernovae affect the stellar distribution 
very little. If we look at the bound fraction of stars shown in 
Figure~\ref{fig_bf} we see that $\approx 60\%$ of the stars remain bound
at the end of the simulation. The escape of stars happens mostly at the 
time of early gas loss, but with a
noticeable delay. Note that the bound fraction plotted
here and for the other runs is an upper limit because this fraction is
determined from the instantaneously bound stars (so at a particular
point in time) and does not account for further unbinding by the
escape of marginally unbound mass (this can account for part of the observed
delay).

Comparing runs with different SFE (models A1,A2 and A3) we see that in the
case of low SFE (run A1) most of the gas is retained up until the moment of
the first supernova (see fig. \ref{fig_bf}). The gas completely unbinds
because of the supernova at 9.54 Myr, and the cluster is essentially
dissolved immediately at that point. The high SFE case A3 
is very similar to the A2 case. The initial gas loss may be somewhat
steeper, but in this  case (fig.~\ref{fig_bf}) the loss of stars of the
cluster proceeds more slowly.  Looking at the evolution before the first SN,
it is clear that the greater  gas mass ($20 \times$ more) gives the A1 run
greater resilience to the  disturbance from stellar winds. Considering the
escape of stars this leads to counteracting effects: when the
fraction of gas is large, the loss of gas can more easily disturb the
stellar distribution. However this loss happens initially much more slowly.
In the A2 and A3 runs on the other hand, the gas starts to be blown away right
from the beginning, but the fraction of gas is small enough not to cause
major disruption for the stellar distribution. When the supernova explodes,
the gas of the A1 run is instantaneously removed, leading to the quick
dissolution of most of the cluster. On the other hand, about 70\% of the 
stars is retained in the A3 run.

If we look at the runs with the lower effective feedback strength (A4, A5
and A6), we see that in the SFE $=0.05$ case (A4) gas is still present after
the first and second supernova. The stellar component reacts completely
differently for the A4 run (compared with A1): in this case $50\%$ of the stars 
remain bound initially, where only $10\%$ remains in the A1 run. In the plot 
of the Lagrangian and core radii of run A4 (fig.~\ref{fig_lr}) the core radius
initially increases after the supernova, but then grows smaller again, as the 
cluster relaxes - markedly different from the continuing expansion in the A1 
case. For the other two runs (A5 and A6), the lower effective feedback strength 
mainly results in a much less violent boil off 
compared with the high $f_{\rm fb}$ cases (e.g. A2 vs A5). Gas is still present
at the time of the first supernova. For the stellar distribution this means
that initially the stellar content remains intact (in particular the A5 run 
shows an almost constant bound fraction for the stellar component up until 
9.5 Myr). The final bound stellar fractions and Lagrangian radii
are qualitatively similar.  The reason for this is that in both cases 
the total feedback energy is sufficiently large to unbind the cluster gas
eventually.  Much of the extra energy in the simulations with high feedback
is carried  away by the escaping gas. In the high feedback case gas loss 
by stellar winds is much more important, whereas in the low feedback case 
gas is driven out by supernovae. The final stellar distribution of the final 
cluster is little affected by the feedback efficiency over the range explored 
here, unless the gas fraction is very high.

\subsection{Different realizations}

\begin{figure}
\centering
\includegraphics[ width=0.32\textwidth]{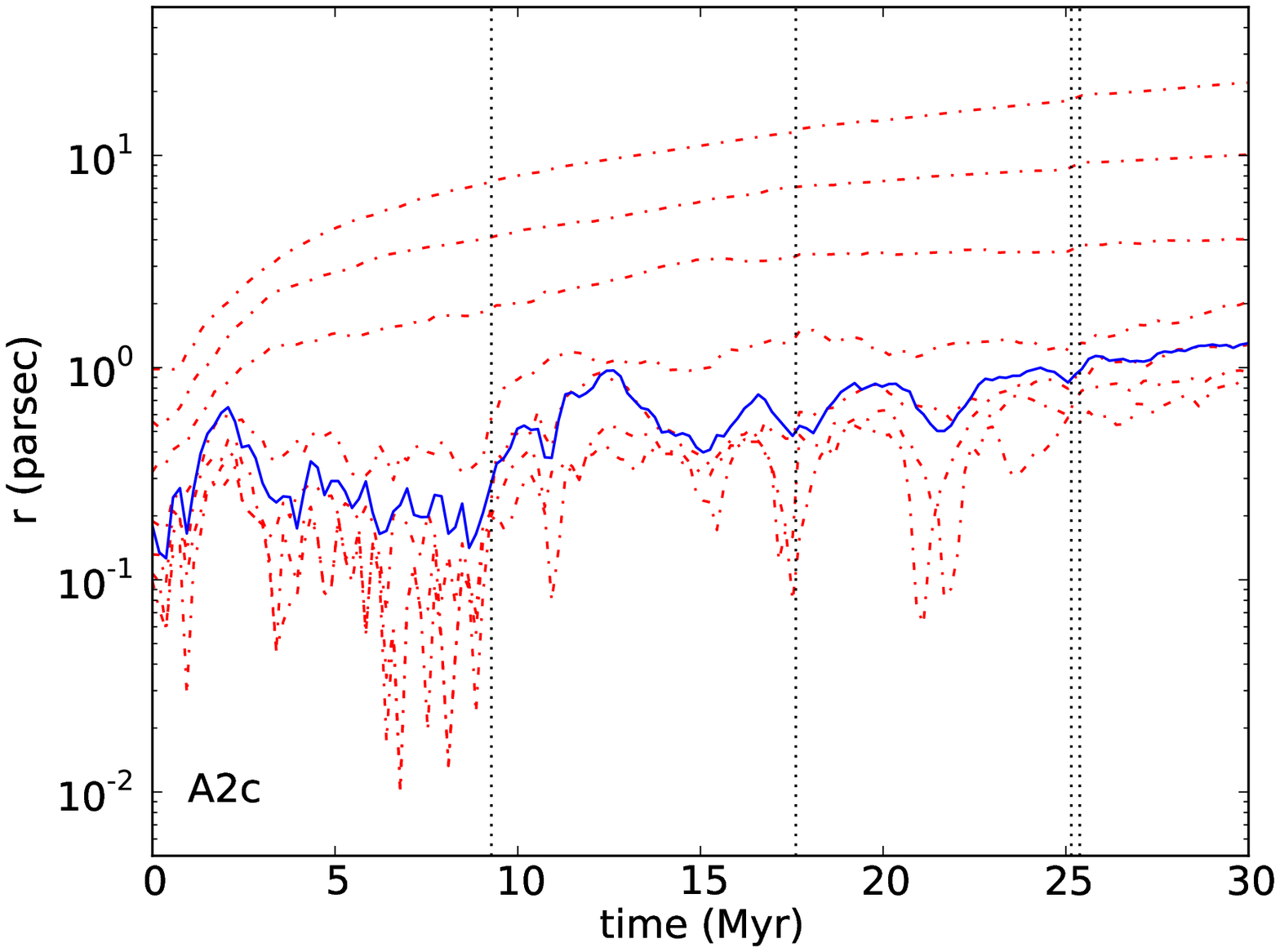}
\includegraphics[ width=0.32\textwidth]{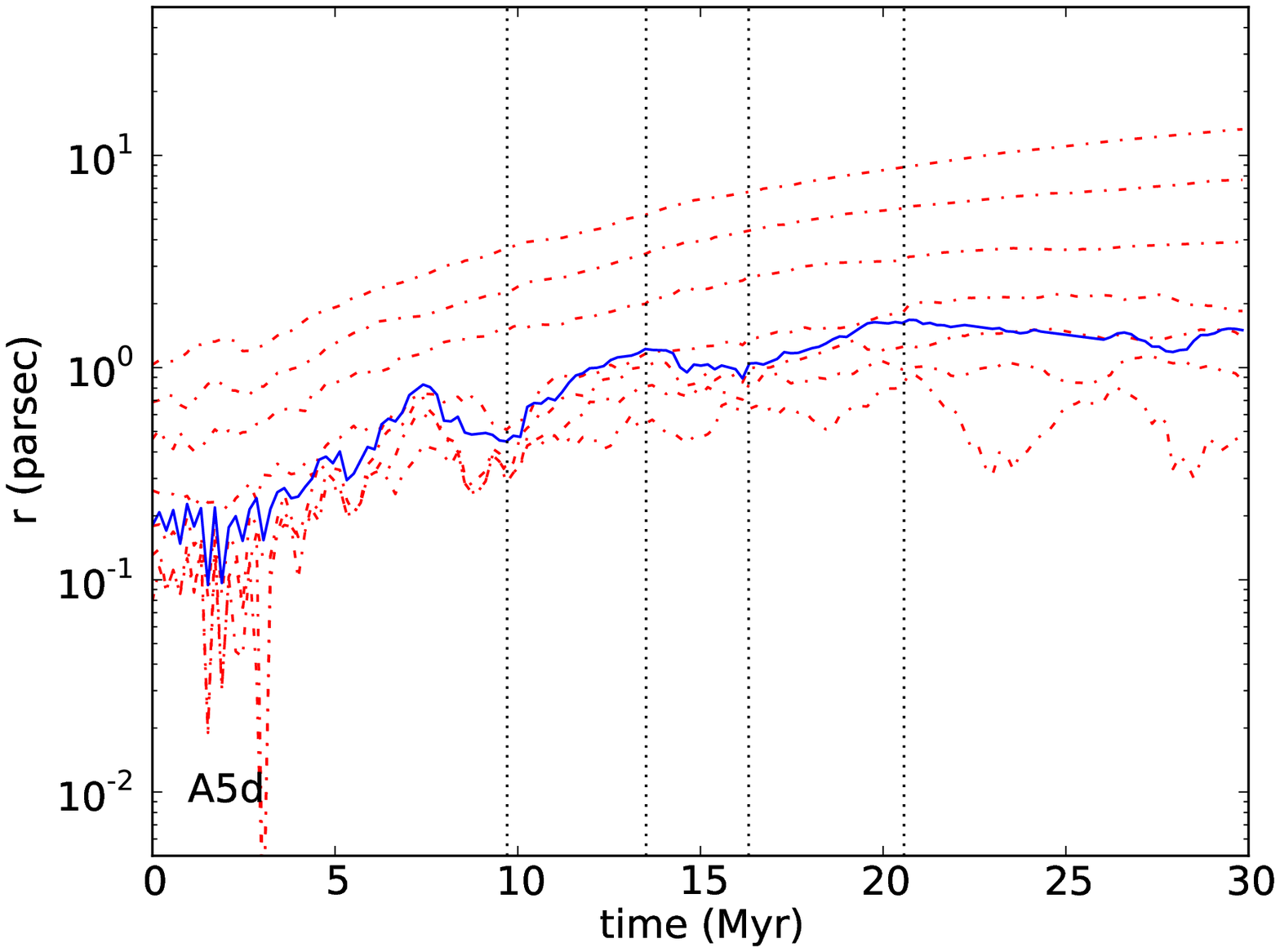}
\caption{Lagrangian radii, core radius and central gas density for the
alternative runs A2c and A5b. These differ from the corresponding runs
in Fig.~\ref{fig_lr} only in the random seed that was used to sample the
density distribution and IMF. Plotted in red dashed lines are the 
(0.02, 0.05, 0.1, 0.2, 0.5, 0.75 and 0.9) Lagrangian mass radii of the 
stellar mass distribution. Blue line is the core radius. Vertical dotted 
lines indicate supernova events.}
\label{fig_lr2}
\end{figure}

\begin{figure}
\centering
\includegraphics[ width=0.32\textwidth]{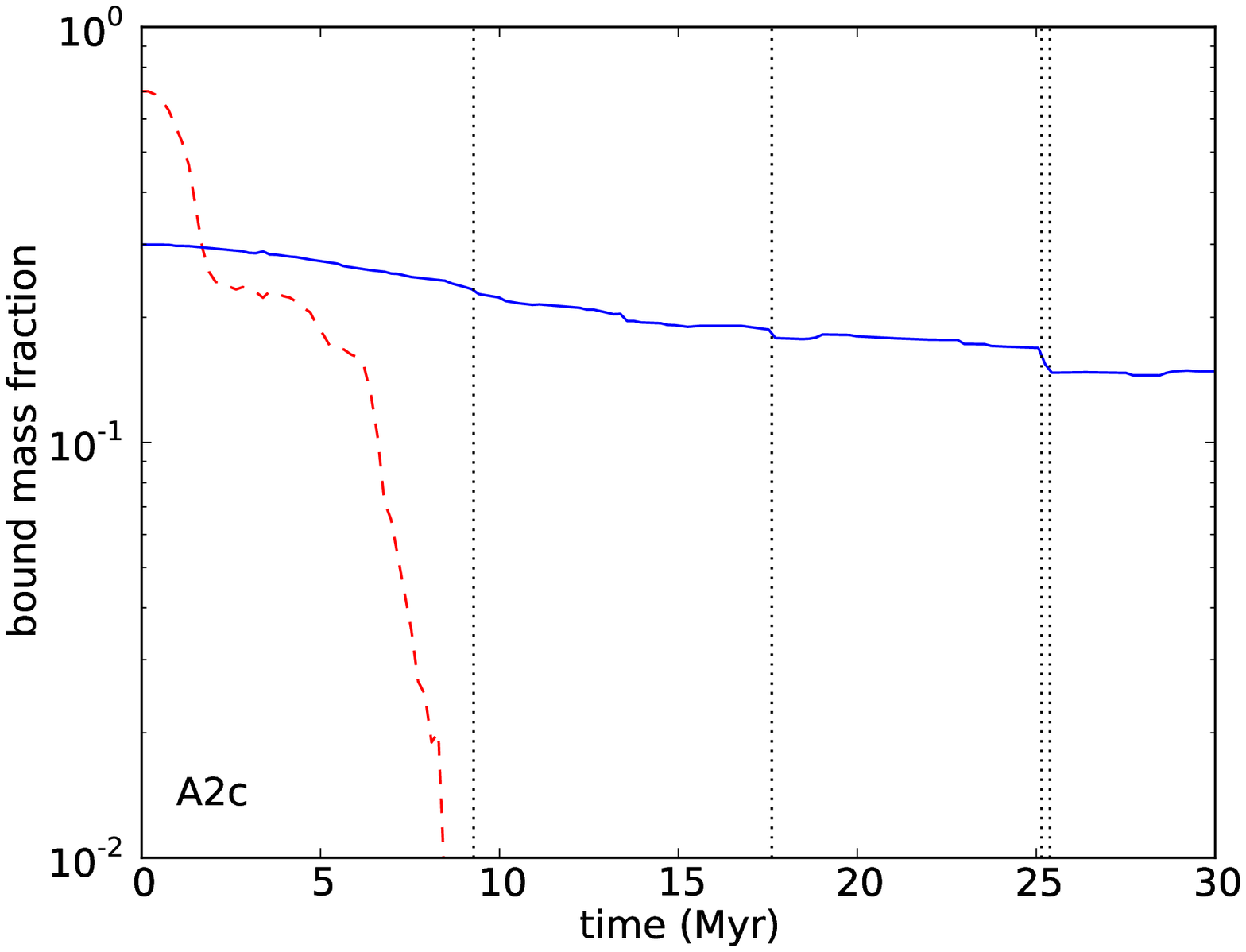}
\includegraphics[ width=0.32\textwidth]{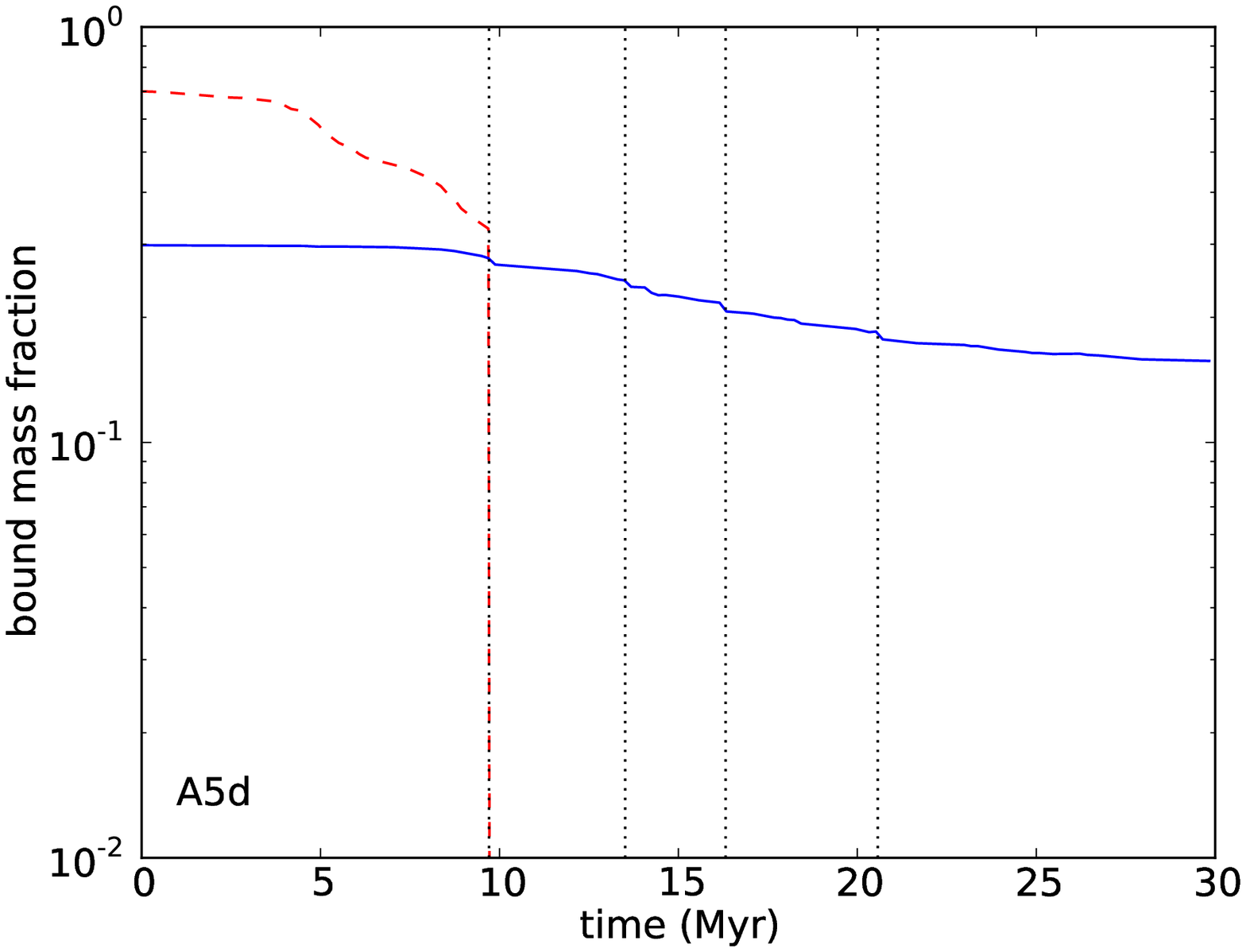}
\caption{Stellar and gaseous bound mass fractions for the
alternative runs A2c and A5b. These differ from the corresponding runs
in Fig.~\ref{fig_lr} only in the random seed that was used to sample the
density distribution and IMF. Plotted is the bound stelllar fraction 
(blue line) and the bound gaseous fraction (red dashed lines)
 as fraction of the total initial system mass.
Vertical lines indicate supernova events.}
\label{fig_bf2}
\end{figure}

For the low mass clusters we examine here different realizations of the
stellar density distribution and the IMF will have very different masses and
orbits for the heavy stars (and thus in the number, timing and location of
supernovae) just due to low number statistics (the number of stars larger
than $10 \Msun$ is $\approx 4$). We have run additional realizations of 
the A2 and A5 simulations to quantify this effect. For the A2 runs the range 
in final stellar bound fractions is $50-70\%$, with the final core radii lying between 0.6 and 
1.2 parsec. For the A5 runs we find similar ranges, 
$50-65\%$ and $0.8-1.5$ parsec. However if we compare the runs in more detail 
we see considerable differences. Two examples of this are given in 
figures~\ref{fig_lr2} (Lagrangian and core radii) and ~\ref{fig_bf2} 
(bound fractions).  The A2c run quickly loses $60\%$ of its mass
within 1.5 Myr, after which the gas loss slows down momentarily. This 
results in a stellar distribution that quickly expands, but then contracts 
again until the remaining gas is expelled. The A5d run seems to follows a 
similar pattern to the A5 run, if one compares the core radius. However if
one looks more closely to the inner Lagrangian radii the A5d run is very
different from the A5 run. In the latter case the (e.g.) $10\%$ Lagrangian
radius hovers around 0.2 parsec after 10 Myr, while in the A5d run it
expands more than $4\times$ further, to $\approx 0.9$ parsec.

\subsection{Sub-virial velocity dispersions} 

\begin{figure*}
\centering
\includegraphics[ width=0.32\textwidth]{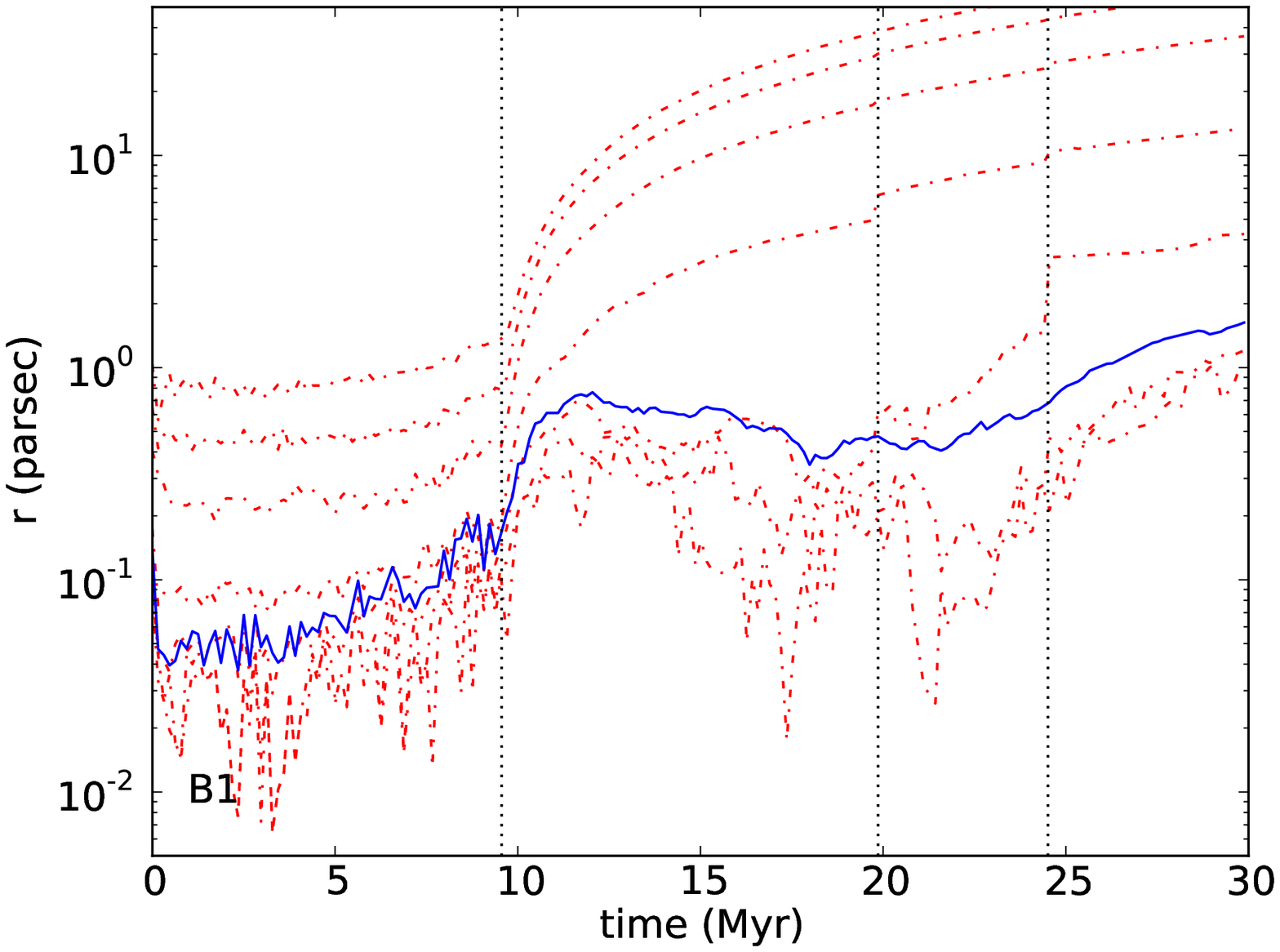}
\includegraphics[ width=0.32\textwidth]{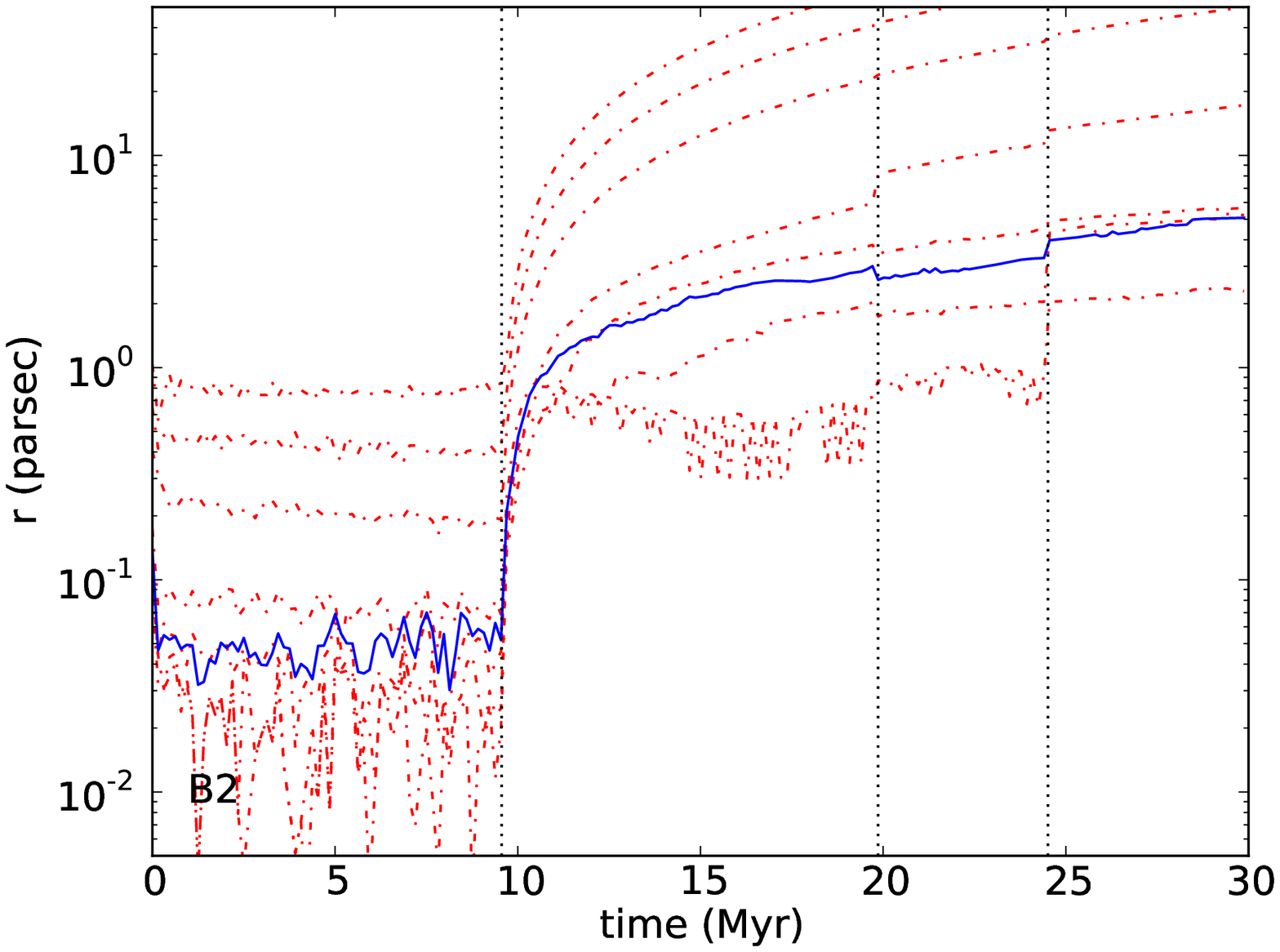}

\includegraphics[ width=0.32\textwidth]{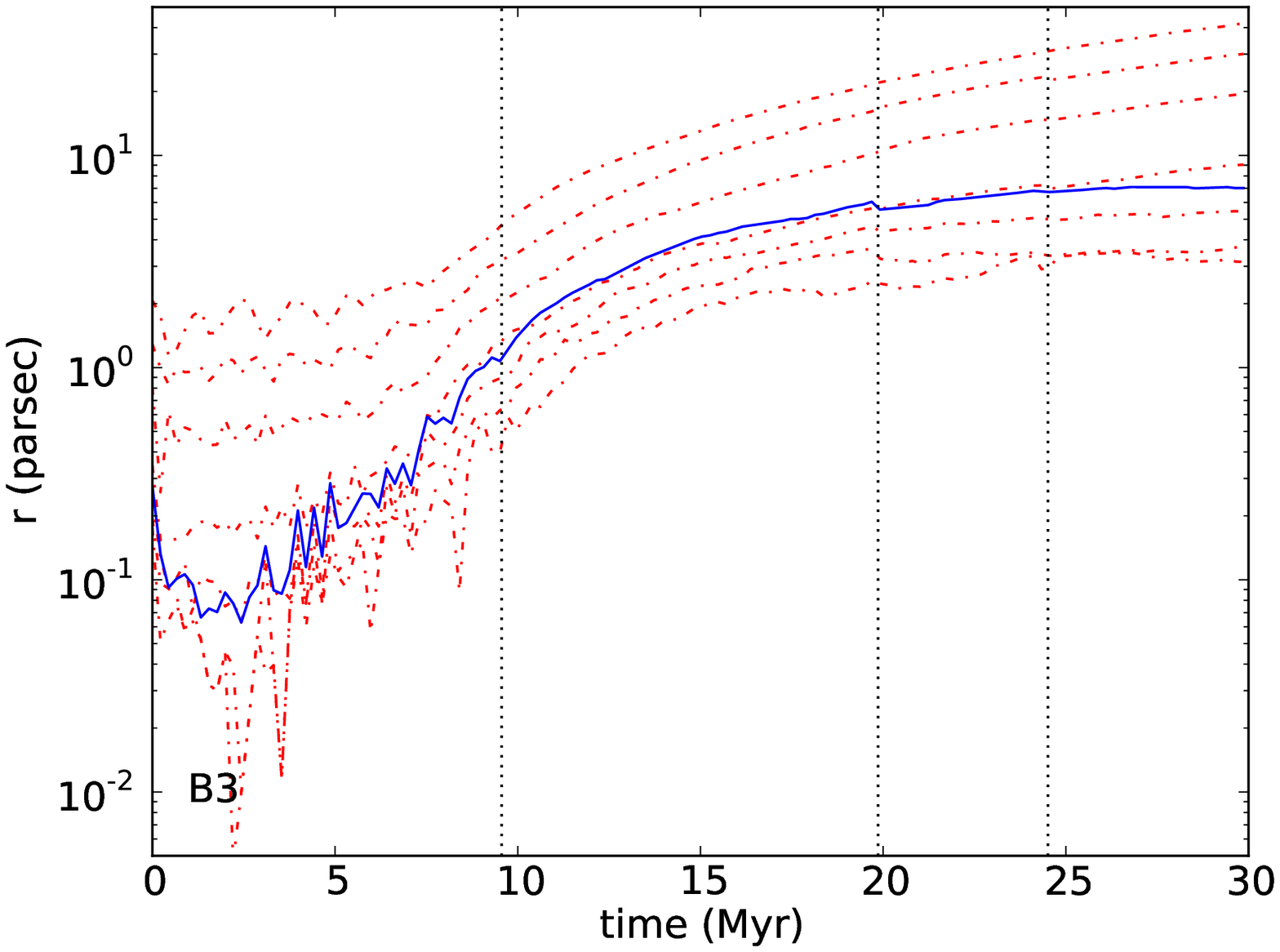}
\includegraphics[ width=0.32\textwidth]{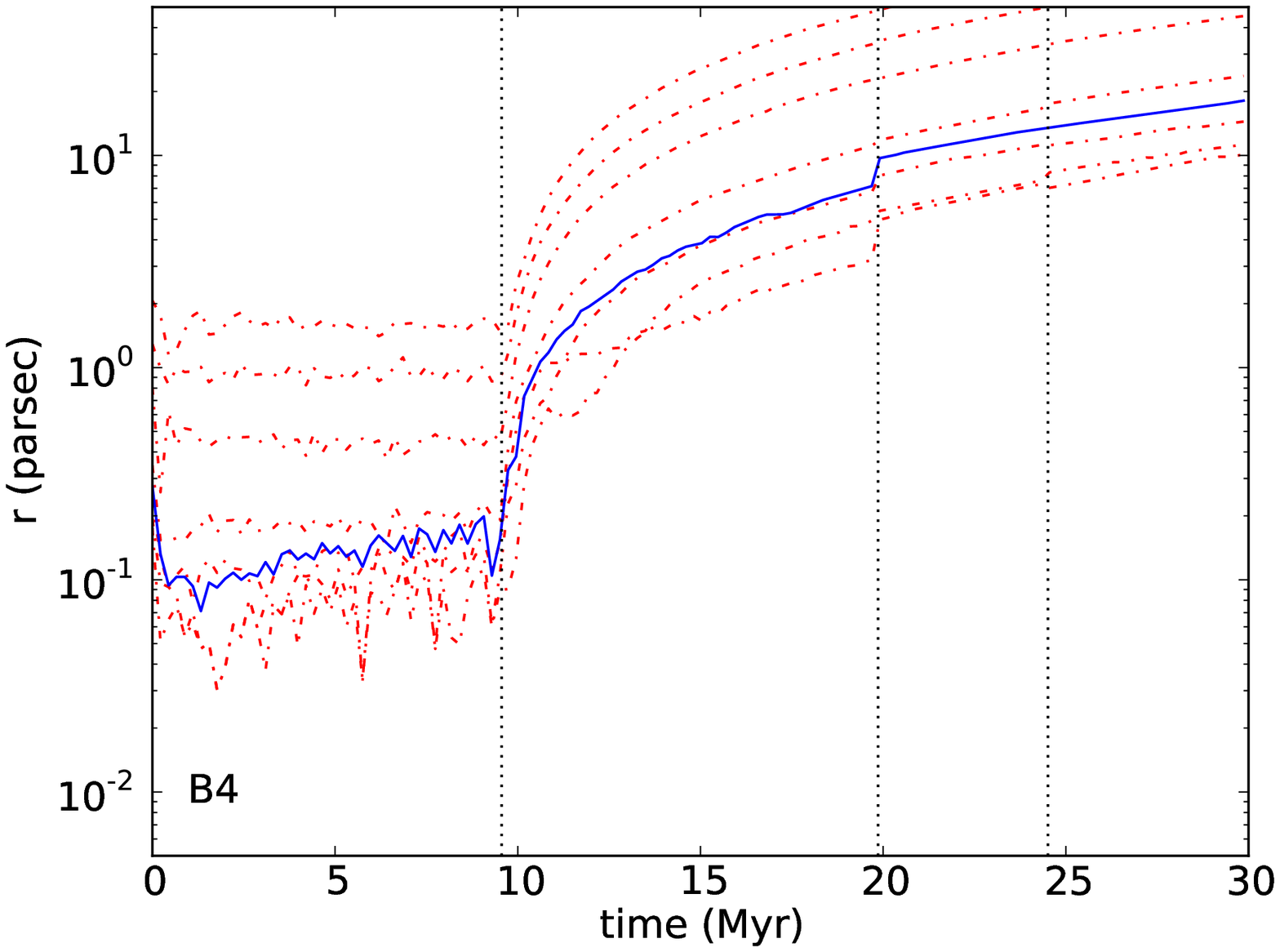}
\caption{Lagrangian radii and core radius for runs
B1-B4. Plotted in red dashed lines are the (0.02, 0.05, 0.1, 0.2, 0.5, 0.75 and 0.9 )
Lagrangian mass radii of the stellar mass distribution. Blue line is 
the core radius. Vertical dotted lines indicate supernova events.}
\label{fig_lrB}
\end{figure*}

\begin{figure*}
\centering
\includegraphics[ width=0.32\textwidth]{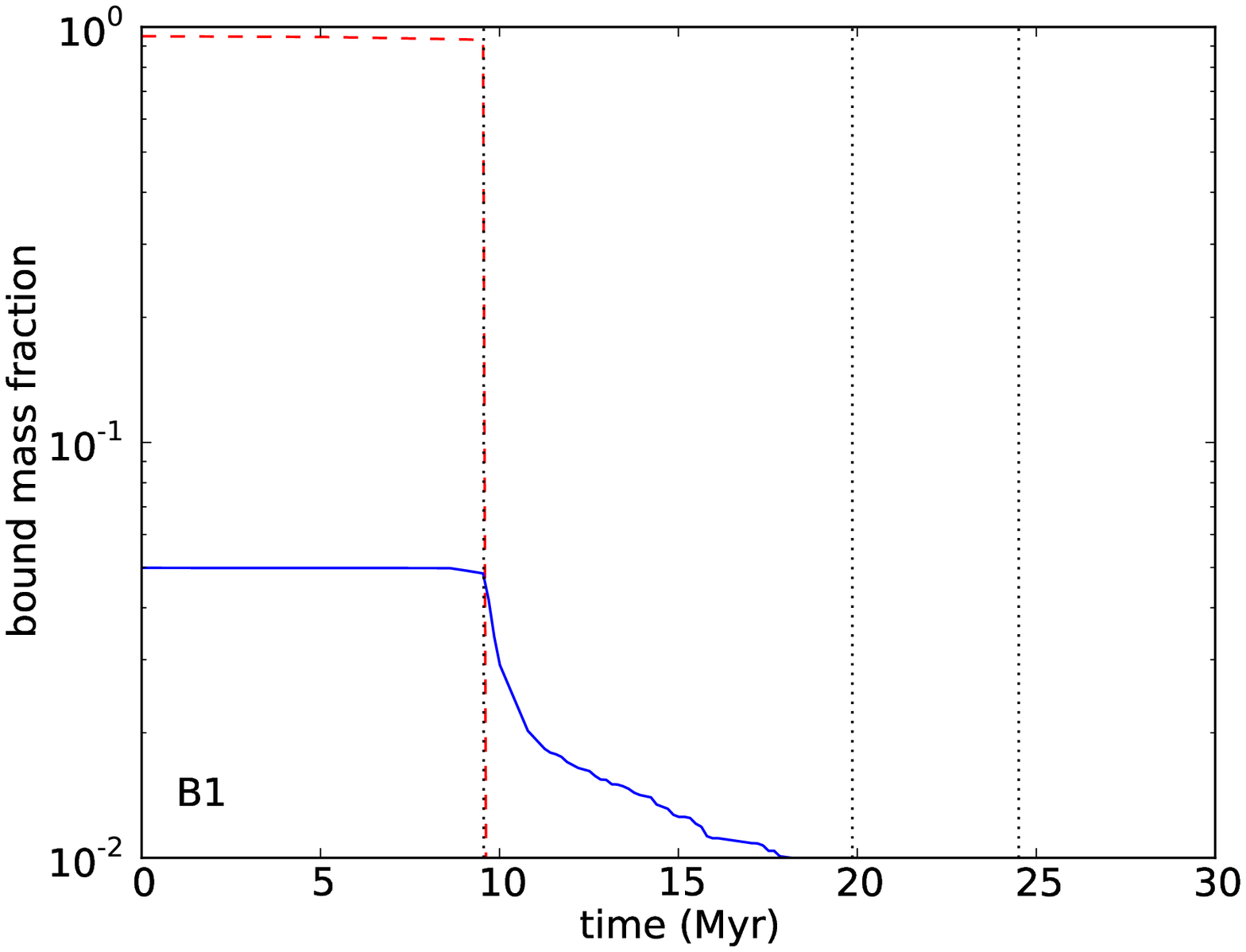}
\includegraphics[ width=0.32\textwidth]{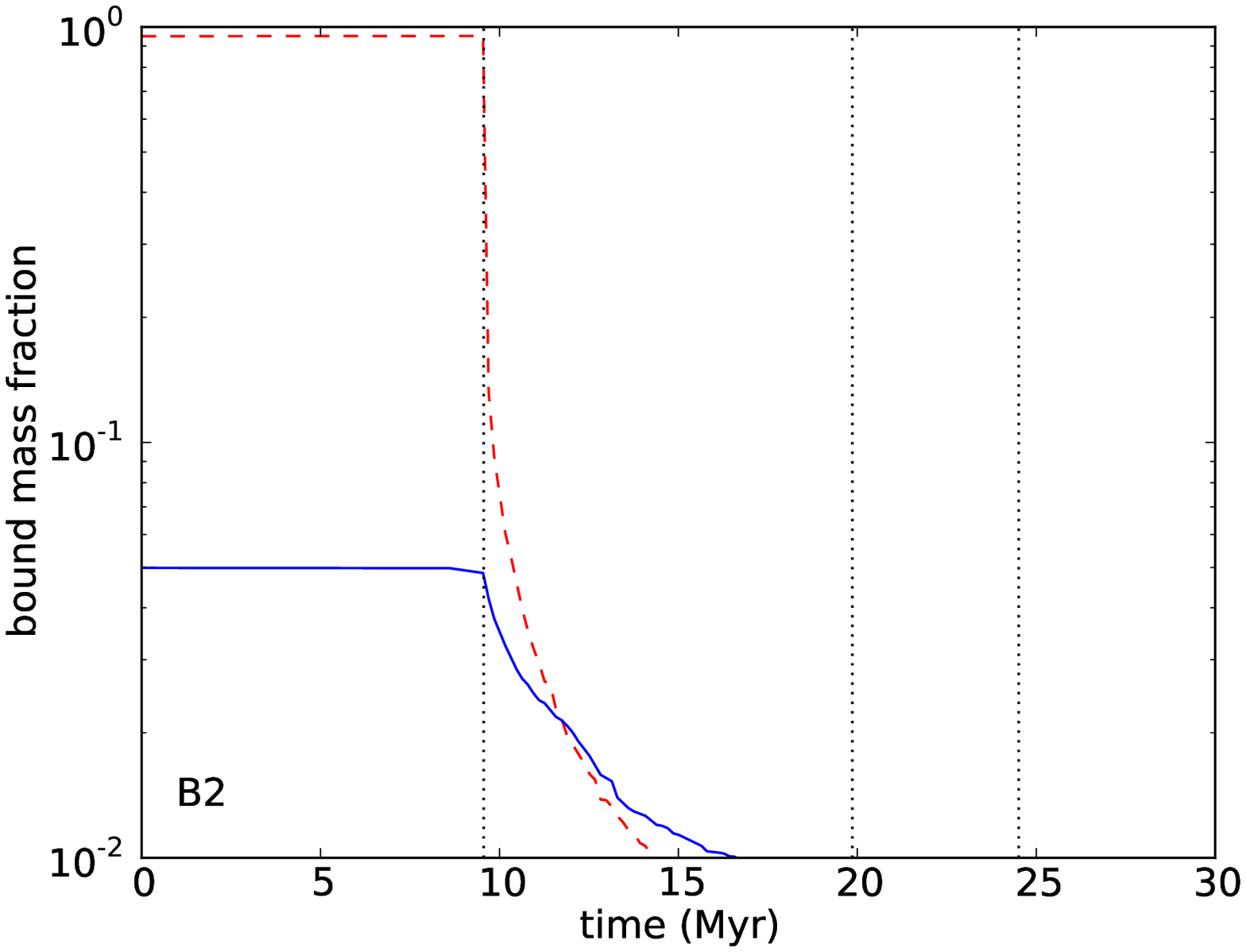}

\includegraphics[ width=0.32\textwidth]{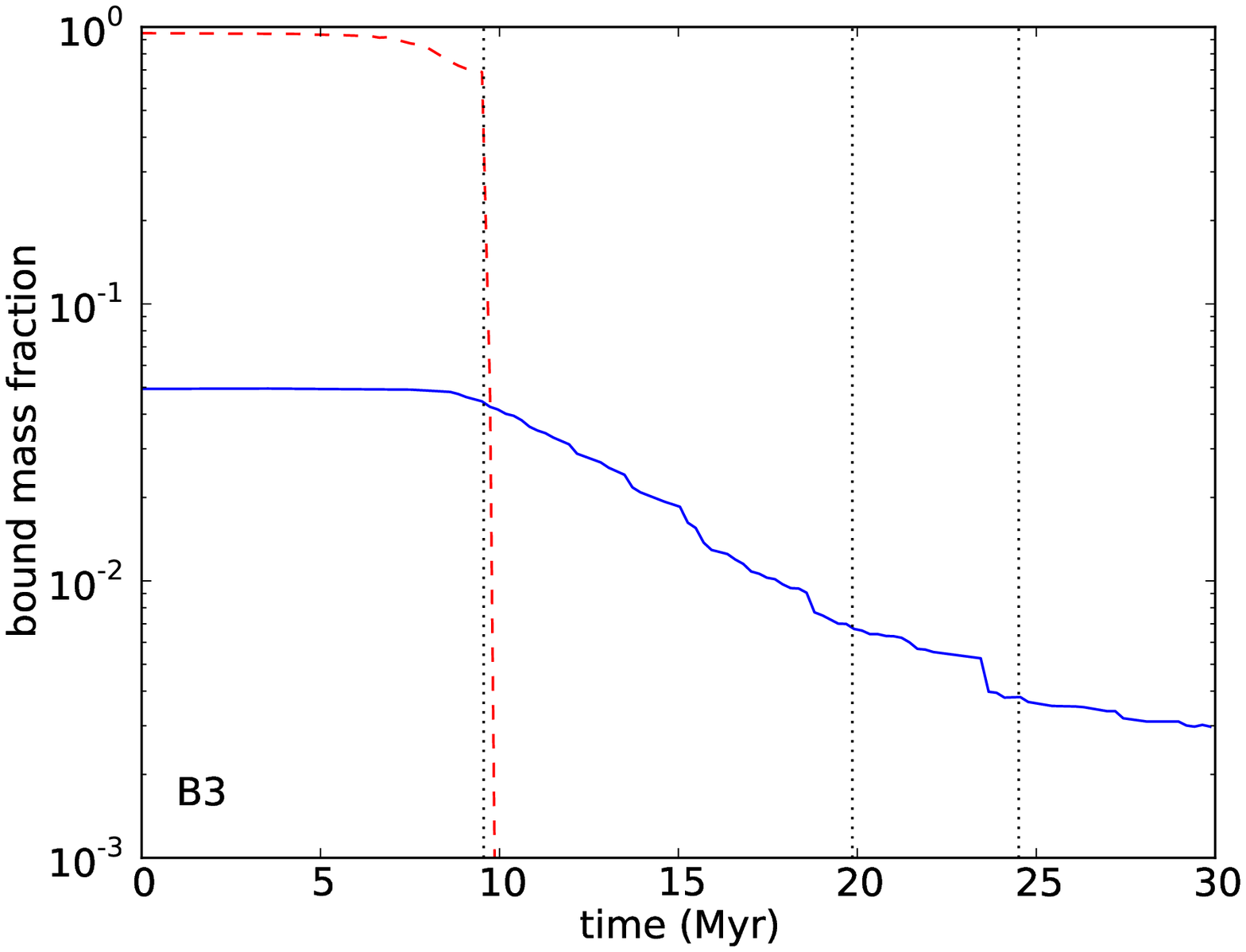}
\includegraphics[ width=0.32\textwidth]{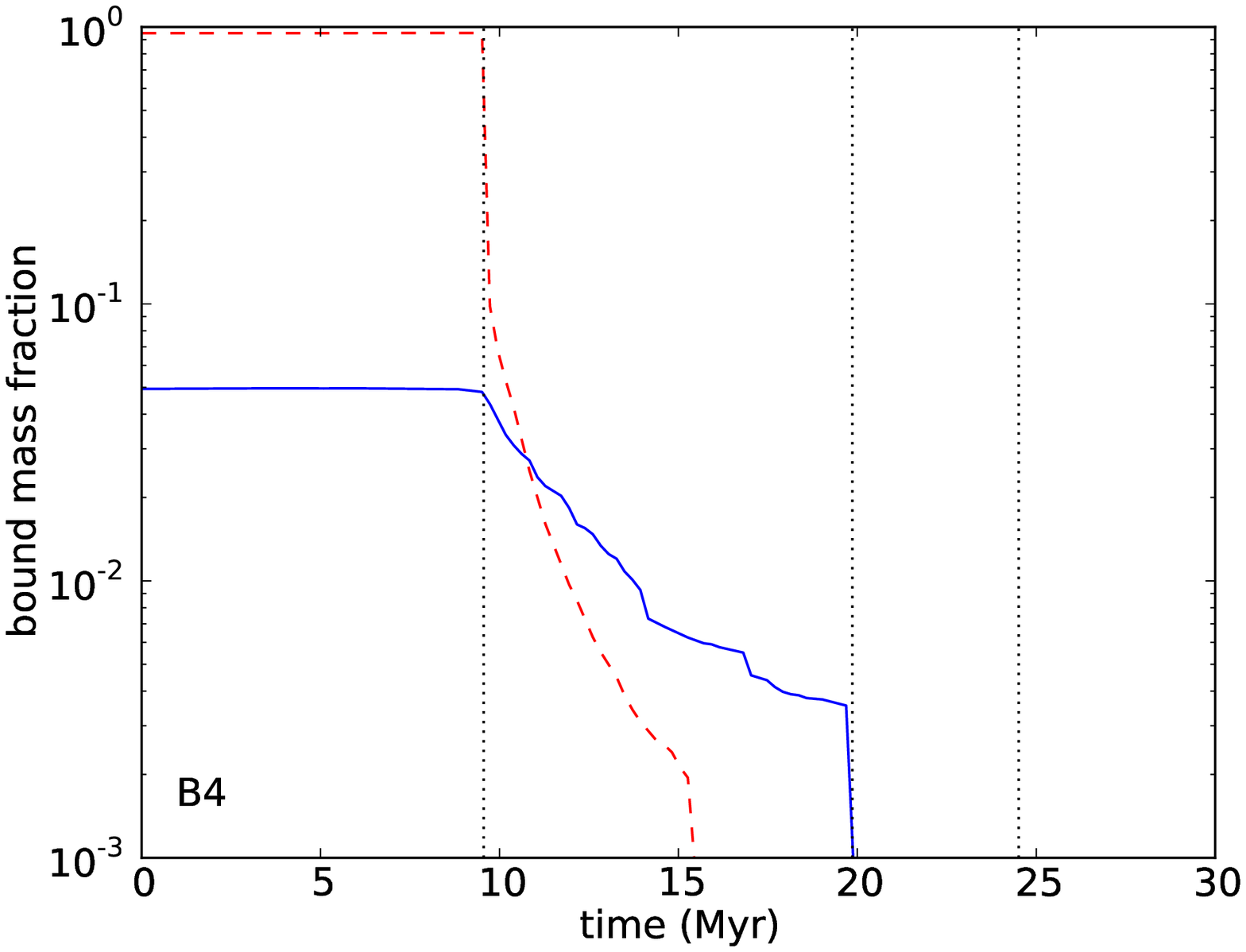}
\caption{Stellar and gaseous bound mass fractions for runs B1-B4. Plotted is
the bound stelllar fraction (blue lines) and the bound gaseous fraction (red
dashed lines)  as fraction of the total initial system mass. 
Vertical lines indicate supernova events.}
\label{fig_bfB}
\end{figure*}

\begin{figure}
\centering
\includegraphics[ width=.55\textwidth]{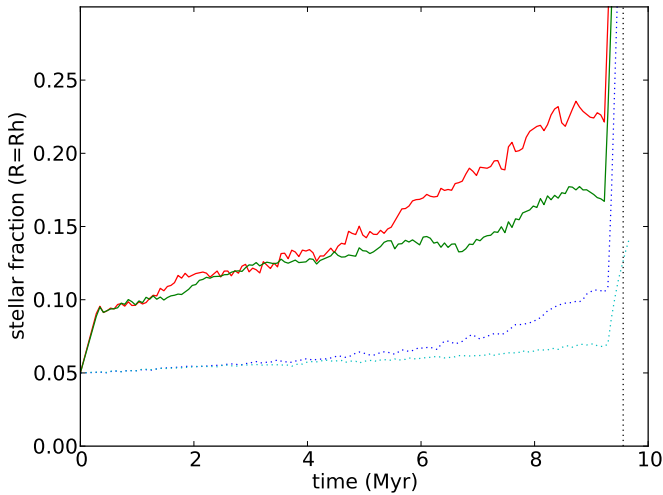}
\caption{Stellar mass fraction at the (stellar) half mass radius. Plotted is
the mass fraction at the half mass radius of the stellar distribution as a
function of time (in the initial stages of the simulations) for the
sub-virial velocity dispersions, runs B1 (red) and B2 (green), and their 
equivalent with full velocity dispersion A1 and A2 (blue and cyan dotted
lines).}
\label{fig_fs}
\end{figure}

The time evolution of the Lagrangian radii and the bound fractions of the
models with sub-virialized velocity dispersion are plotted in
figures~\ref{fig_lrB} and~\ref{fig_bfB}. Figure~\ref{fig_lrB} shows that the
stellar distributions quickly shrink compared to the virialized cases:
e.g. for B1 the 0.5 Lagrangian radius drops roughly a factor 2 within less than 0.5
Myr. This collapse of the stellar distribution can increase the survival of 
the stellar cluster, because the collapse increases the local
SFE \citep{Smith2011}. The B1 model shows this effect, because compared
with the A1 model it shows a much less drastic expansion of the core radius
(fig.~\ref{fig_lrB}), although in the end the cluster still becomes
completely unbound (fig.~\ref{fig_bfB}).  However, surprisingly, if we look at
the B2 model with a lower $f_{\rm fb}$ (which is the sub-virialized version
of the A4 run) we see exactly the opposite: in this case the cluster
dissolves quickly, where as one may expect on the basis of the A1 and A4 run
that the B2 run would dissolve much more slowly. Runs
B3 and B4 show a similar pattern. The reason for this becomes clear if we
examine figure~\ref{fig_fs}, where we have plotted the stellar mass fraction
(the ratio of stellar mass to total mass) within the stellar half mass
radius as function of time up to the time of the first supernova. The dotted
lines here show the A1 and A4 runs, while the drawn lines are the B1 and B2
runs. The B1 and B2 runs initially show an increase of a factor of 2 in stellar
fraction (which is smaller than expected from the contraction of the half
mass radius shown in fig.~\ref{fig_lrB}, indicating the gas distribution
also contracts). After that they both show a slow increase, but crucially 
the B1 run shows a stronger increase to around $23\%$
vs the B2 run. Although the increase is modest, it is probably the cause of 
the difference between the B1 and B2 run. This is consistent with the
\citet{Baumgardt2007} results, which show a very sensitive dependence of the
bound fraction on SFE around a value of 0.25.

Note that Figure~\ref{fig_fs} shows that the stellar fraction of the A2 run 
is almost the initial value (SFE\,$=0.05$). In this case 
the survival of the cluster after the first supernova is not due to a 
high stellar mass fraction, but due to the fact that not all gas is expelled
(fig.~\ref{fig_bf}). The effective SFE hence increases to 
SFE\,$\approx 0.05+0.2$ (measured after 1 Myr) if one takes into account the
gas that is not expelled. Hence the B1 and A2 run show an increased
survivability due to completely different mechanisms: the B1 stellar
component clears its environment of gas slowly through stellar winds and
adapts to the slowly changing potential, while the stellar component of the A2 
run is held together after the first supernova by remnant gas (in spite
of this only being a very minor fraction of the initial gas component). 

\subsection{Mass segregation and selective mass loss}

\begin{figure}
\centering
\includegraphics[ width=.55\textwidth]{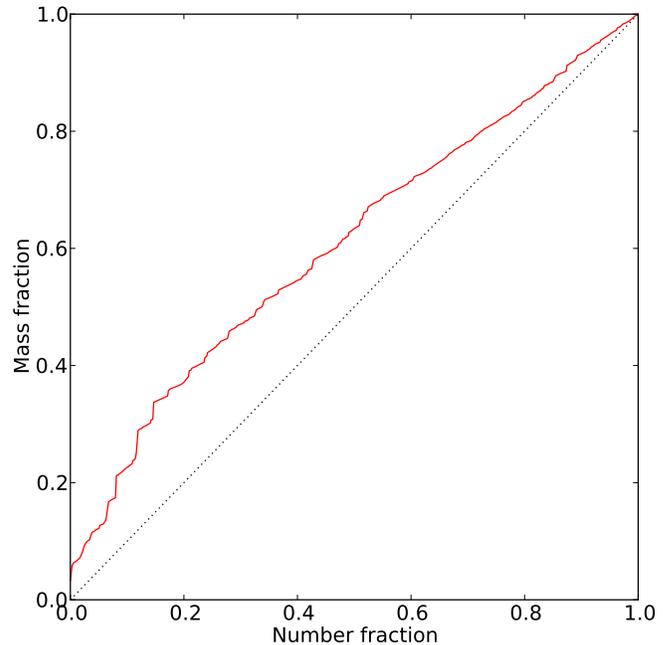}
\caption{Mass fraction versus number fraction for the A2 run. Plotted 
is the mass fraction at given radii versus vs the number fraction at 
those same radii (at 30 Myr). A cluster without mass segregation
would follow the dotted diagonal line. Following \citet{Converse2008} we take
the area between the curve and the diagonal as a measure for the mass
segregation (the Gini coefficient for the cluster).}
\label{fig_gini}
\end{figure}

\begin{figure}
\centering
\includegraphics[ width=.55\textwidth]{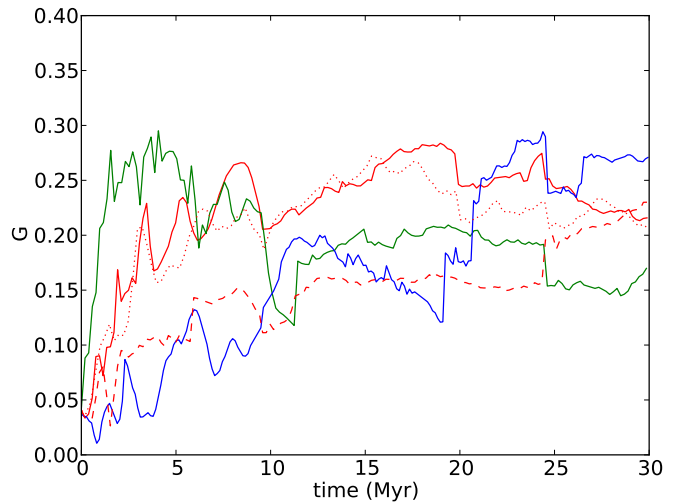}
\caption{Plot of the Gini coefficient G as a function of time for different
runs. The solid lines (red, green and blue) are runs with different 
initial half mass radius (A5, A8 and A10 runs respectively). Red dashed and
dotted lines are the A2 run and A6 run. G is the area ($\times2$) between 
the graph of mass fraction vs number fraction and the diagonal 
(Fig.~\ref{fig_gini}).}
\label{fig_gt}
\end{figure}

\begin{figure*}
\centering
\includegraphics[ width=0.32\textwidth]{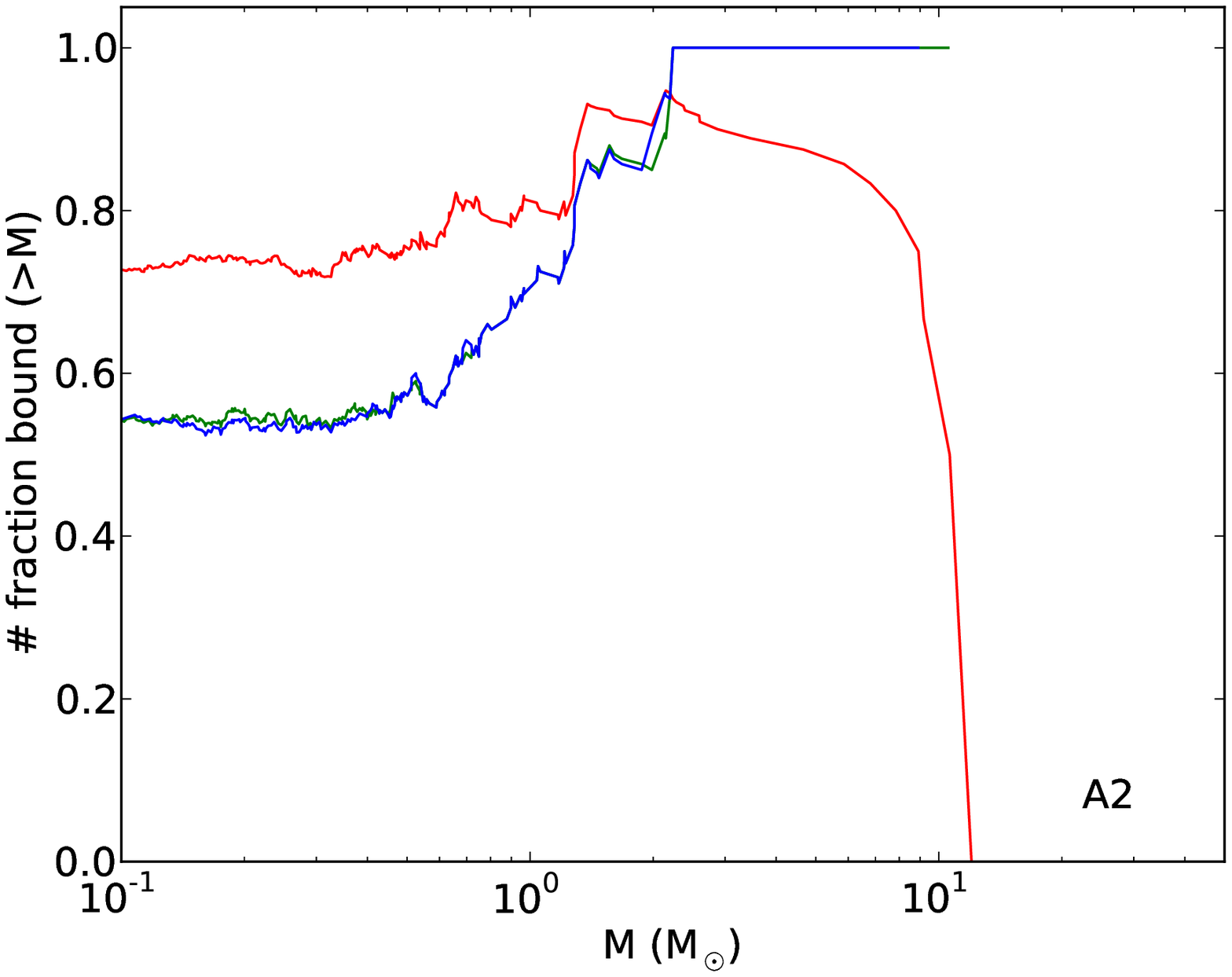}
\includegraphics[ width=0.32\textwidth]{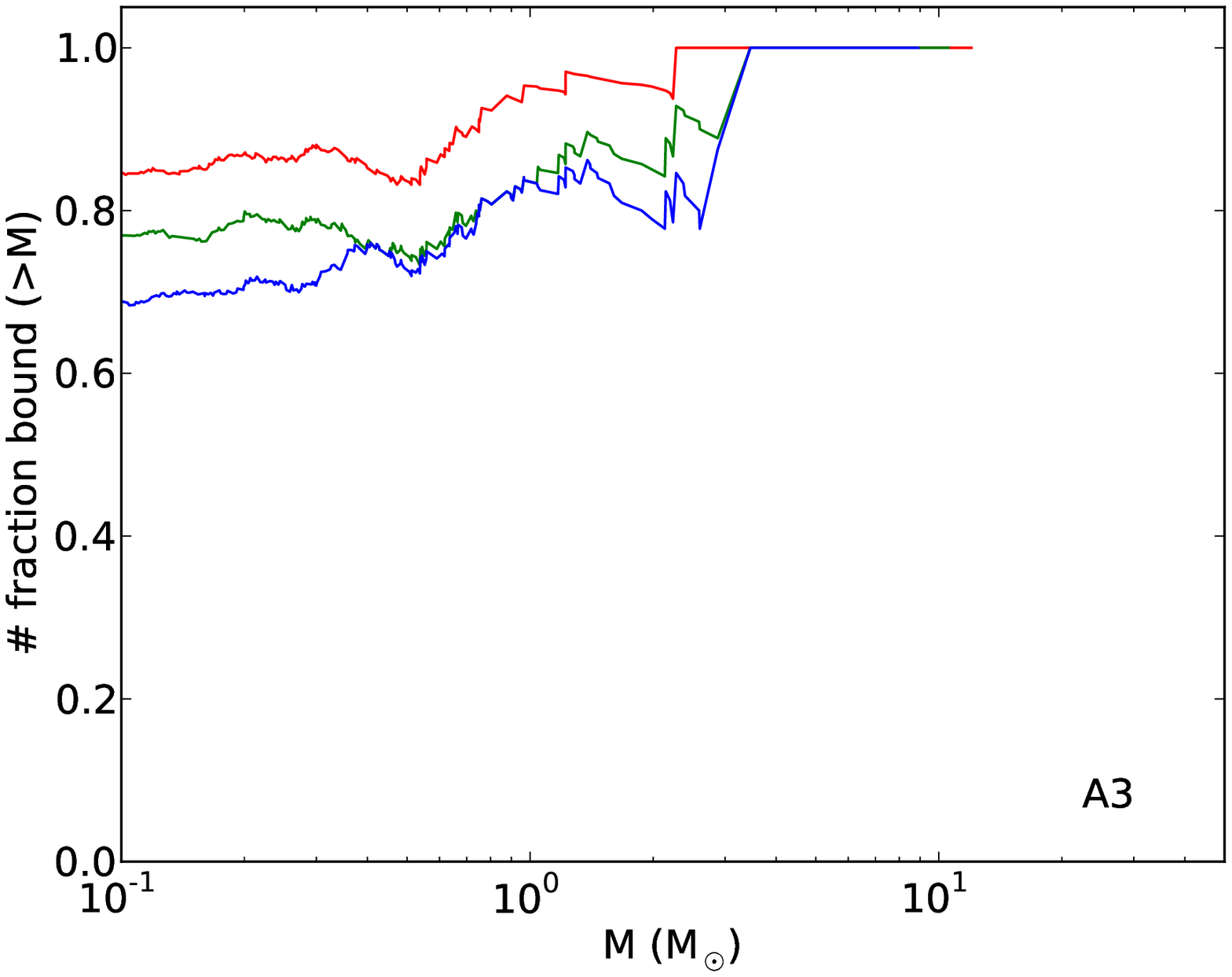}
\includegraphics[ width=0.32\textwidth]{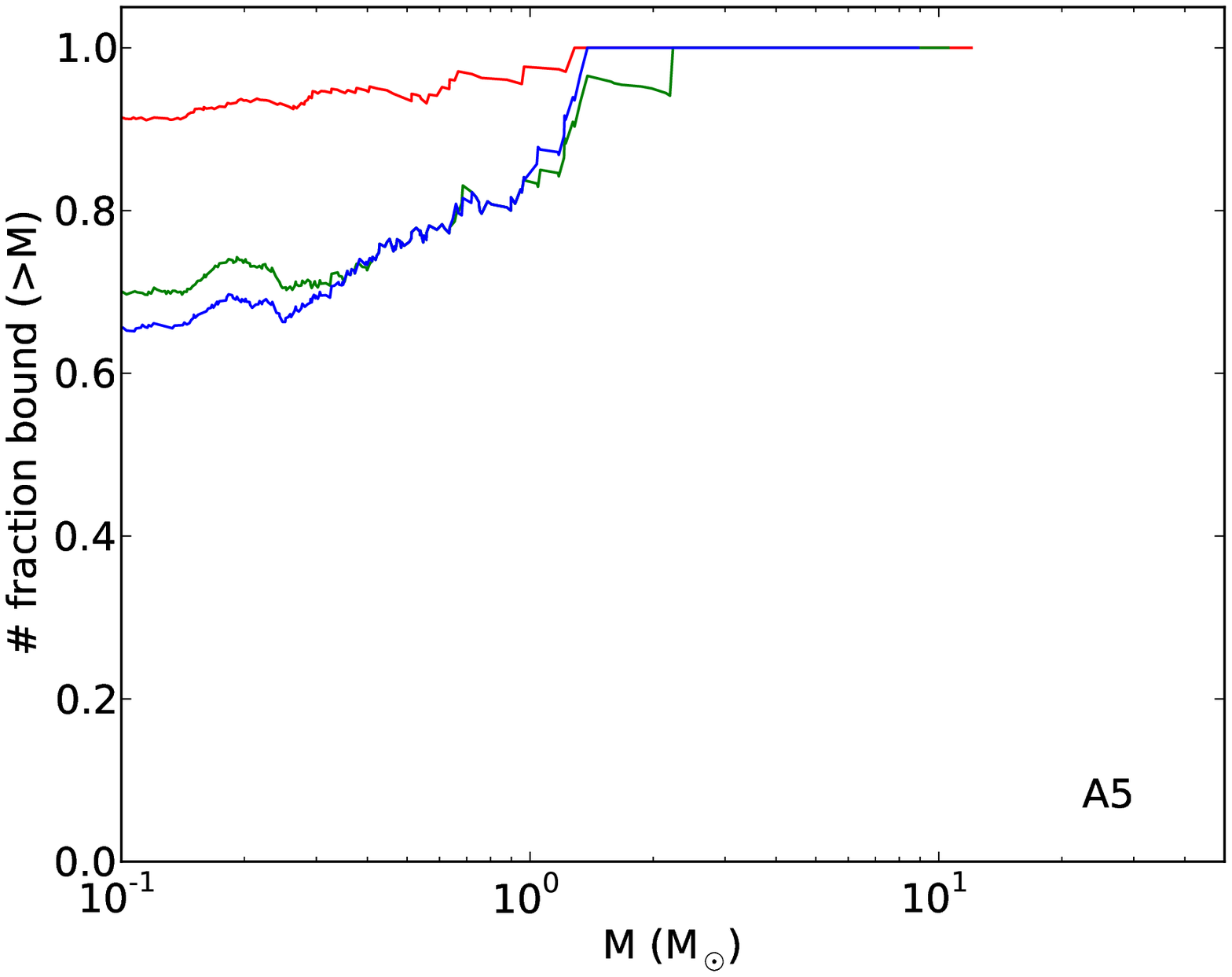}
\includegraphics[ width=0.32\textwidth]{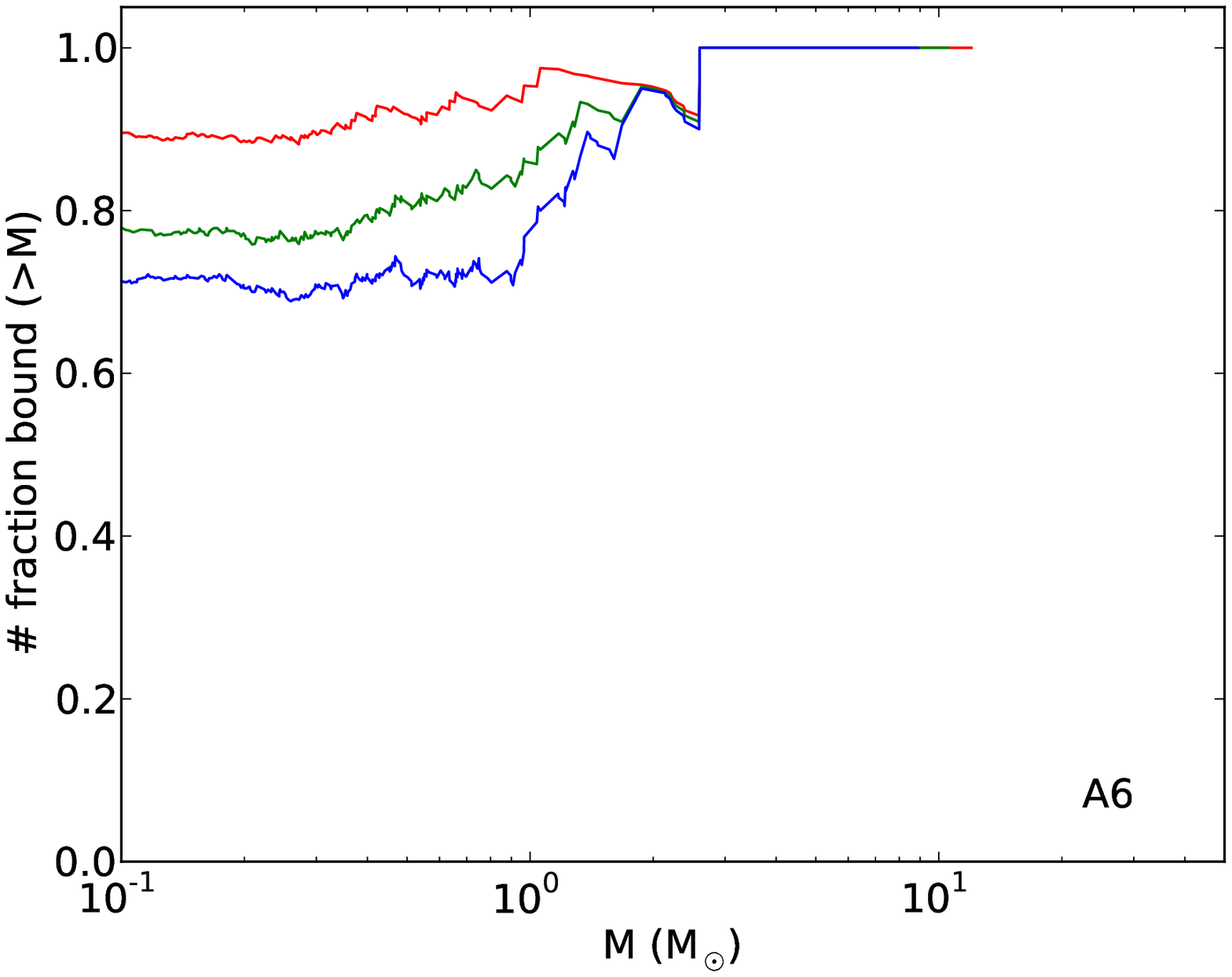}
\includegraphics[ width=0.32\textwidth]{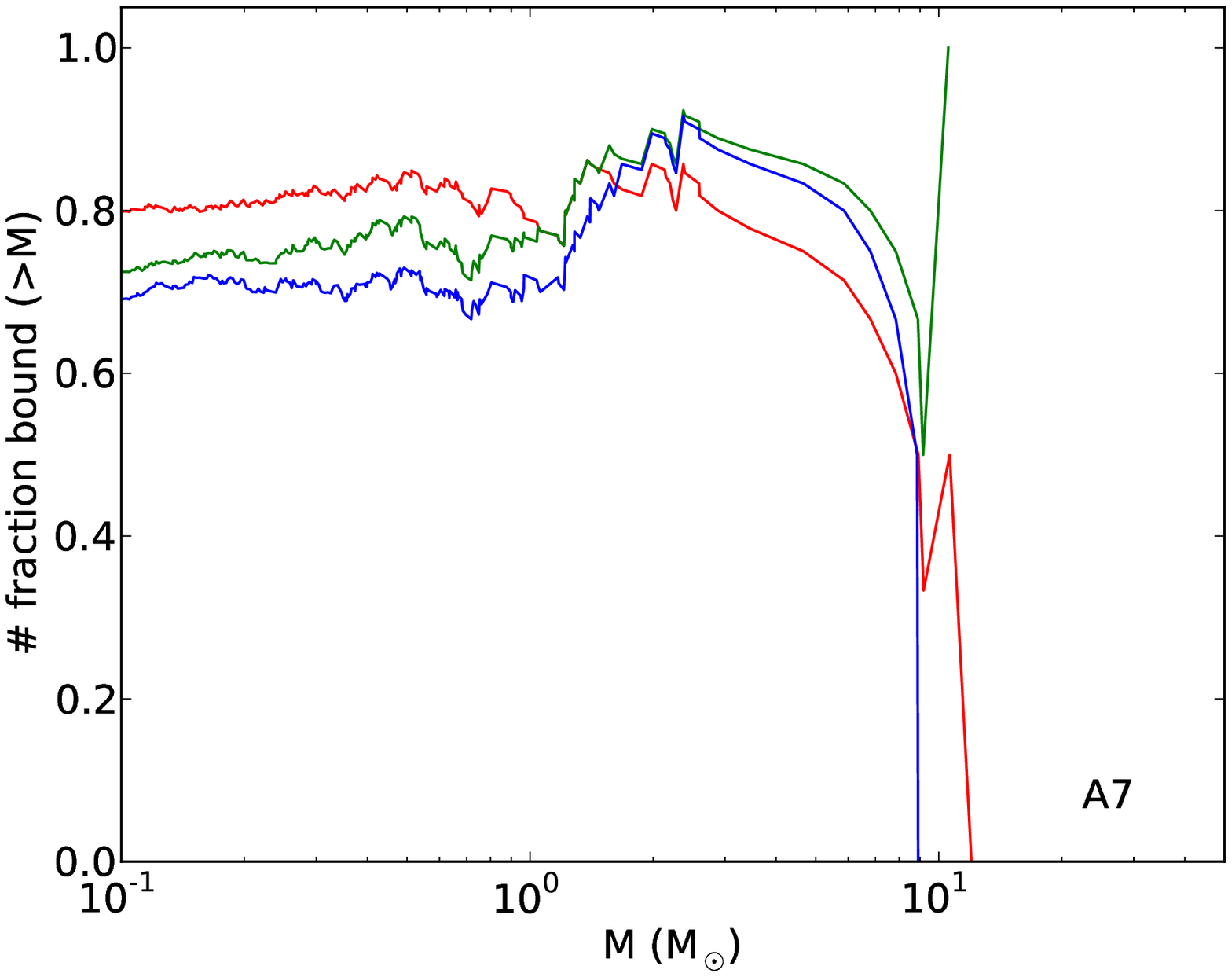}
\includegraphics[ width=0.32\textwidth]{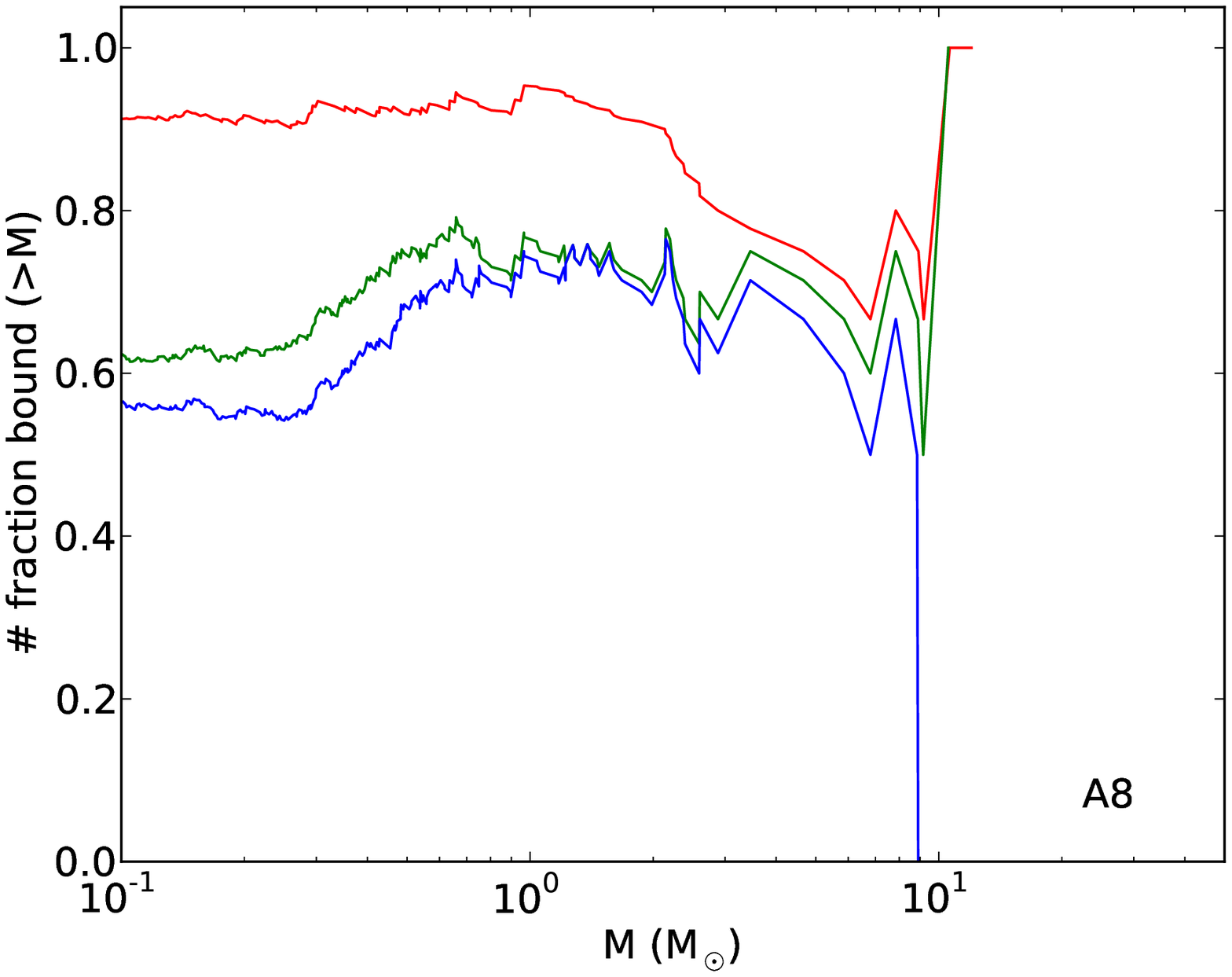}
\includegraphics[ width=0.32\textwidth]{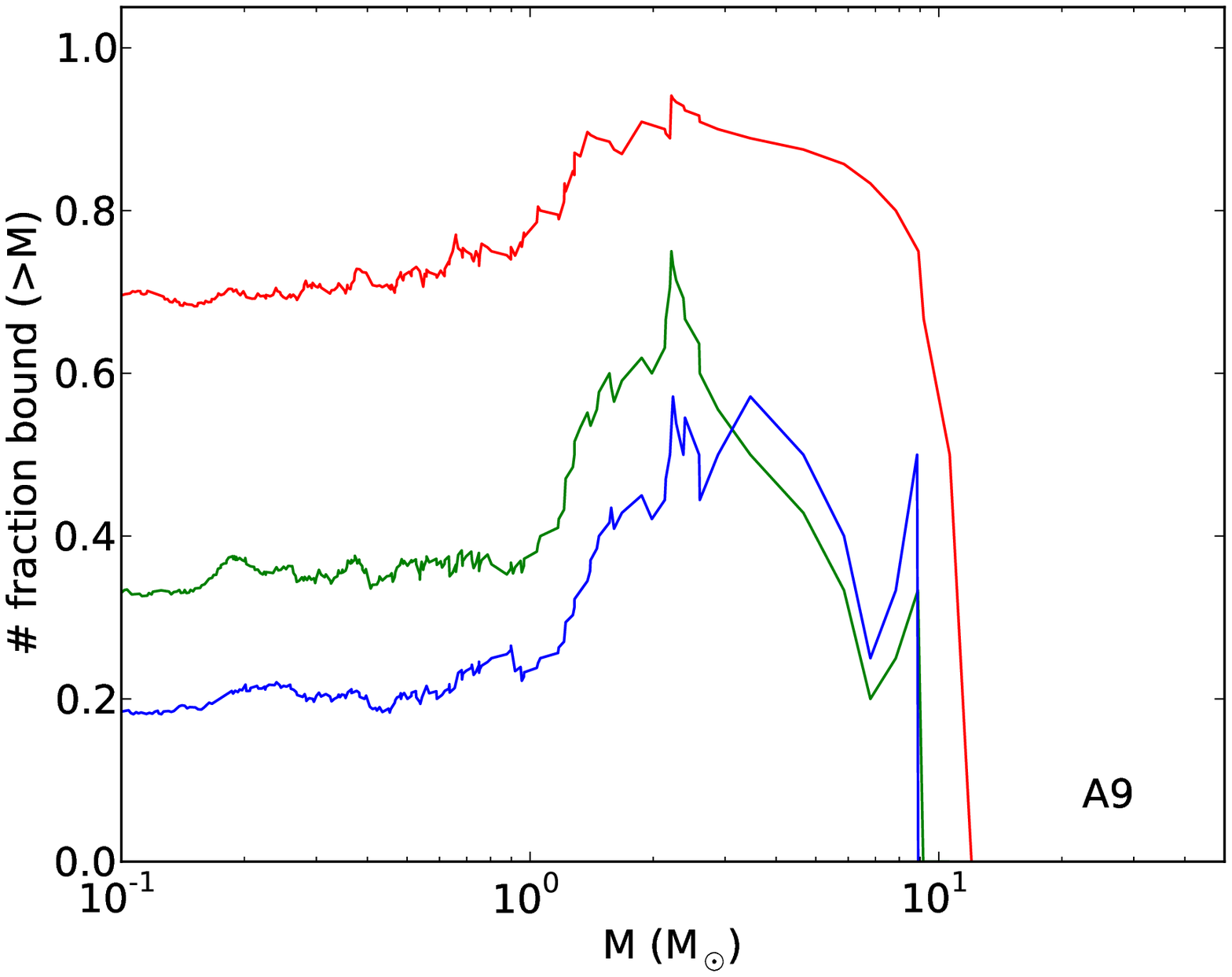}
\includegraphics[ width=0.32\textwidth]{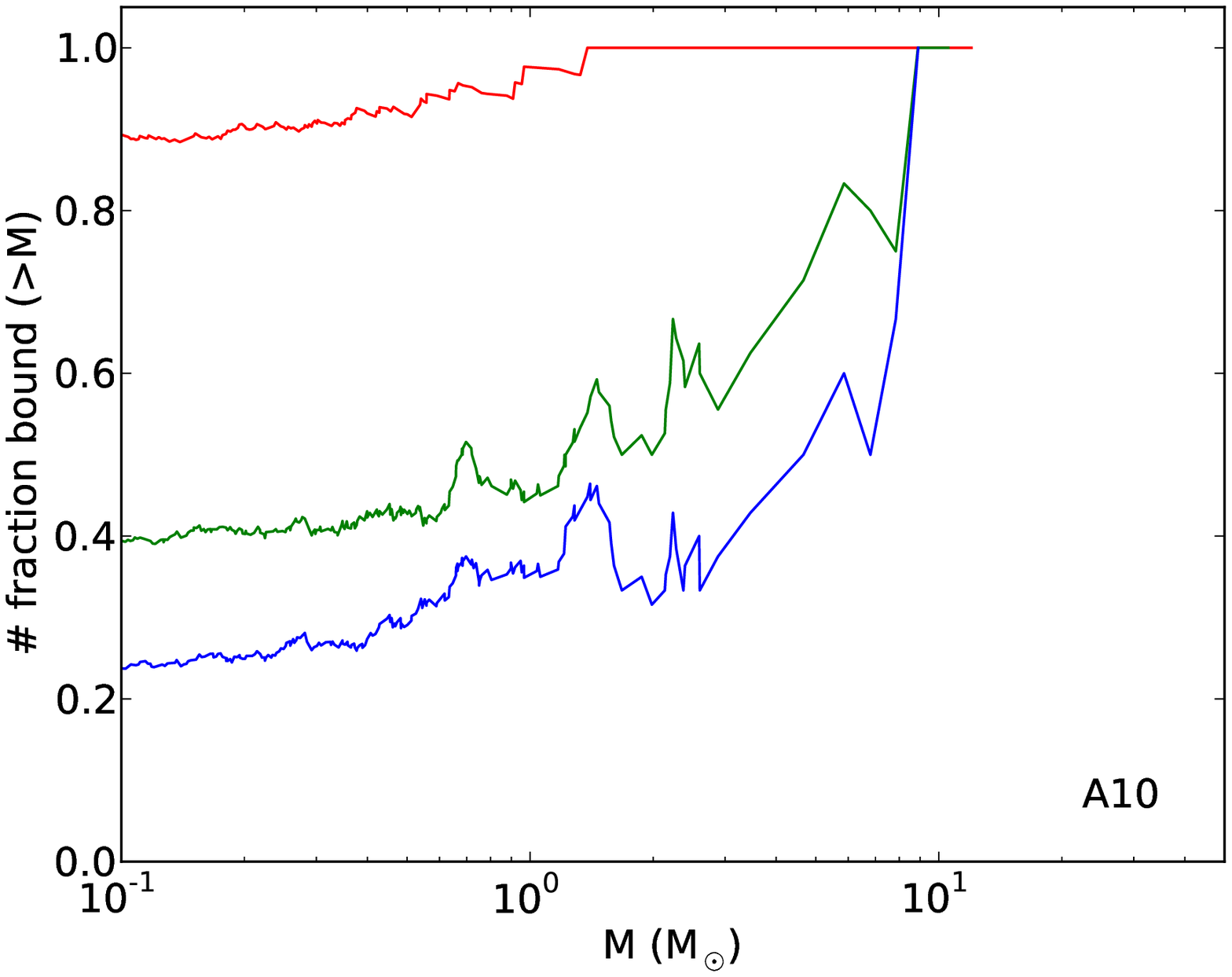}
\includegraphics[ width=0.32\textwidth]{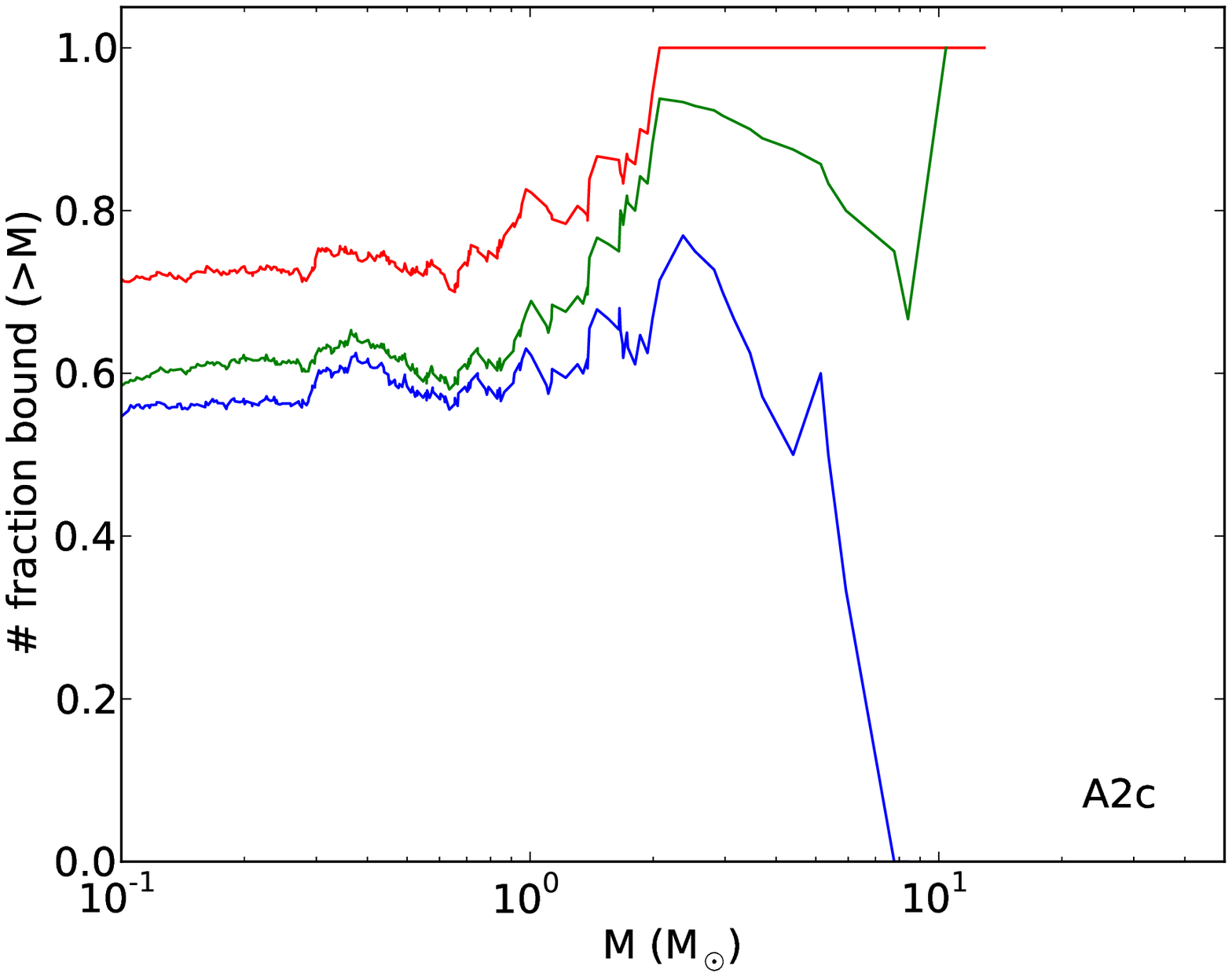}
\caption{
Cumulative bound number fraction as a function of mass.
Plotted is, as a function of mass, the number fraction of stars heavier than
that mass that are still bound at 10, 20 and 30 Myr. Shown are models from
the basis set A1-A10 which result in a bound cluster (and the A2c run). Note that errors on
these are not plotted - these increase from left to right from $\approx 0.03$
to $1$ for the right most point (because the right most point is based on a
single data point).
}
\label{fig_cbf}
\end{figure*}

Visual inspection of the maps such as those in figure~\ref{fig_maps} shows
that the resulting remnant cluster are mass segregated. A useful way to
quantify the mass segregation was presented by  \cite{Converse2008} - they
plotted the number fraction enclosed at given radii $f_N$ against the
enclosed mass fraction at those radii ($f_M$).  For mass segregated clusters
(e.g. the Pleiades cluster in their paper)  $f_M$ will lie above $f_N$ (an
example of this procedure for our simulation A2 is given in fig. \ref{fig_gini}). 
A measure of the mass segregation can then be given by calculating the area
between the graph of $f_M$ and line of equality, which Converse \& Stahler
labeled the Gini coefficient $G$ for the cluster. 

We have plotted $G$ for various runs as a function of time in figure
\ref{fig_gt}. Looking at run A5 (solid red line) we  see that in the
beginning $G$ rises quite quickly, within 5 Myr; after that $G$ is more or
less constant, $G\approx 0.2-0.3$. At the end of the  simulations the
cluster is hence in a mass segregated state. A similar picture emerges for
run A6 (higher SFE, dotted line in Fig.~\ref{fig_gt}) and for the higher
feedback run (A2, dashed line). In the latter case $G$ plateaus at a
somewhat lower value. The other two drawn lines in figure~\ref{fig_gt} show
the runs with a higher and lower initial half mass radius, runs A10 and A8.
As is expected in these cases the initial rise  in $G$ is slower (A10) and
faster (A8) respectively. In all cases the final $G$ settles around
$G\approx 0.2$.

It is  interesting to note that \cite{Converse2010} tried to model the
Pleiades  cluster in some detail by running pure N-body models, and they
could not  explain the observed $G$ unless they started their simulations
with a nonzero Gini coefficient of at least $G\approx 0.2$. Although our
models do not  try to specifically reproduce the Pleiades cluster, this
seems consistent with Figure \ref{fig_gt}: \cite{Converse2010} started their
simulation after the dispersion of gas and our models suggest that the
nonzero $G$ is a remnant of the fast evolution before gas dispersal. After
gas dispersal the cluster is left in a much more dispersed state (the models
of \cite{Converse2010} had virial radii $\ge 3$ pc), with a $G$ close to its
present value. This suggests that the mass segregation in open clusters such
as the Pleiades is not a result of the secular evolution of its present
configuration, but a remnant of its embedded evolution stage.

The fact that mass segregation in our simulations occurs early in the 
embedded stage of the cluster evolution also has a consequence for the 
stellar population of the remnant cluster. In Figure~\ref{fig_cbf} we plot 
the number of stars heavier than a given mass which end up in the remnant 
cluster (i.e. are bound) as a function of star mass. The panels in 
figure~\ref{fig_cbf} show that in most cases where there results a bound 
cluster, the heavier
stars are preferentially retained (11 out of 15 runs show this pattern).
A number of models do not show this pattern, especially A7 and A8. Note these 
clusters have drastically expanded. Note also that while the A7 and A8 runs 
do not show strong preferential retention of heavy stars, they do show mass segregation.
It was previously noted \citep{Boily2003} that simple virial theorem 
arguments underestimate the size of clusters after gas clearing, because 
there are always a number of stars in low velocity orbits.  
The preferential retention of heavy stars can be understood by noting 
that the mass segregation during the initial evolutionary stages happens
because of the partitioning of stars in equal energy orbits: the heavy stars 
end up in low velocity orbits that are much more difficult to unbind. 
The heavy stars that are more easily retained may skew the IMFs that are 
derived from  intermediate age clusters (as long as they are not observed in 
the initial stage) and this may result in an observable signature. Note that 
this could be tested for the Pleiades cluster, which is already known to
have an excess of heavy stars in the center \citep[e.g.]{Converse2010}.

\subsection{Ratio of age and crossing time}

\begin{figure*}
\centering
\includegraphics[ width=0.32\textwidth]{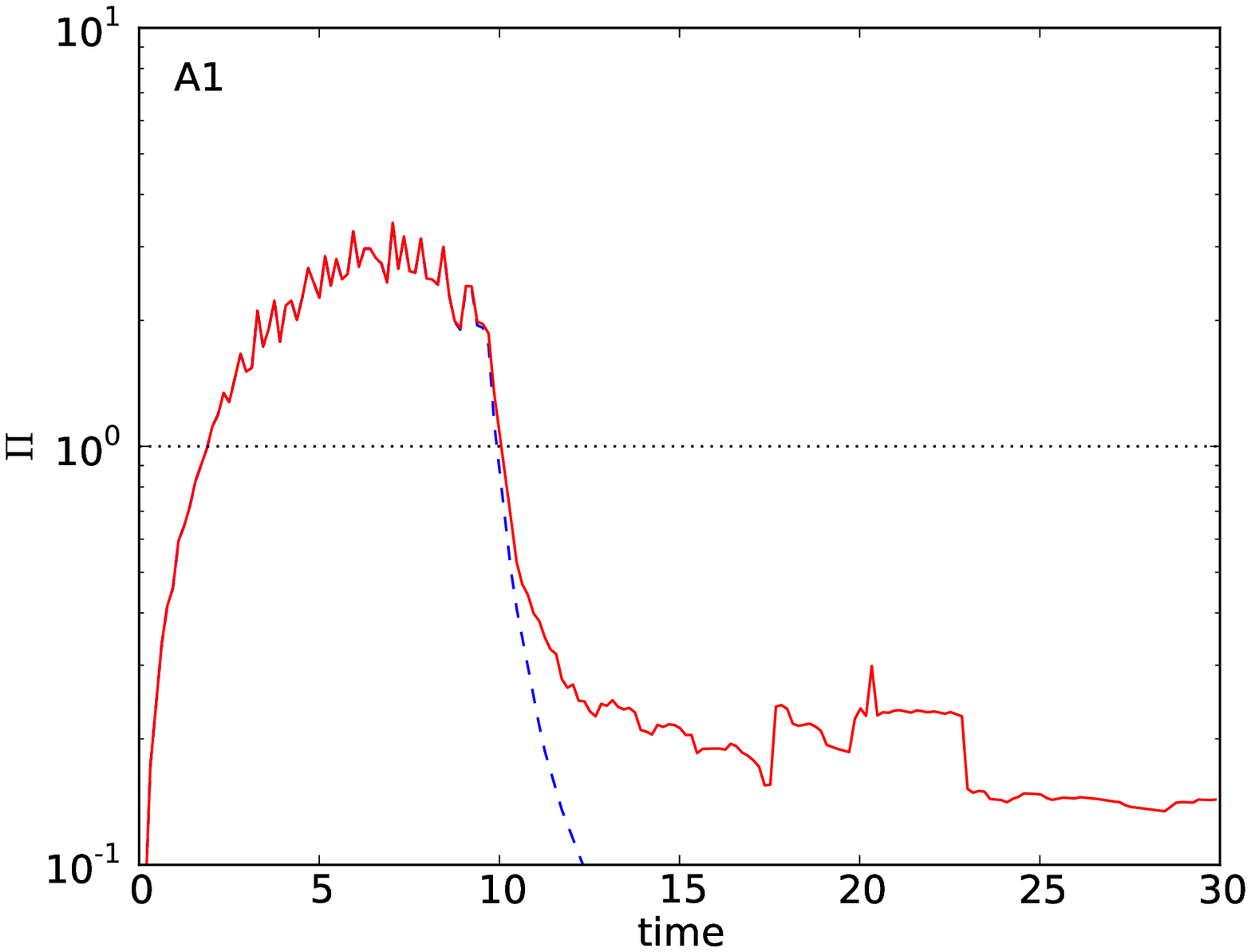}
\includegraphics[ width=0.32\textwidth]{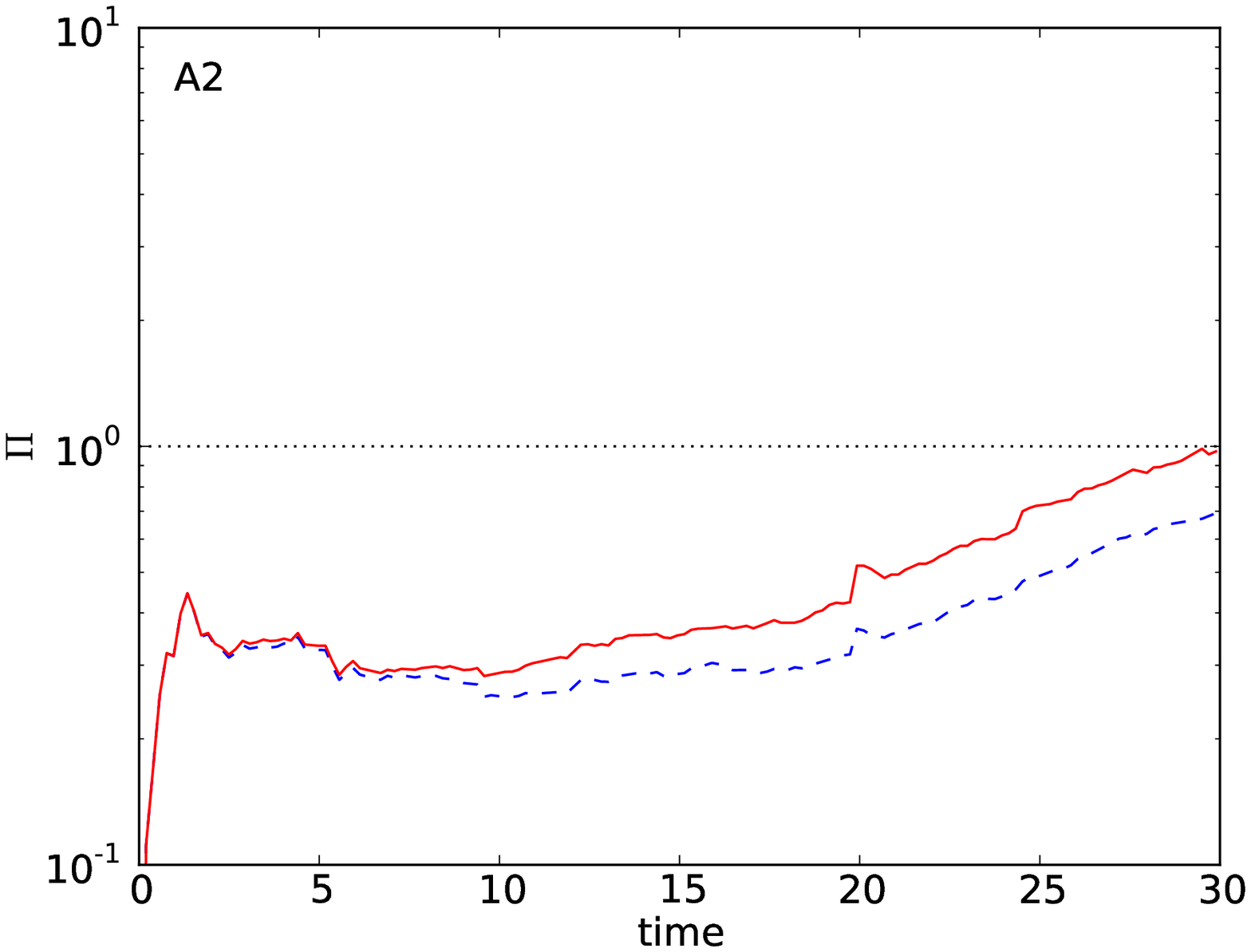}
\includegraphics[ width=0.32\textwidth]{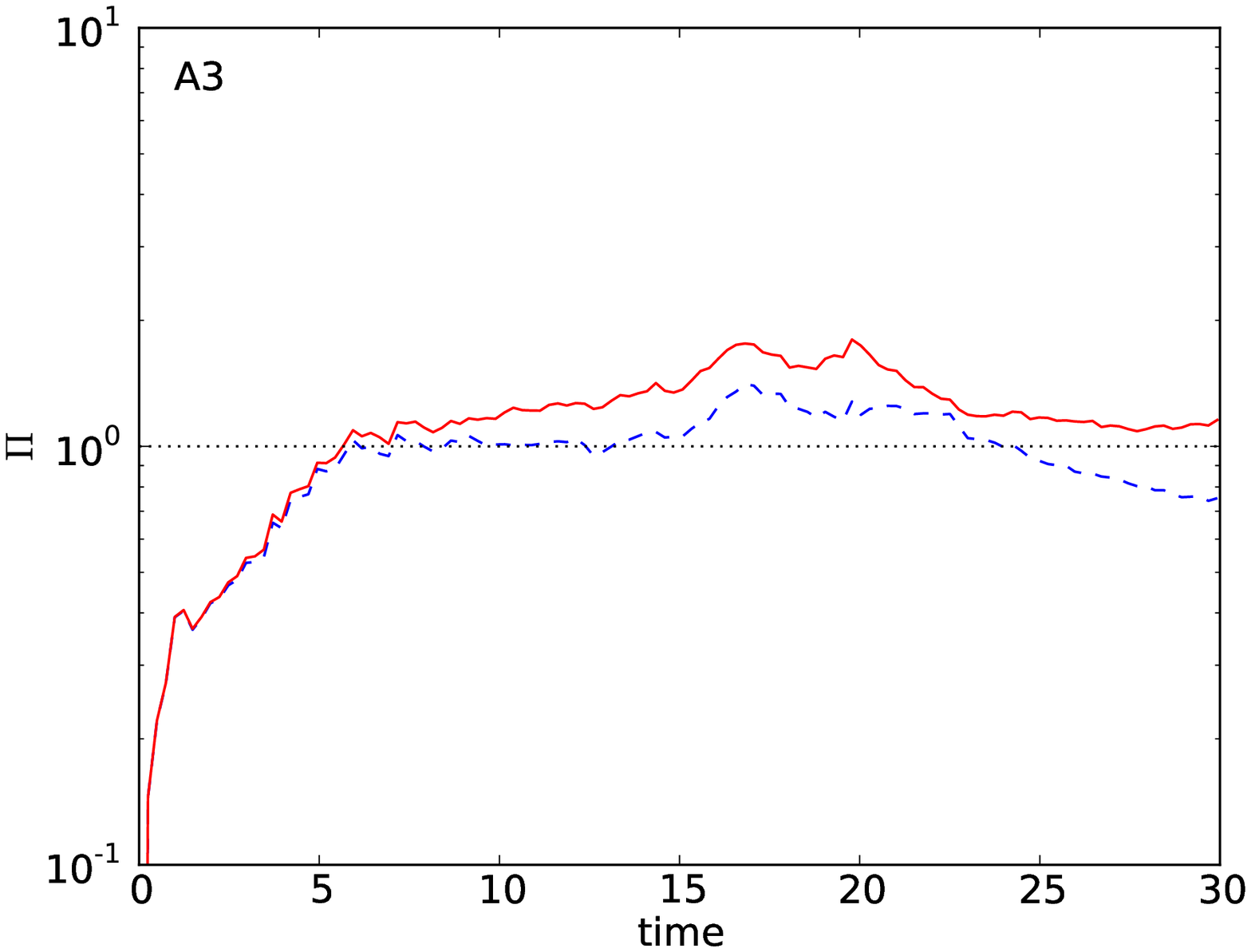}
\includegraphics[ width=0.32\textwidth]{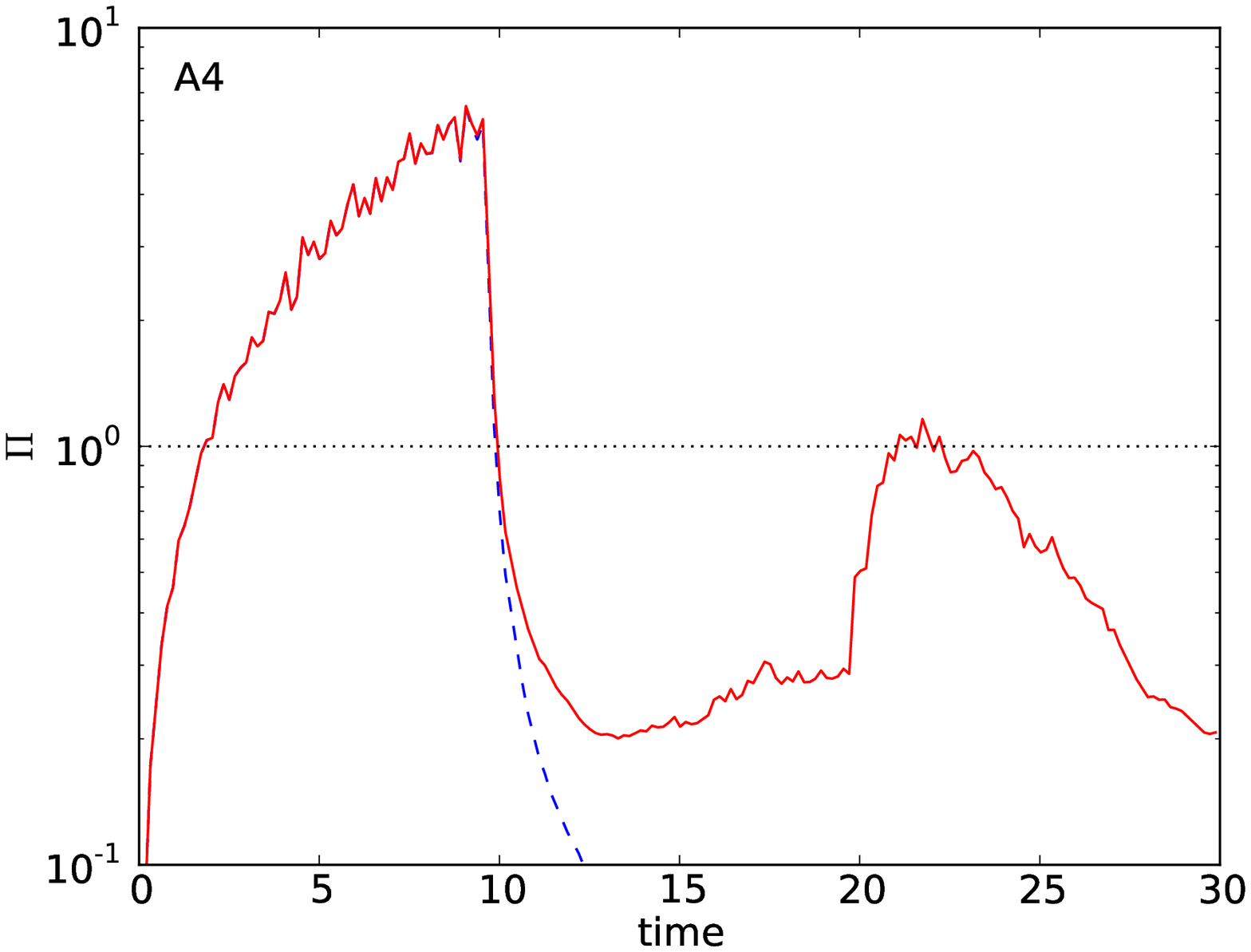}
\includegraphics[ width=0.32\textwidth]{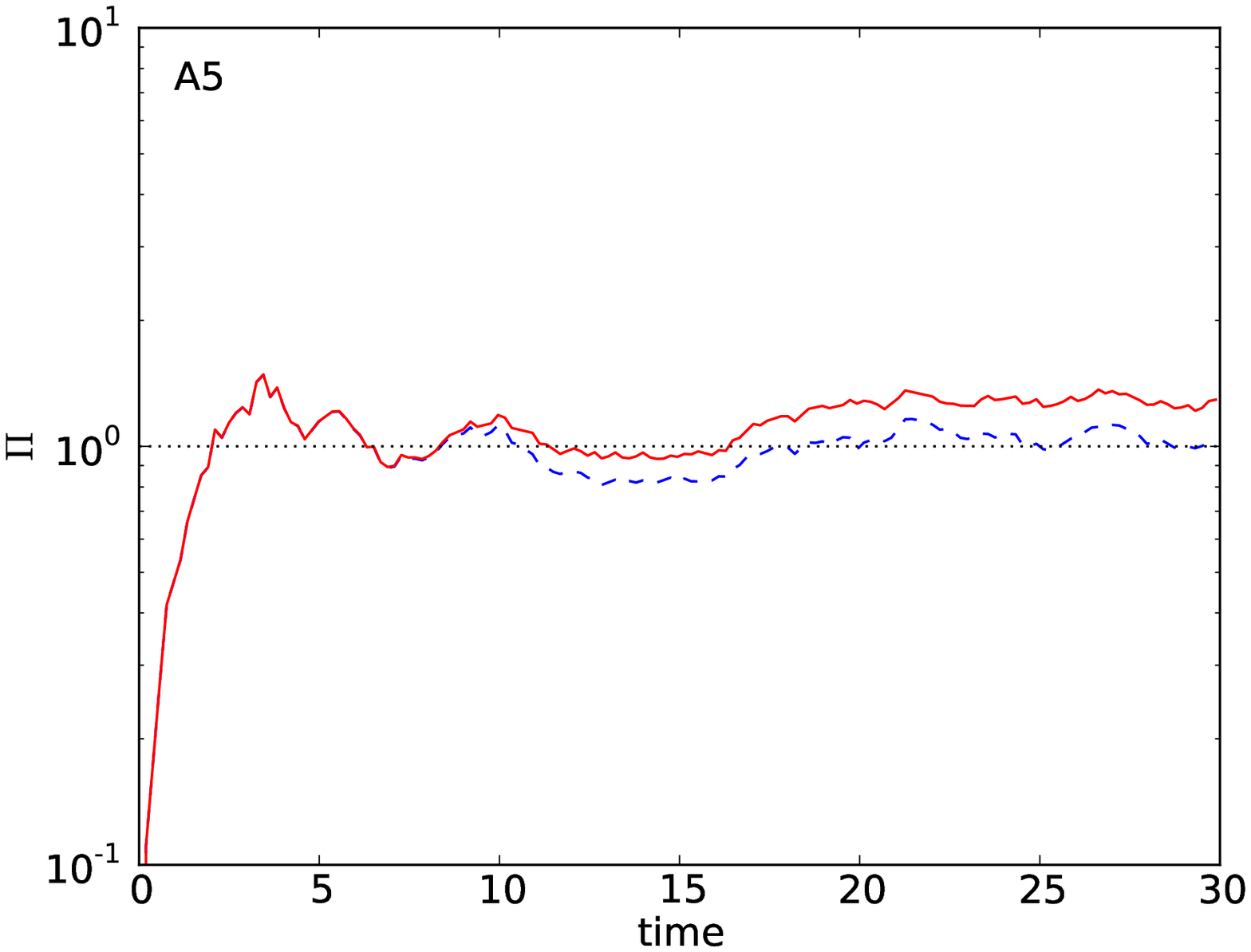}
\includegraphics[ width=0.32\textwidth]{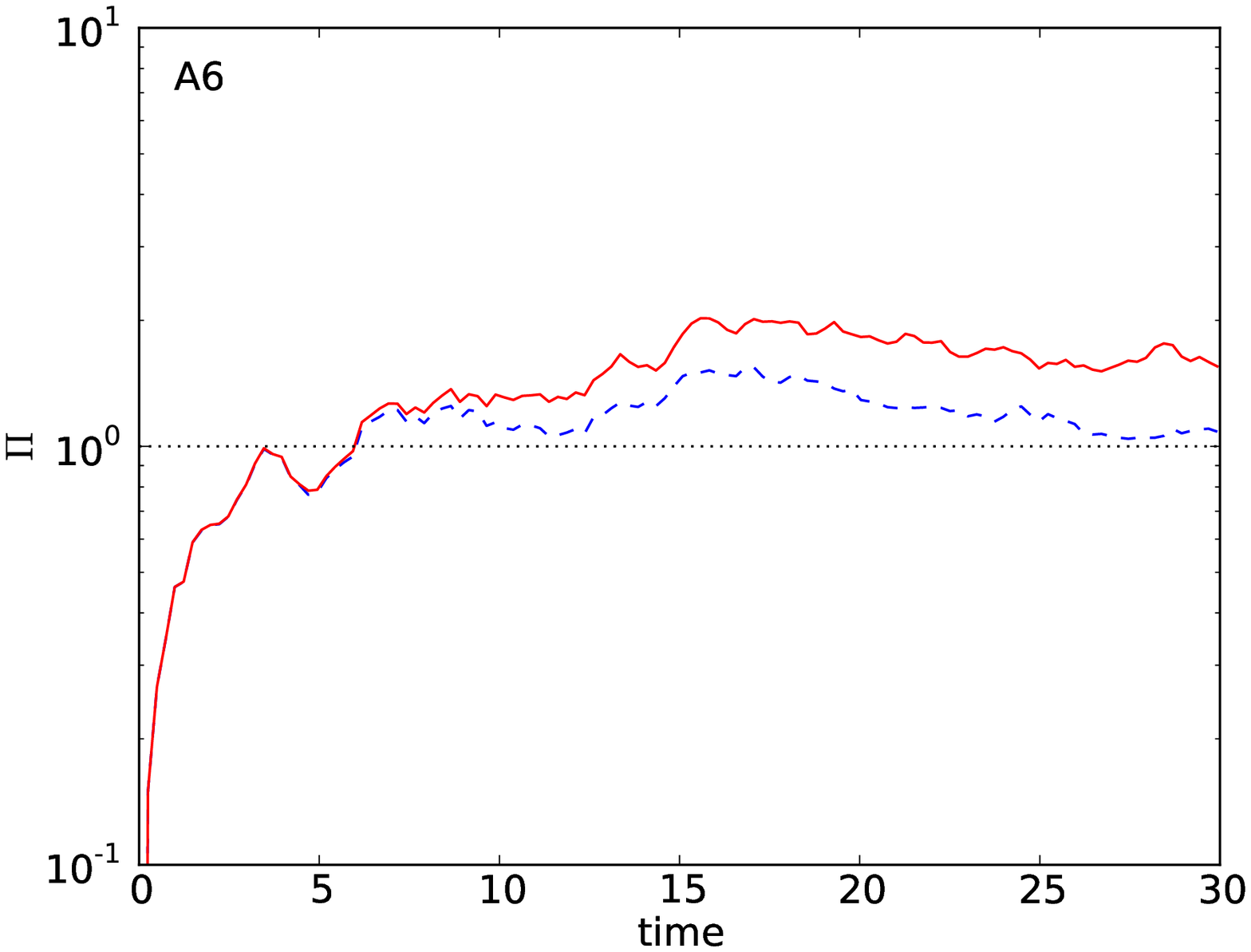}

\caption{
Plot of $\Pi={\rm age}/ T_{\rm cr}$ for models A1-A6. $T_{\rm cr}$ is determined as in
Eq.~\ref{eq_pi} where $T_{\rm cr}$ depends on 
$R_{\rm eff}$ and $M$ \citep{Gieles2011}. Two different cases are plotted: taking into 
account bound stars only (red line) or alternatively including all the stars 
(blue dashed lines). }
\label{fig_pi}
\end{figure*}

Loosely bound associations can be distinguished from `true' stellar clusters 
using limited observables by considering the ratio $\Pi$ of age $T_{\rm cl}$ 
and the crossing time $T_{\rm cross}$ \citep{Gieles2011}:
\begin{equation}
\label{eq_pi}
\Pi \equiv T_{\rm cl}/T_{\rm cross}
\end{equation}
where the $T_{\rm cross}$ is derived from
\begin{equation}
\label{eq_tcross}
T_{\rm cross} \approx 10 \left( \frac{R_{\rm eff}^3}{G M} \right)^{1/2}
\end{equation}
where the effective radius $R_{\rm eff}$ and mass $M$ are easily estimated
observable quantities. We determined the time evolution of $\Pi$ using 
Eq.~\ref{eq_tcross} for our simulations by measuring the $R_{\rm eff}$ 
and $M$. The resulting $\Pi$ are plotted in figure~\ref{fig_pi}. Note that
we plot the $\Pi$ for the case where we identify the cluster as being 
composed of all stars initially present or alternatively where only the 
bound stars are taken into account. The evolution  of the $\Pi$ of the
clusters in Figure~\ref{fig_pi} can be understood as follows: in
the  beginning $\Pi$ increases linearly but as the cluster is disturbed the 
$\Pi$ ratio is ``reset,'' as the $R_{\rm eff}$ increases and the mass decreases.
The tracks of our clusters end mostly with either a $\Pi \approx 1$ (bound case)
or with a very low $\Pi$ in which case they should be dissolved (and no longer
recognized as clusters). This is consistent with a possible explanation for
the distribution of $\Pi$ as a function of mass noted by \citep{Gieles2011}.
They found that for young stellar associations both low and high $\Pi$
values are present. While older stellar associations were exclusively found
with  $\Pi \ge 1$ (and identified as proper clusters). 

\section{Discussion}
\label{sec_disc}

We performed simulations of relatively small clusters of 1000 stars
with gas. Our simulations self-consistently take the dynamical
evolution of the stars, their internal nuclear evolution and the
dynamics of the embedded gas and the gas liberated by stellar
evolution into account. Each of the physical domains, stellar
dynamics, stellar evolution and hydrodynamics, is resolved with
well tested numerical solvers which have been developed
independently. The coupling between the domain-specific solvers is 
realized with the AMUSE framework, which enables us to couple codes 
written in a wide variety of languages in a homogeneous, transparent 
and self-consistent fashion. The final code has the form of a relatively
simple Python script, which is general in its purpose of simulating 
embedded star clusters.

Our model clusters start with a mix of stars and gas in virial equilibrium
at a hypothetical moment we call $t=0$, at which all stars are on the
zero-age main-sequence. The distribution of gas and stars was taken
identical with a star-formation efficiency to control the relative mass
fraction of the gas with respect to the stars. The feedback strength is
parametrized using a parameter $f_{\rm fb}$. We constrain the initial mass
function to follow a Salpeter IMF with a minimum mass of $0.1\,\Msun$ and an
upper limit of $100\,\Msun$, but we fixed the maximum mass of a cluster star
to $22\,\Msun$ in order  to prevent random fluctuations from the
coincidental generation of a  more massive star, because it turns out that
the most massive star in our cluster dominates the out-gassing by its efficient
segregation to the cluster center, the strong stellar wind and the moment
the star experiences a supernova.

In our simulations a star formation efficiency SFE\,$\ge 0.05$ results in a
bound cluster, which is consistent with earlier simulations
\citep[e.g.][]{Baumgardt2007}.  However, a more detailed  analysis of our
simulation results reveals that the process is  considerably more
complicated. It turns out that the qualitative result with respect to final
bound fraction can be approximated reasonably well with the parametrization  of
the SFE, but that the entire process is much richer. 
\citet{Baumgardt2007} found that the most important parameters
affecting the final cluster were the SFE and the gas loss timescale.
Our simulations convincingly demonstrate that the simple parametrization in 
terms of the SFE and/or gas loss timescale is a gross oversimplification of 
the process of stellar-supported out-gassing of the young cluster.  
For example for the low feedback A4 run the initial supernova is not
enough to disperse all cluster gas. After the first  supernova  $15\%-20\%$
of the gas remains bound. From the standpoint of  the dynamics of the stars
this remnant gas increases the effective SFE. Hence 
the cluster initially relaxes (between 10-20 Myr) with a $\approx 50\%$ bound 
fraction. Comparing with
the  Baumgardt \& Kroupa results (Figure 1), we see that this is a
reasonable value if we consider their SFE$ =0.2$ curves: at these SFE a bound
fraction  of  $50\%$ is quite possible (even not accounting for any radial
dependence of the gas loss). Further supernova events blow away the
remaining gas, but do so only after the cluster has adapted itself
dynamically.  
On the other hand, the B1 run shows
if the gas loss occurs more gradually, the cluster can also adapt to gas
loss (albeit in our simulations not enough to give rise to a bound cluster).

The time scale on which the gas is expelled cannot be described as a 
simple function, but depends on the distribution of the masses and positions 
of the stars, when they start blowing a strong stellar wind and the details
regarding the moment of the supernovae (and the adopted feedback strength).
Considering the wealth of variation in cluster parameters that result 
from this complex process the question remains why young star clusters 
show a relatively narrow range of properties, whereas
we would naively expect a much wider range of cluster radii, and
masses. At this moment we are unaware what key parameter may play a
moderating role in the range of observables for young star clusters. Note
that we did not take into account external limitations, such as the galactic
tidal fields or the environment of the parent cloud.

In our simulations we do not model the stellar feedback fully self-consistently.
The main uncertainties in our prescription  concern the uncertainty in the
efficiency of the coupling of the mechanical luminosity, the constraints of 
the numerical method and the lack of radiative feedback in our models.  In
future simulations we expect to deal with this latter point by resolving 
the radiative feedback using a radiative transfer code,
but for now this remains part of the uncertainty encoded in the  free
parameter for the feedback strength we introduced in our models. This
efficiency with which the output of stars couples to  the surrounding gas
may also depend on the mass  of the star itself \citep{Freyer2003,
Freyer2006}, and on the state of the gas: the density, temperature 
but also on its clumpiness (an aspect we have not modeled).  We
do demonstrate, however, that the core radii and mass of the surviving
cluster does not depend strongly on this efficiency parameter. 

 On the numerical side, it is well known that SPH shows relatively poor 
performance in the regime of explosive shock waves: for example the peak 
densities are generally smoothed out. This could be important for the 
energy budget of the feedback, as the cooling happens mainly in these dense 
regions. On the other hand, the speed and post-shock states are well
reproduced  \citep{Springel2005, Pelupessy2005} in SPH, so with the
limitations we imposed  on the details of the thermal evolution (the fact
that we did not try to account self-consistently for the energy budget of
the feedback) this has probably not a major impact. In future work, AMR
employing Godunov-type solvers for the gas dynamics may be employed to
improve this aspect.  This would necessitate very high resolution
simulations to reach a converged solution and the inclusion of much more
physics to follow the (non-equilibrium)  heating and cooling processes in
the gas.  

\section{Conclusions}
\label{sec_concl}

\begin{itemize}

\item  We find that a SFE\,$>0.05$ is necessary to result in bound clusters.
\item  The actual dispersion of clusters at SFE\,$=0.05$ may occur gradual 
or sudden depending on the details of feedback. Before eventual dissolution, 
stable phases in the cluster evolution may occur if not all gas is expelled.
Contributing to this is the increase in the effective SFE from 
the point of view of the stellar dynamics if feedback does not blow away all 
gas.  
\item We find that statistical variations can have a 
considerable effect on the internal structure of the remnant cluster.
\item We find different modes of gas loss depending on the feedback coupling
strength adopted. A high coupling strength results in a gradual loss of gas
on timescales of $\approx 5$ Myr. For the low feedback coupling stellar 
winds are not able to disperse the gas: the timescale of gas loss in this 
case is very short (due to supernova explosions).
\item More detailed modeling is necessary to put constraints on the energy
budget and thermal evolution of the gas, especially for the early phases
(where we do not take into account radiative feedback).
\item During the embedded phase mass 
segregation can develop which persists after gas has been removed and the
cluster has become less dense. This suggests that the observed mass
segregation of open clusters such as the Pleiades is a remnant rather than
the result of later secular evolution. 
\item Mass segregation followed by partial dissolution of the cluster tends
to lead to an excess of heavy stars in the cluster. This is due to the fact 
that low velocity orbits are preferentially retained and the energy 
equipartition that forms the basis of mass segregation means that heavier 
stars preferentially populate the low velocity orbits. 
\item The dependence of the observed distribution of the ratio of the age of
an association and its crossing time ($\Pi$) is consistent with the 
``resetting of the $\Pi$ clock'' we find in our models. The short
crossing times of young dense clusters increase due to stirring effect of
mass loss. Unbound clusters are seen to have low $\Pi$ values. These two 
effects together mean that clusters with both $\Pi<1$ and $\Pi>1$ are 
expected to be present as long as the gas dispersal and mass loss of young 
stars happens (before 30 Myr). This is consistent with what is observed 
(although more precise modeling should determine whether the observed 
distributions can be matched).  
\end{itemize}

The fact that we find mass segregation and the preferential retention of
high mass stars has consequences for the interpretation of cluster
observations. In fact, observed clusters like the Pleiades show mass
segregation. This mass segregation is usually attributed to the secular
evolution in the pure N-body regime after the short lived embedded phase. For
the Pleiades primordial mass segregation is necessary
\citep{Converse2008}. Our models are able to reproduce the expected
amount of initial mass segregation for the Pleiades cluster  (although our
models are not specifically build to reproduce the Pleaides). The fact that
high mass stars are better able to withstand the disruptive effect of the
loss of the natal cloud may lead to an excess number of unbound low mass stars
in the vicinity of young clusters, which may be testable by expanding the 
search for cluster members to include kinematically unbound stars of the
right age. We have demonstrated that it is important to consider
both the embedded phase evolution and the gas clearing phase of cluster
evolution together. Future studies will aim to include radiative feedback
effects and should also take into account more realistic initial conditions
for the cluster and gas distribution. 

{\bf Acknowledgements}
This work was supported by the Netherlands Research Council NWO 
(grants \#643.200.503, \#639.073.803 and \#614.061.608) and by 
the Netherlands Research School for Astronomy (NOVA).

\bibliographystyle{mn2e}

\end{document}